\let\ifarxiv=\iftrue     
\documentclass[12pt,a4paper]{article}

\ifarxiv\ifnum\pdfoutput=1\else
\PassOptionsToPackage{hypertex}{hyperref}
\PassOptionsToPackage{draft}{graphicx}
\fi\fi

\usepackage{amsmath,amssymb}
\usepackage[bookmarks=true,hyperfigures=true]{hyperref}
\usepackage{graphicx}
\usepackage[nosort]{cite}
\usepackage{feynmp}
\usepackage{multirow}
\usepackage{listings}
\usepackage{longtable}
\usepackage{multicol}
\usepackage{hhline}
\graphicspath{{figures/}}
\usepackage{cleveref}
\usepackage{color}
\usepackage[tight]{subfigure}
\usepackage{bm}

\usepackage[a4paper,text={173mm,216mm},centering]{geometry}

\allowdisplaybreaks[3]

\numberwithin{equation}{section}

\usepackage[font=small,labelfont=bf,width=0.85\textwidth]{caption}

\makeatletter
\let\old@startsection=\@startsection
\renewcommand{\@startsection}[6]{\old@startsection{#1}{#2}{#3}{#4}{#5}{#6\mathversion{bold}}}
\makeatother

\makeatletter
\newlength{\apb@width}
\newcommand{\autoparbox}[2][c]{\settowidth{\apb@width}{#2}\parbox[#1]{\apb@width}{#2}}

\makeatother

\let\oldPhi=\Phi
\let\oldPsi=\Psi
\let\oldGamma=\Gamma
\let\oldDelta=\Delta
\let\oldSigma=\Sigma
\let\oldLambda=\Lambda
\let\oldTheta=\Theta
\let\oldPi=\Pi
\let\oldXi=\Xi
\let\oldUpsilon=\Upsilon
\let\oldOmega=\Omega
\renewcommand{\Phi}{\mathnormal{\oldPhi}}
\renewcommand{\Psi}{\mathnormal{\oldPsi}}
\renewcommand{\Gamma}{\mathnormal{\oldGamma}}
\renewcommand{\Sigma}{\mathnormal{\oldSigma}}
\renewcommand{\Delta}{\mathnormal{\oldDelta}}
\renewcommand{\Theta}{\mathnormal{\oldTheta}}
\renewcommand{\Lambda}{\mathnormal{\oldLambda}}
\renewcommand{\Pi}{\mathnormal{\oldPi}}
\renewcommand{\Xi}{\mathnormal{\oldXi}}
\renewcommand{\Upsilon}{\mathnormal{\oldUpsilon}}
\renewcommand{\Omega}{\mathnormal{\oldOmega}}

\ifx\genfrac\sdflkaj\else\fi
\newcommand{\sfrac}[2]{{\textstyle\frac{#1}{#2}}}

\newcommand{\be}{\begin{equation}}
\newcommand{\ee}{\end{equation}}

\newcommand{\Tr}{\mathop{\mathrm{Tr}}}

\newcommand{\defi}{\mathrel{\raisebox{0.012cm}{:}\hspace{-0.11cm}=}}
\newenvironment{rcase}{\left.\begin{aligned}}{\end{aligned}\qquad\right\rbrace}

\makeatletter
\def\mr@ignsp#1 {\ifx\:#1\@empty\else #1\expandafter\mr@ignsp\fi}%
\newcommand{\multiref}[1]{\begingroup
\xdef\mr@no@sparg{\expandafter\mr@ignsp#1 \: }%
\def\mr@comma{}%
\@for\mr@refs:=\mr@no@sparg\do{\mr@comma\def\mr@comma{,}\ref{\mr@refs}}%
\endgroup}
\makeatother

\ifx\href\asklfhas\newcommand{\href}[2]{#2}\fi

\newcommand{\hypref}[2]{\ifx\href\asklfhas #2\else\href{#1}{#2}\fi}

\renewcommand{\eqref}[1]{(\multiref{#1})}

\newcommand{\1}{\ensuremath{\mathds{1}}}

\newcommand{\indexfett}[2][f]{\if#1f{\index{#2|textbf}}\else{\index{#2}}\fi}

\newcommand{\qb}{\bar{q}}
\renewcommand{\r}{\text{\begin{small}$\mathcal{R}$\end{small}}}

\ifx\hypersetup\sadfkjashdfkxja\newcommand\hypersetup[1]{}\newcommand{\texorpdfstring}[2]{#1}\fi

\hypersetup{plainpages=false}
\hypersetup{pdfpagemode=UseNone}
\hypersetup{bookmarksnumbered=true}
\hypersetup{pdfstartview=FitH}
\hypersetup{colorlinks=false}
\hypersetup{citebordercolor={.5 1 .5}}
\hypersetup{urlbordercolor={.5 1 1}}
\hypersetup{linkbordercolor={1 .7 .7}}

\renewcommand{\=}{\mathrel{\phantom{=}}}

\newcommand{\+}{\negthinspace+\negthinspace}

\allowdisplaybreaks

\begin{document}
\thispagestyle{empty}

\begin{flushright}
HU-EP-10/57
\end{flushright}

\begingroup\centering
{\Large\bfseries\mathversion{bold}
Color ordering in QCD\par}%
\hypersetup{pdfsubject={}}%
\hypersetup{pdfkeywords={}}%
\ifarxiv\vspace{15mm}\else\vspace{15mm}\fi

\begingroup\scshape\large 
Theodor Schuster
\endgroup
\vspace{5mm}

\begingroup\ifarxiv\small\fi
\textit{Institut f\"ur Physik, Humboldt-Universit\"at zu Berlin, \\
Newtonstra{\ss}e 15, D-12489 Berlin, Germany}\\[0.2cm]
\ifarxiv\texttt{theodor.schuster@physik.hu-berlin.de\phantom{\ldots}}\fi
\endgroup
\vspace{1cm}

\textbf{Abstract}\vspace{5mm}\par
\begin{minipage}{14.7cm}
We derive color decompositions of arbitrary tree and one-loop QCD  amplitudes into color ordered objects called primitive amplitudes. Furthermore, we derive general fermion flip and reversion identities spanning the null space among the primitive amplitudes and use them to prove that all color ordered tree amplitudes of massless QCD can be written as linear combinations of  color ordered tree amplitudes of $\mathcal{N}=4$ super Yang-Mills theory.
\end{minipage}\par
\endgroup
\newpage


\setcounter{tocdepth}{2}
\hrule height 0.75pt
\tableofcontents
\vspace{0.8cm}
\hrule height 0.75pt
\vspace{1cm}

\setcounter{tocdepth}{2}


\section{Introduction} 

A precise theoretical understanding of the Standard Model contributions to the multiple jet events observed at the Large Hadron Collider (LHC) is crucial for the discrimination of new physics as well as for precision measurements of Standard Model parameters. Since the scattering processes observed at the LHC are dominated by strong interactions, perturbative QCD is of major interest. Calculating high multiplicity next to leading order (NLO) QCD scattering amplitudes, the technique of color decomposition \cite{Berends:1987cv,Mangano:1988kk,Bern:1990ux,Bern:1994zx,Bern:1994fz} has become an essential tool.  Color decomposition provides a systematic way to treat the color degrees of freedom in a scattering process by separating them from the kinematical parts, called partial amplitudes. Since the color structures appearing in a certain amplitude are straight forward to  identify, the nontrivial part of the color decomposition of an amplitude is to express the partial amplitudes in terms of gauge invariant, color ordered objects called primitive amplitudes. The fixed ordering of the external legs, results in a simpler analytical structure of the primitive amplitudes. The development of powerful non-diagrammatic methods for their computation, such as the BCFW on-shell recursion
relation~\cite{Britto:2004ap,Britto:2005fq} at tree level and the generalized
unitarity techniques at one-loop level \cite{Bern:1994zx,Bern:1994cg,Anastasiou:2003kj} (see \cite{Bern:2007dw} for a review) has been key to the recent progress in calculating NLO QCD corrections. State-of-the-art are the recent computations of NLO QCD corrections to the four \cite{Bern:2011ep,Badger:2012pf}, and five jet production \cite{Badger:2013yda}.

Decompositions of QCD amplitudes into primitive amplitudes are known for all one-loop amplitudes with up to one quark--anti-quark pair \cite{Bern:1990ux,Bern:1994fz}. To obtain decompositions for amplitudes with more than one quark line a diagram based algorithm has been developed in references \cite{Ellis:2008qc,Ellis:2011cr,Ita:2011ar,Badger:2012pg} which involves performing the color decomposition of a sufficiently large set of Feynman diagrams for quarks transforming in the fundamental as well as for quarks in the adjoint representation of the gauge group. The resulting linear equations can be solved for the partial amplitudes, leading to the desired decomposition into primitive amplitudes. If the algorithm incorporates the Furry relations between different diagrams, as has been done in \cite{Badger:2012pg}, it delivers all diagram based identities among the primitive amplitudes as well.

The aim of this paper is to provide an alternative to the diagram based algorithm by directly constructing the decomposition of arbitrary tree-level and one-loop QCD amplitudes. Our construction is purely combinatorial and gives insight into the structure of the partial amplitudes of QCD. Furthermore we thoroughly investigate the identities among the primitive amplitudes, finding general fermion flip and reversion identities which generalize the identities observed in \cite{Ita:2011ar} and are a consequence of the symmetries of the color ordered vertices. The derived identities span the null space among the primitive amplitudes and allow to analytically compare different expression containing primitive amplitudes. Accompanying the paper is a freely available {\tt Mathematica} package {\tt QCDcolor} which contains implementations of the color decompositions as well as the flip and reversion identities.

As a nontrivial application of the fermion flip identities we use them to prove the conjecture made in \cite{Dixon:2010ik} that all color ordered tree level amplitudes of massless QCD can be obtained from color ordered tree level amplitudes of $\mathcal{N}=4$ super Yang-Mills theory.

The paper is organized as follows. In the remainder of this section we give a pedagogical introduction to the technique of color decomposition before deriving the tree-level and one-loop color decomposition of QCD in \cref{section:QCDcolorTree,section:QCDcolorLoop}. In \cref{section:SymPrim} we derive the general fermion flip and reversion identities of the primitive amplitudes which we exploit in \cref{section:checks} to perform analytical checks of the derived color decompositions against known results. In \cref{section:FromN=4toQCD} we prove that all color ordered tree-level amplitudes of massless QCD can be obtained from color ordered tree-level amplitudes of $\mathcal{N}=4$ super Yang-Mills theory. 
Finally, \cref{appendix:QCDcolor} is
devoted to a documentation of the {\tt Mathematica} package {\tt QCDcolor}. 
\subsection{The general idea of color decompositions}\label{section:ColorDecomposition}
Within a brute force diagrammatic calculation of a gauge theory scattering amplitude the gauge dependence of the individual Feynman diagrams leads to gauge redundancies. As a consequence a diagrammatic calculation is ill suited to obtain compact analytical expressions for the scattering amplitudes, which are gauge independent. One possibility to reduce the complexity within an amplitude calculation is to decompose the amplitudes into gauge independent pieces that allow for a non-diagrammatic calculation.  This can be accomplished by factoring the amplitudes into color structures $C_{n,k}$, depending solely on the gauge group and the representations of the gauge group the partons are transforming in, and the gauge independent partial amplitudes $P_{n,k}$ only depending on the flavors $f_i$, helicities $h_i$ and momenta $p_i$ of the scattered particles. To be more precise, at each loop order a scattering amplitude of some gauge theory can be written as
\begin{equation}\label{eq:defColorStructure}
 \mathcal{A}^{L\text{-loop}}_n(\{p_i,f_i,h_i,r_i\})= g^{\scriptscriptstyle n+2(L-1)}\sum_k C^{L\text{-loop}}_{n,k}(\{r_i\},N) P^{L\text{-loop}}_{n,k}(\{p_i,f_i,h_i\})\,,
\end{equation}
where $\{r_i\}$ are the gauge group indices of the partons. Hence, the calculation of a scattering amplitude boils down to the determination of the partial amplitudes. Depending on the gauge theory and particular color structure under consideration the partial amplitudes can have additional symmetries which they inherit from the color structures. Note that we adopt the convention to include all powers of $N$ and $\frac{-1}{N}$ as well as additional free parameters such as e.\,g.~the number of fermion flavors $n_f$ into the color structures, resulting in partial amplitudes independent of these parameters.

\begin{figure}
\begin{align*}
&\raisebox{-2.0cm}{\includegraphics[height=3.7cm]{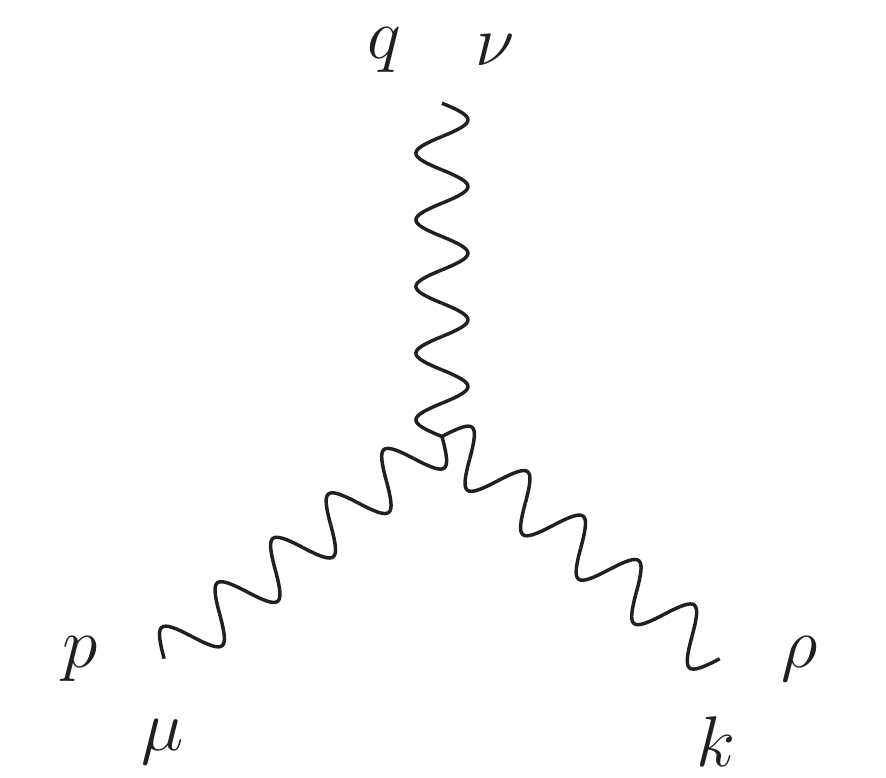}}=\begin{aligned}[t]&(p-q)^\rho \eta^{\mu\nu}\\+&(q-k)^\mu \eta^{\nu\rho}\\+&(k-p)^\nu \eta^{\mu\rho}\end{aligned}&\raisebox{-1.15cm}{\includegraphics[height=2.5cm]{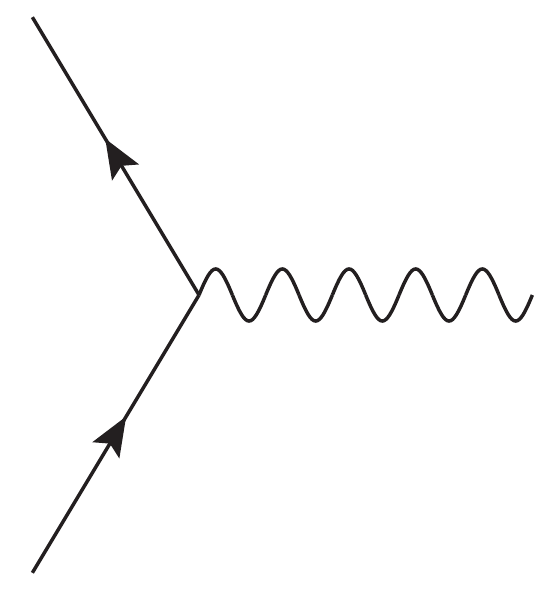}}&=\gamma^\mu\\
&	\raisebox{-1.6cm}{\includegraphics[height=3.5cm]{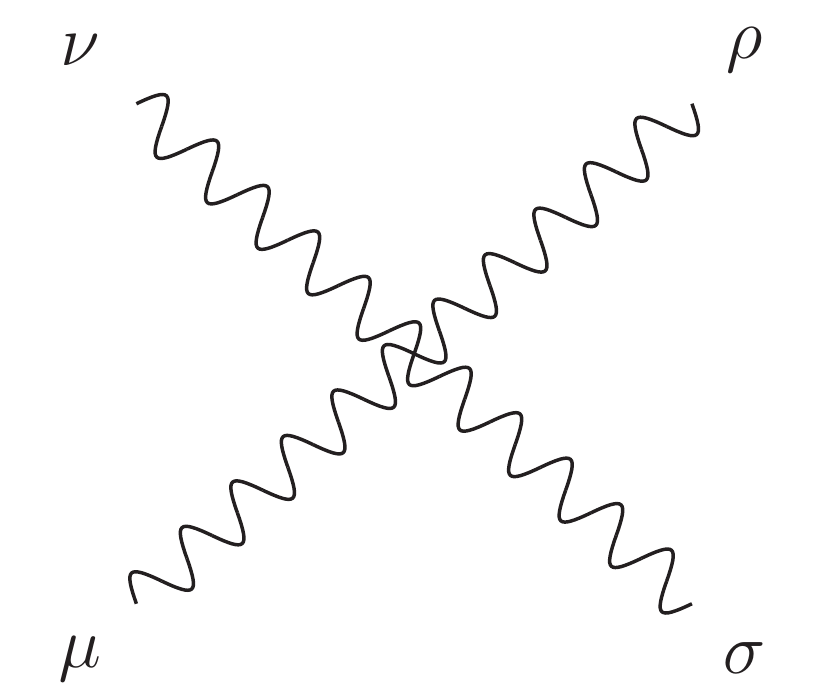}}=\begin{aligned}[t]2&\eta_{\mu\nu}\eta_{\rho\sigma}\\-&\eta_{\mu\rho}\eta_{\nu\sigma}\\-&\eta_{\mu\sigma}\eta_{\nu\rho}\end{aligned}&\raisebox{-1.15cm}{\includegraphics[height=2.5cm]{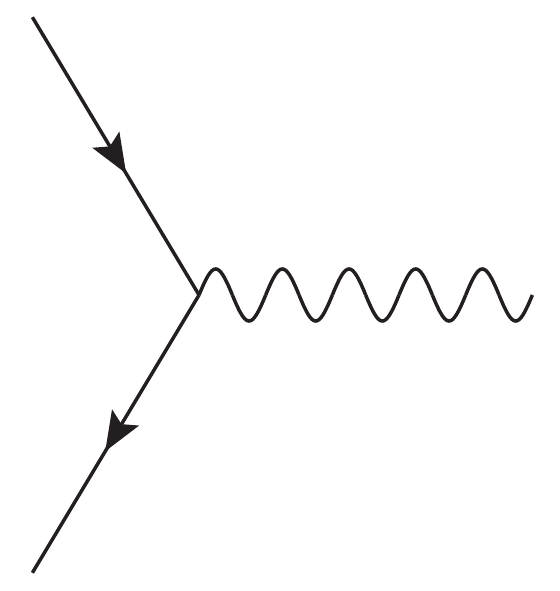}}&=-\gamma^\mu
\end{align*}
\caption{Color ordered Feynman rules for pure gluon and gluon-fermion vertices.}\label{fig:ColorOrderedRules}
\end{figure}

Of particular interest are the partial amplitudes appearing in the large $N$ limit of a $U(N)$ or $SU(N)$ gauge theory with only adjoint particles. The color decomposition of such a theory reads
\begin{equation}
{\cal A}^{L\text{-loop}}_{n}(\{p_{i},h_{i},f_i,a_{i}\})
 = g^{\scriptscriptstyle n+2(L-1)}N^{\scriptscriptstyle L}\!\!\left(\sum_{\sigma\in S_{n}/Z_{n}}\!\!
 \Tr(T^{a_{\sigma(1)}}\ldots T^{a_{\sigma(n)}})\, 
A^{L\text{-loop}}_{n}(\sigma)+\mathcal{O}(\sfrac{1}{N})\right)\!,
\label{eq:colorOrdered}
\end{equation}
and the leading color partial amplitudes $A^{L\text{-loop}}_{n}$ are called color ordered amplitudes. The color ordered amplitudes are specified by a particular ordering of the external legs. As will become clear in \cref{section:AdjointColor}, color ordered amplitudes can be calculated applying the color ordered Feynman rules of \cref{fig:ColorOrderedRules} to all planar diagrams with the particular ordering of the external legs.
The fixed ordering of the external particles leads to a simpler analytic structure, i.\,e.~tree level color ordered amplitudes only receive poles when sums of successive momenta go on-shell. Color ordered amplitudes are subject to a large number of identities, both at tree and at loop level. A detailed investigation of their symmetries can be found in \cref{section:SymPrim}. However, the most striking feature of the color ordered amplitudes is that they can be calculated in a non-diagrammatic way using on-shell methods, that is the BCFW recursion \cite{Britto:2004ap,Britto:2005fq} at tree level and generalized unitarity at loop level \cite{Bern:1994zx,Bern:1994cg,Anastasiou:2003kj}. By modifying all particles to transform in the adjoint representation of the gauge group we can properly associate color ordered amplitudes to every gauge theory, e.\,g.~the color ordered QCD amplitudes are the leading color partial amplitudes in a modified gauge theory where the quarks transform in the adjoint representation.

Given the simplicity of the color ordered amplitudes it is convenient to try to decompose partial amplitudes into linear combinations of color ordered amplitudes. 
However, relating the partial amplitudes to color ordered amplitudes is in general a nontrivial task.  In the case of QCD we are going to show, that it is indeed possible to construct an integer valued matrix $M^{\text{tree}}$ that is solving the equations
\begin{equation}\label{eq:colorMatrix}
 P^{\text{tree}}_{n,k}=\sum_{j}M^{\text{tree}}_{k\, j}\,A^{\text{tree}}_{n,j}\,.
\end{equation}
Besides the obvious computational advantages, such decompositions into color ordered tree amplitudes open up the possibility to relate different gauge theories. On the level of color ordered amplitudes the differences between gauge theories reduce to the helicities and flavors of the matter fields and the pure matter interactions present in the theories. All interactions with the gauge field $A_\mu$ are universal as they are induced by the covariant derivative $D_\mu=\partial_\mu-igA_\mu$. This universality results in a surprisingly large number of equivalent color ordered tree amplitudes between differing gauge theories.

In order to be able to write down similar expressions for all one-loop partial amplitudes of QCD we have to divide the one-loop color ordered QCD amplitudes involving quarks into smaller gauge invariant pieces, the primitive amplitudes. In a primitive amplitude each quark line has a definite orientation with respect to the loop, called the routing of the quark. A particular quark can either turn to the left or to the right of the loop. We specify the primitive amplitudes by assigning a routing label to each of the quarks and anti-quarks.
\begin{equation}\label{eq:primitives}
A(q_1,\ldots,\bar{q}_1,\dots)= A(q^L_1,\ldots,\bar{q}_1^L,\dots)+A(q^R_1,\ldots,\bar{q}_1^R,\dots)+\dots
\end{equation}
Speaking in terms of diagrams,  \cref{eq:primitives} is a disjoint decomposition of all contributing color ordered Feynman diagrams into gauge invariant subsets of all possible routings of the quarks. The gauge invariance of the primitive amplitudes follows from the observation that they can be interpreted as the partial amplitudes of some special gauge theory \cite{Bern:1994fz}. Finally, we have to further divide the primitive amplitudes into the part containing a pure fermion loop and the part with a pure gluon or mixed quark-gluon loop.

In summary, the goal of a tree-level or one-loop color decomposition of QCD is to give an expression for the amplitude in terms of color structures and color ordered amplitudes or primitive amplitudes. Hence, dropping the one-loop label we will derive expressions of the form  
\begin{equation}
 \mathcal{A}^{\text{tree}}_n=g^{n-2}\sum_k C^{\text{tree}}_{n,k}\sum_{j}M^{\text{tree}}_{k\, j}\,A^{\text{tree}}_{n,j}\,,
\end{equation}
and
\begin{equation}
 \mathcal{A}_n=g^{n}\sum_k C_{n,k}\left(\sum_{j}M_{k\, j}\,A_{n,j}+n_f\sum_{j}(M_f)_{k\, j}\,(A_f)_{n,j}\right)\,,
\end{equation}
for arbitrary tree and one-loop QCD amplitudes, with $A_{n,j}$ denoting the non-fermion loop part and $(A_f)_{n,j}$ denoting the fermion loop part of primitive amplitudes. As a consequence, an arbitrary leading order or next to leading order QCD calculation reduces to the calculation of color ordered tree amplitudes and primitive amplitudes.
\subsection{Color decomposition for adjoint particles}
Since it is instructive for the preceding derivation of the color decomposition of QCD and will yield further insight into the definition and calculation of color ordered amplitudes, we are going to give a derivation of the known results \cite{Bern:1990ux} for the color decomposition of scattering amplitudes containing only adjoint particles. Obviously this includes the pure gluon sector, common to all non Abelian gauge theories.
\subsubsection{Tree level}\label{section:AdjointColor}
At tree level there are no subleading color contributions and the color decomposition of an $n$-parton tree amplitude reads 
\begin{equation}
{\cal A}^{\text{tree}}_{n}(\{p_{i},h_{i},f_i,a_{i}\})
 = g^{n-2}\sum_{\sigma\in S_{n}/Z_{n}}
 \Tr(T^{a_{\sigma(1)}}\ldots T^{a_{\sigma(n)}})\, 
A^{\text{tree}}_{n}(\sigma(1),\ldots ,\sigma(n))\, ,
\label{eq:colororderingGluons}
\end{equation}
with the argument $i$ of the color ordered amplitude $A^{\text{tree}}_{n}$ 
denoting an outgoing parton of light-like momentum $p_{i}$, helicity $h_i$ and flavor $f_i$, $i\in[1,n]$. The $su(N)$ traceless Hermitian generator matrices
$T^{a_i}$ are in the fundamental representation, and are normalized such
that ${\rm Tr}(T^a T^b) = \delta^{ab}$.

It is straight forward to deduce this decomposition by determining the color structures that can appear in a Feynman diagram. This can be accomplished making use of the Lie algebra the generator matrices are fulfilling
\begin{equation}\label{eq:LieAlgebra}
 [T^a,T^b]=if^{abc}T^c\,.
\end{equation}
Given an arbitrary Feynman diagram, we choose one external particle. Its adjoint index $a$ appears in the structure constants of the vertex it is connected to. We replace these structure constants by
\begin{equation}
 f^{abc}=-i\Tr(T^a [T^b,T^c])\,.
\end{equation}
The other adjoint indices in these traces are either belonging to other external legs or are contracted with a structure constant of another vertex. With the help of \cref{eq:LieAlgebra} these contractions can be replaced by commutators $f^{dec}T^c=-i[T^d,T^e]$. Continuing this step we end up with a trace of nested commutators of the $n$ generator matrices associated to the external legs. Expanding the commutators we end up with the desired decomposition of the diagram, thereby proving \cref{eq:colororderingGluons}.
\subsubsection{Color ordered diagrams and color ordered Feynman rules}
Form the above analysis it follows immediately that there is a one to one correspondence between the kinematical contributions of the Feynman diagrams to the color ordered amplitudes and all the planar diagrams whose external legs follow the ordering specified by the color ordered amplitude. The kinematical contributions associated to these color ordered diagrams can be calculated by applying the color ordered Feynman rules listed in \cref{fig:ColorOrderedRules} to them. In general, the color ordered Feynman rules of a particular gauge theory can be obtained by expressing the contractions of structure constants present in the ordinary Feynman rules of the associated gauge theory with only adjoint fields by sums of traces of generators. The coefficients of these traces are the color ordered Feynman rules for the cyclic ordering specified by the traces. As a consequence, the color ordered fermion-gluon vertex is antisymmetric with respect to the ordering of the legs.
\subsubsection{One-loop level}
At one-loop level there are subleading double trace color structures as well as the leading color single trace color structures we encountered already at tree level. The decomposition of a one-loop amplitude of $n$ adjoint particles into color structures and color ordered amplitudes $A(\sigma(1),\dots,\sigma(n))$ reads
\begin{align}
{\cal A}_{n}(\{p_{i},h_{i},f_i,a_{i}\})
 &= g^{n}\Biggl(\sum_{\sigma\in S_{n}/Z_{n}}
 N \Tr(T^{a_{\sigma(1)}}\ldots T^{a_{\sigma(n)}})\, 
A(\sigma)\\
&\=\phantom{g^{n}\Biggl(}+\sum_{k=2}^{\left[\sfrac{n}{2}\right]}\sum_{\sigma\in S_{n,k}}
 C_{n,k}(\sigma)\, 
P_{n,k}(\sigma)\biggr)\, ,
\label{eq:colorDecAdjLoop}
\end{align}
with the subleading color structures $C_{n,k}$ and subleading partial amplitudes $P_{n,k}$ being given by
\begin{equation}\label{eq:doubleTrace}
  C_{n,k}(\sigma)=\Tr(T^{a_{\sigma(1)}}\ldots T^{a_{\sigma(k)}})\Tr(T^{a_{\sigma(k+1)}}\ldots T^{a_{\sigma(n)}})\,,
\end{equation}
and
\begin{equation}\label{eq:PartialAmpAdjoint}
 P_{n,k}(\sigma)=(-1)^{k}\sum_{\tau\in \text{COP}_{k}(\sigma)}A(\tau)\,.
\end{equation}
The permutations $S_{n,k}$ are defined as the permutation group $S_n$ modulo the symmetry group of the double traces $C_{n,k}$ which is isomorphic to $Z_k\times Z_{n-k}$. Hence, $S_{n,k}\simeq S_n/(Z_k\times Z_{n-k}$). The cyclically ordered permutations $\text{COP}_{k}(\sigma)$ are defined as
\begin{equation}
 \text{COP}_{k}(\sigma)\defi\text{COP}\{\sigma(k),\dots,\sigma(1)\}\{\sigma(k+1),\dots,\sigma(n)\}\;,
\end{equation}
where $\text{COP}\{\alpha_1,\dots,\alpha_a\}\{\beta_1,\dots,\beta_b\}$ denotes all permutations of the $\alpha_i$ and $\beta_i$ that preserve their cyclic ordering and start with $\alpha_1$. For example $\text{COP}\{1,2,3\}\{4,5\}$ is given by the twelve permutations
\begin{align}
 &\{1, 2, 3, 5, 4\}, &&\{1, 2, 3, 4, 5\}, &&\{1, 2, 5, 3, 4\}, &&\{1, 2, 4, 3, 5\},\notag\\
&\{1, 2, 5, 4, 3\}, &&\{1, 2, 4, 5, 3\}, &&\{1, 5, 2, 3, 4\}, &&\{1, 4, 2, 3, 5\},\notag\\ 
 &\{1, 5, 2, 4, 3\}, &&\{1, 4, 2, 5, 3\}, &&\{1, 5, 4, 2, 3\}, &&\{1, 4, 5, 2, 3\}.
\end{align}

In order to prove \cref{eq:colorDecAdjLoop} we perform the color decomposition of an arbitrary one-loop diagram. This will yield the color structures for the amplitudes as well as a characterization of the Feynman diagrams contributing to the partial amplitudes.
Besides \cref{eq:LieAlgebra} we need the Fierz identity of the $su(N)$ generator matrices
\begin{equation}\label{eq:colorFlow}
\sum_{a=1}^{N^2-1}(T^a)_{i_1 j_1}(T^a)_{i_2 j_2}=\delta_{i_1 j_2}\delta_{i_2 j_1}-\frac{1}{N}\delta_{i_1 j_1}\delta_{i_2 j_2 }\,.
\end{equation}
The term proportional to $N^{-1}$ is reflecting that the generator matrices are traceless.

We start with the basic observation that all one-loop diagrams can be drawn in planar fashion. Given a particular one-loop diagram we simply cut one of the loop propagators. From the previous section we know, that the color parts of the obtained tree diagram are traces of nested commutators. Making use of \cref{eq:colorFlow} we can contract the adjoint index of the cut loop propagator. Due to the commutators in the trace the $\frac{1}{N}$ term in \cref{eq:colorFlow} does not contribute and can be omitted. Expanding the commutators we are left with either a contraction of adjacent or a contraction of non-adjacent generators, leading to a single trace of generators multiplied by $N$ or a product of two traces of generators. Using the double line formalism, it is instructive to analyze the color flows corresponding to both types of color structures. In the case of the single trace we have a closed color loop leading to the prefactor of $N$ and by definition \cref{eq:colorOrdered} the partial amplitude is simply 
the color ordered amplitude whose external legs are ordered according to the generators in the trace.
Up to a factor of $(-1)^k$, the contribution of a Feynman diagram to a double trace color structure \cref{eq:doubleTrace} is given by a color ordered diagram whose external legs are ordered such that the cyclic ordering of the legs $\{\sigma(k),\dots,\sigma(1)\}$ and $\{\sigma(k+1),\dots,\sigma(n)\}$ are preserved. Furthermore, each tree attached to the loop contain only legs of one of the traces. The reason the order of the legs  $\{\sigma(k),\dots,\sigma(1)\}$ is reversed with respect to the trace is that the color is flowing counter clockwise around the loop opposed to clockwise color flow connecting the legs $\{\sigma(k+1),\dots,\sigma(n)\}$. The remaining task is to express this set of color ordered diagrams as a linear combination of color ordered amplitudes. Obviously the linear combination in \cref{eq:PartialAmpAdjoint} contains all contributing color ordered diagrams. Within the sum over the permutations $\text{COP}_{n,k}(\sigma)$ all diagrams containing trees that mix the traces cancel out due to 
the symmetry properties of the color ordered Feynman rules \cref{fig:ColorOrderedRules}. 
\section{Color decomposition of QCD amplitudes}
The decomposition of an arbitrary QCD tree or one-loop amplitude with $k$ quark anti-quark pairs $\{q_i,\bar{q}_i\}$ of distinct flavors and $n$ gluons into color structures and color ordered or primitive amplitudes has a much richer structure than in the case of only adjoint particles. The reason being the large number of different subleading color structures that are present already at tree level. 

Note that we exploit the fact that the number of flavors $n_f$ is a free parameter of QCD. Hence, unequal flavor amplitudes with $k>6$ are well defined. The color decomposition of QCD amplitudes containing equally flavored quark--anti-quark pairs can be straightforwardly obtained by summing up the appropriate single flavor amplitudes. Furthermore, we keep the number of colors $N$ of the $SU(N)$ gauge symmetry a free parameter as well. 
\subsection{Tree-level}\label{section:QCDcolorTree}%
 Let now $i_1,\ldots,i_k$ and $\bar{j}_1,\ldots,\bar{j}_k$ be the color indices of the quarks and anti-quarks. We introduce the short hand notation
\begin{equation}
 \left(n_m\right)_{i\,\bar{j}}=\left(\prod_{l=n_{m-1}+1}^{n_m}T^{a_{\sigma(l)}}\right)_{i\,\bar{j}}
\end{equation}
where $\sigma\in S_n$ is some arbitrary permutation of the gluons and $0=n_0\leq n_1\leq n_2\leq\ldots\leq n_k=n$ is some arbitrary partition. As we are going to prove in the following,  all color structures are given by
\begin{equation}\label{eq:colorstructuresQuarks}
 \left(\frac{-1}{N}\right)^{p}\left(\prod_{\alpha=1}^k\left(n_\alpha\right)_{i_\alpha\,\bar{j}_{\tau(\alpha)}}\right)\,,
\end{equation}
where $\tau\in S_k$ is some permutation of the anti-quarks,  and 
\begin{equation}\label{eq:power}
p= p(\tau)=c(\tau)-1\,,
\end{equation}
with $c(\tau)$ being the number of cycles of the permutation $\tau$. Hence, modulo permutations of the gluons we have $k!{\binom{n+k-1}{k-1}}$ different color structures leading to a total of $k\cdot (n+k-1)!$ terms in the color decomposition of an arbitrary quark-gluon tree amplitude. We recall that a permutation is uniquely determined by its cycles and that a cycle of length $l$ of a permutation $\tau$ can be represented by a sequence of the form $\{i,\tau(i),\tau^2(i),\dots,\tau^{l-1}(i)\}$, with all elements being distinct and $i=\tau^{l}(i)$. Since we have $l$ choices for $i$, there are $l$ such representatives, all related by cyclic permutations $\pi\in Z_l$.

Similar to the case of only adjoint particles we are going to prove \cref{eq:colorstructuresQuarks} by determining the color structures appearing in an arbitrary Feynman diagram.
With regard to the structure of the partial amplitudes it is convenient to include the gluons into the definition of a cycle. Hence, without any reference to a permutation of anti-quarks, we define a cycle to be a sequence that
\begin{itemize}
 \item starts with a quark and ends with its anti-quark,
 \item contains additional quarks only as the successors of their anti-quarks, and
 \item contains gluons only between successive quarks and anti-quarks of different flavor.
\end{itemize}
Cycles related by a cyclic permutation are considered equal. Disjoint cycles have neither common quarks nor common gluons.
An example of a cycle with three quarks and four gluons is $c_1=\{q_1,1,2,\bar{q}_2,q_2,\bar{q}_3,q_3,3,4,\bar{q}_1\}$. The integers in the definition of a cycle represent the gluons. It is obvious that there is a one to one correspondence between the sets of $p+1$ disjoint cycles with a total of $k$ quarks and $n$ gluons and $\tau, \sigma$ and the gluon partition $\{n_i\}$ in \eqref{eq:colorstructuresQuarks}, e.\,g.~the color structure corresponding to the example cycle $c_1$ above is just $C_1=(T^{a_1}T^{a_2})_{i_1\bar{j}_2}\delta_{i_2\bar{j}_3}(T^{a_3}T^{a_4})_{i_3\bar{j}_1}$. The importance of the concept of cycles becomes evident when analyzing the partial amplitude multiplying a color structure corresponding to a single cycle, such as our example $C_1$. Each Feynman diagram contributing to such a partial amplitude can be drawn in planar fashion such that the external legs are ordered as in the cycle. In other words, the partial amplitude is given by the color ordered amplitude whose external legs are 
ordered as in the cycle. This observation is crucial for the succeeding analysis of the color decomposition of QCD at tree and one-loop level and a basic fact following from the color flow of a Feynman diagram.

We remark that \cref{eq:colorstructuresQuarks} differs from the color structures given in ref.~\cite{Mangano:1990by} by the power of $\frac{-1}{N}$ associated with the permutation of the anti-quark indices. According to ref.~\cite{Mangano:1990by} the power $p$ should be equal to $k-1$ for the identity permutation and equal to the number of fixed points of the permutation else. However, this is only valid for up to three quark lines. As will be shown in the following, \cref{eq:power} is the correct generalization to an arbitrary number of quark lines. 

\begin{figure}[t]\centering
\raisebox{-1.5cm}{\includegraphics[height=3.5cm]{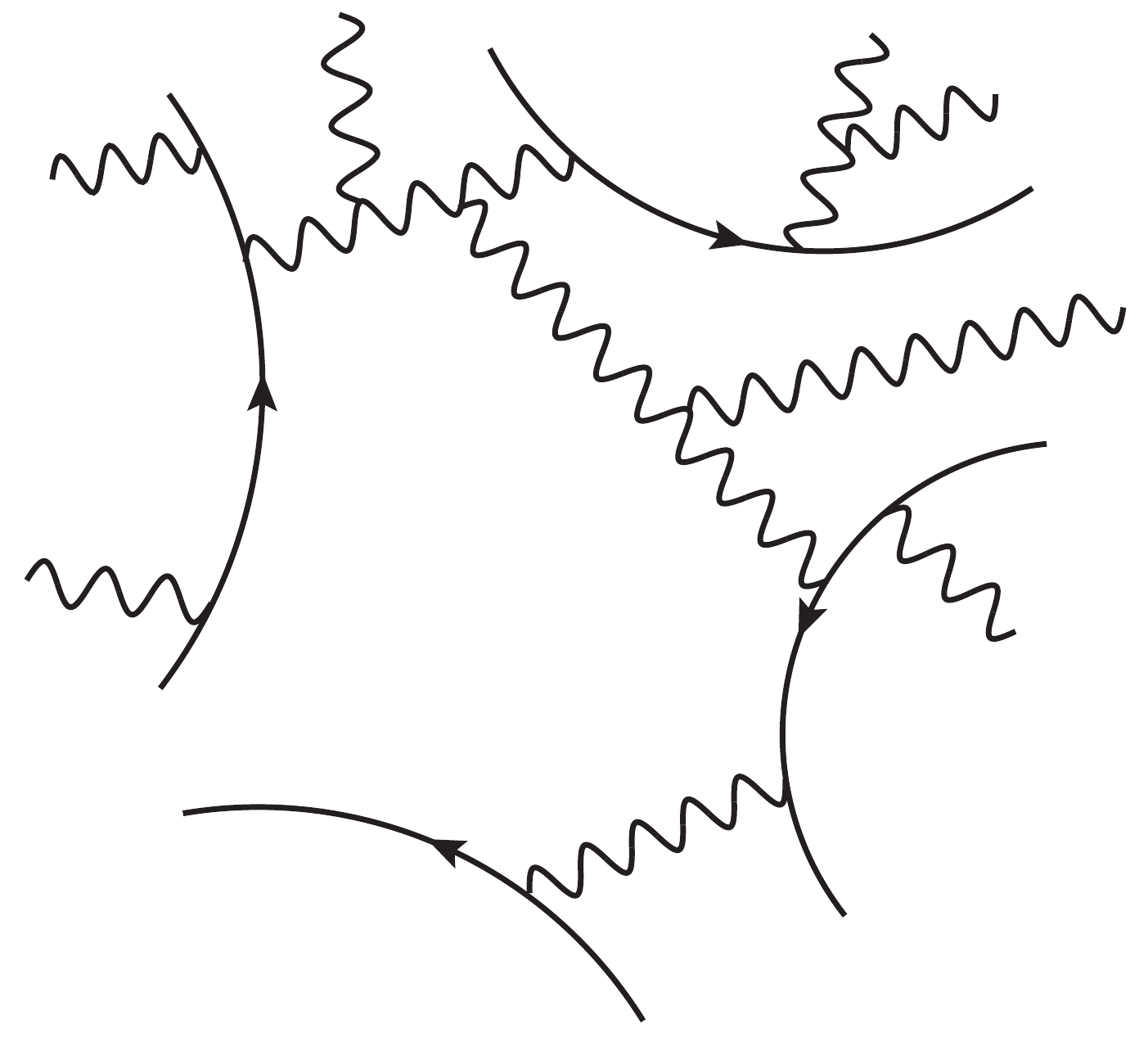}}=\raisebox{-1.4cm}{\includegraphics[height=3cm]{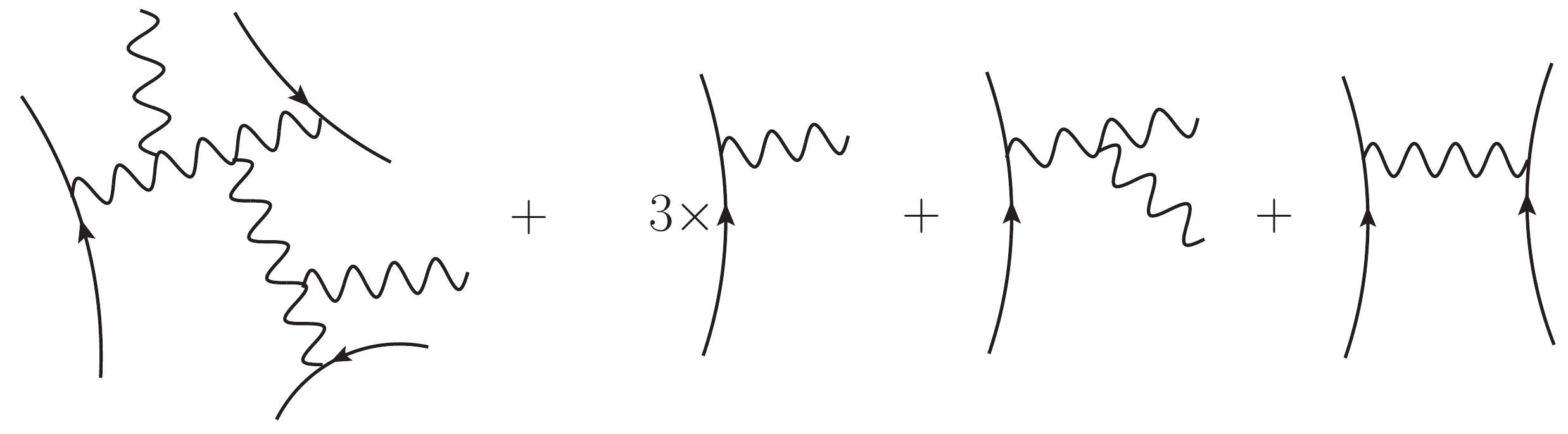}}
\caption{Unique partition of a Feynman diagram into cropped diagrams obtained by cutting quark lines between quark-gluon vertices.}\label{fig:partitionOfDiagrams}
\end{figure}

\begin{figure}[t]\centering
\subfigure[Case 1.1]{\includegraphics[height=4cm]{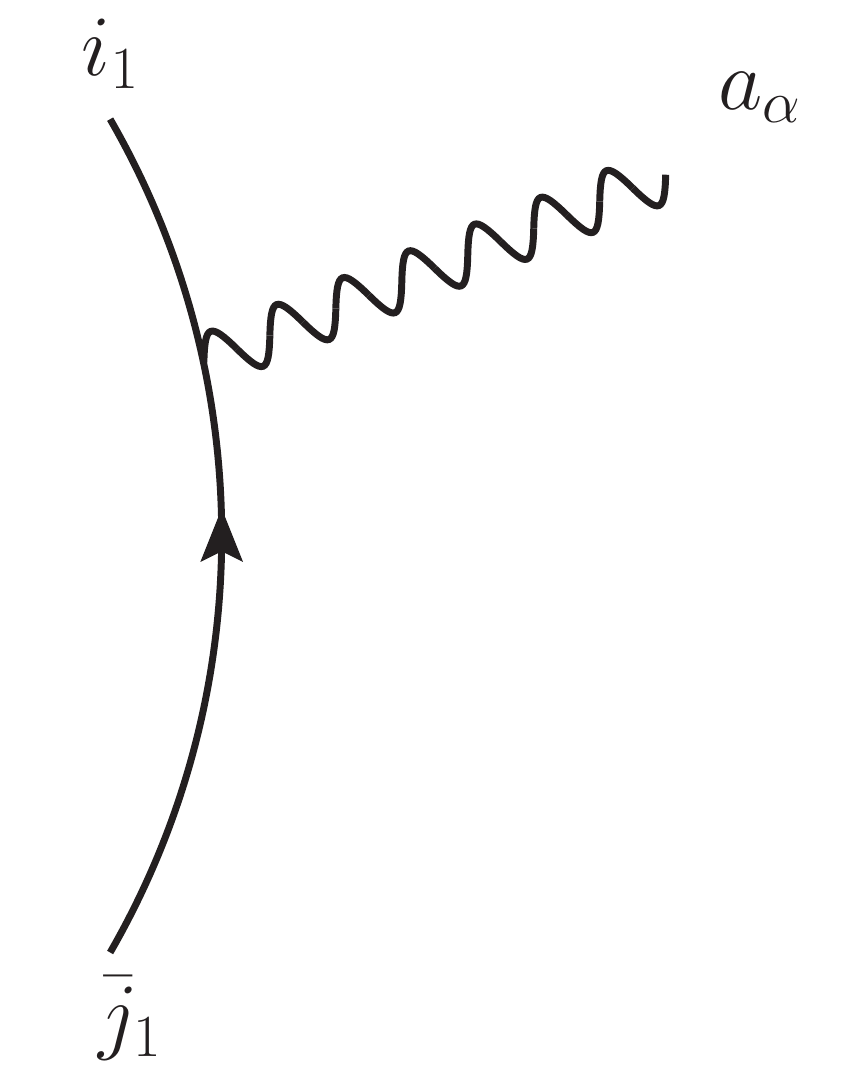}}\hfill
\subfigure[Case 1.2a]{\includegraphics[height=4cm]{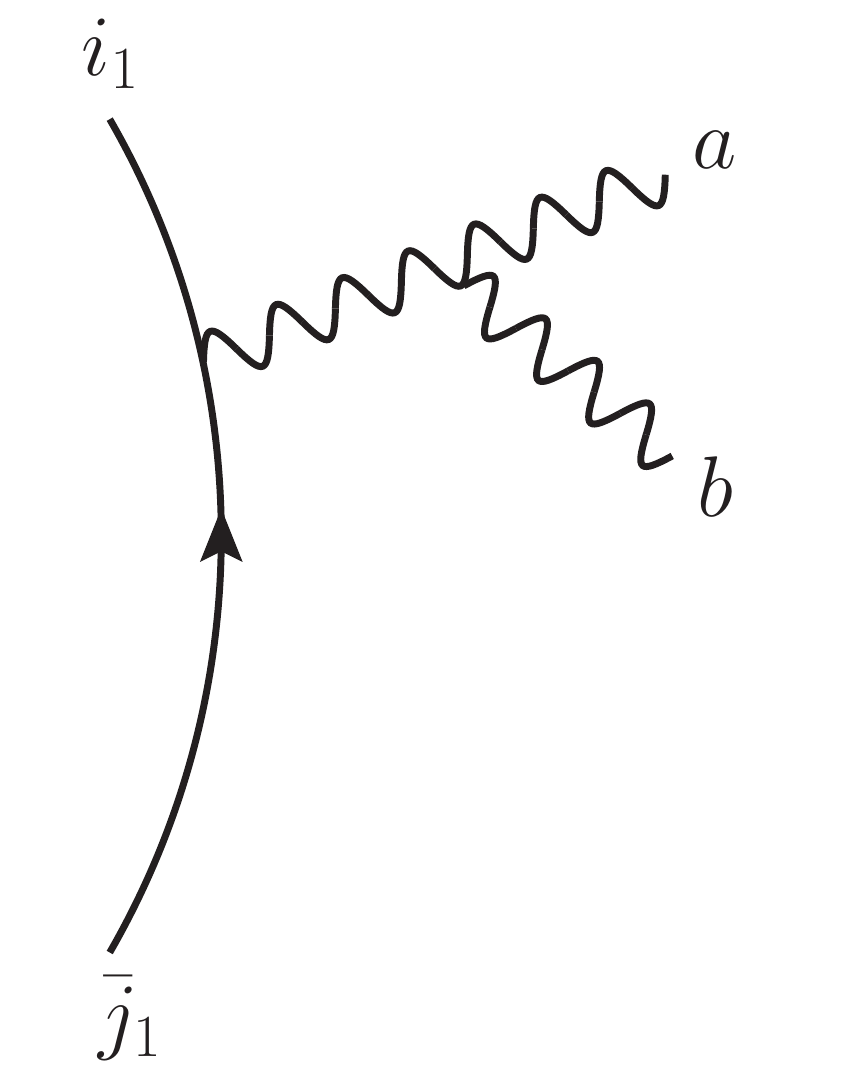}}\hfill
\subfigure[Case 1.2b]{\includegraphics[height=4cm]{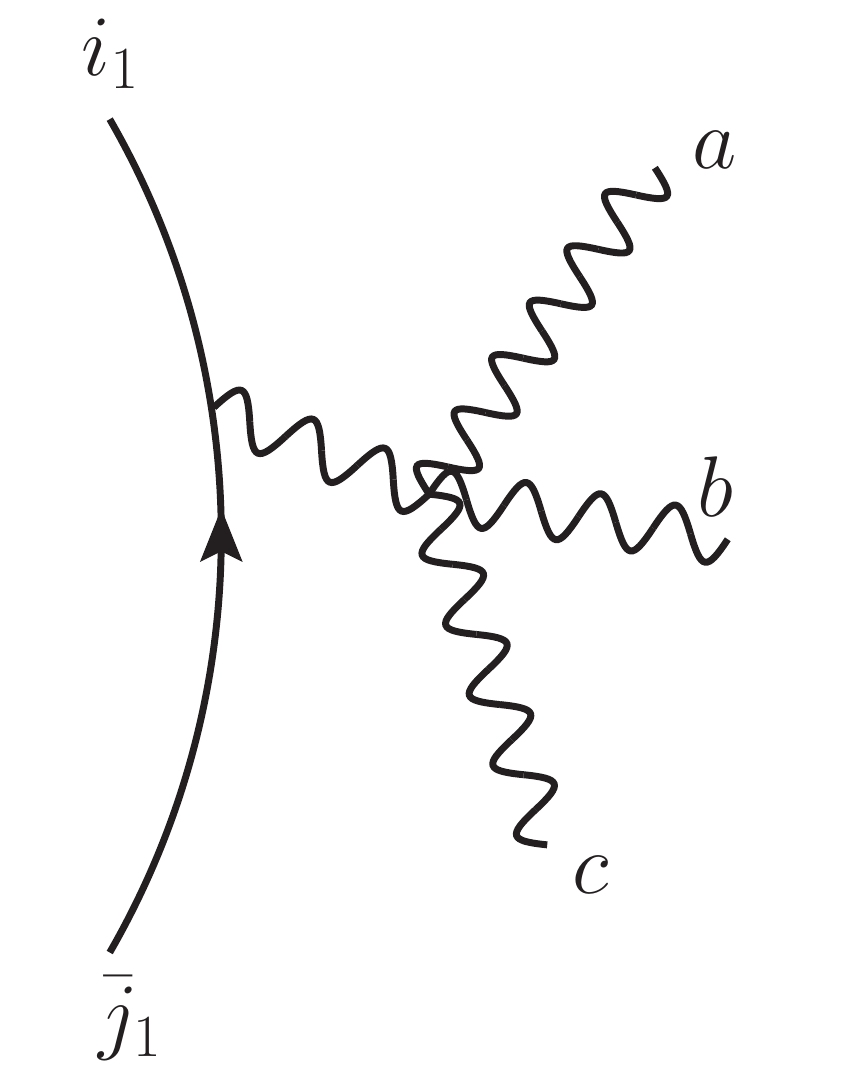}}\hfill
\subfigure[Case 1.3]{\includegraphics[height=4cm]{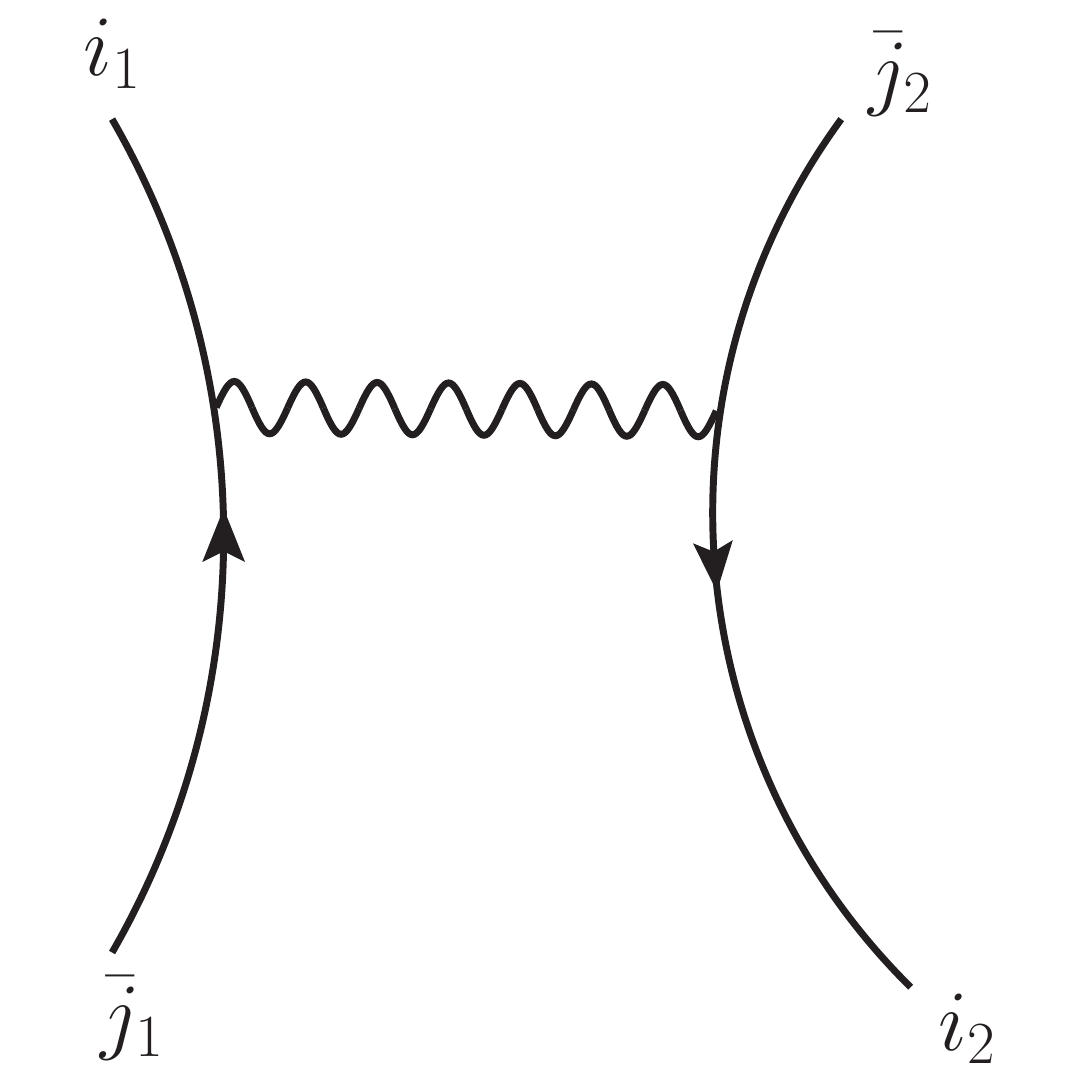}}
\caption{Determination of the color structure of a cropped quark-gluon Feynman diagram.}\label{fig:colordecompositionQuarks}
\end{figure}

\begin{figure}[t]
\subfigure[Case 2.1a]{\includegraphics[height=4cm]{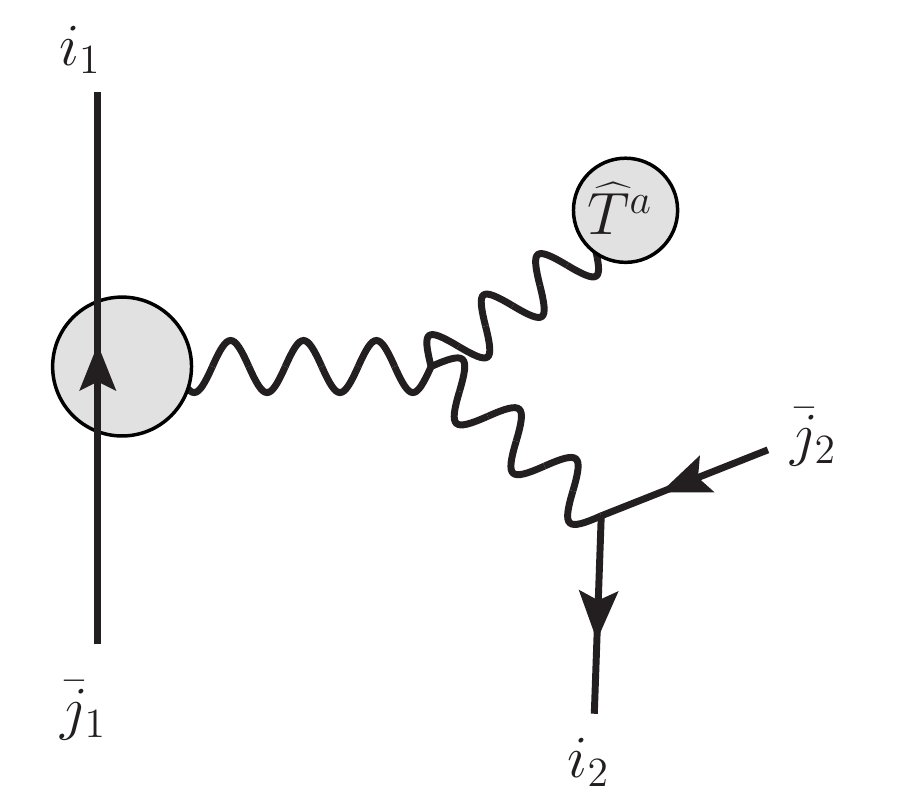}}\hfill
\subfigure[Case 2.1b]{\includegraphics[height=4cm]{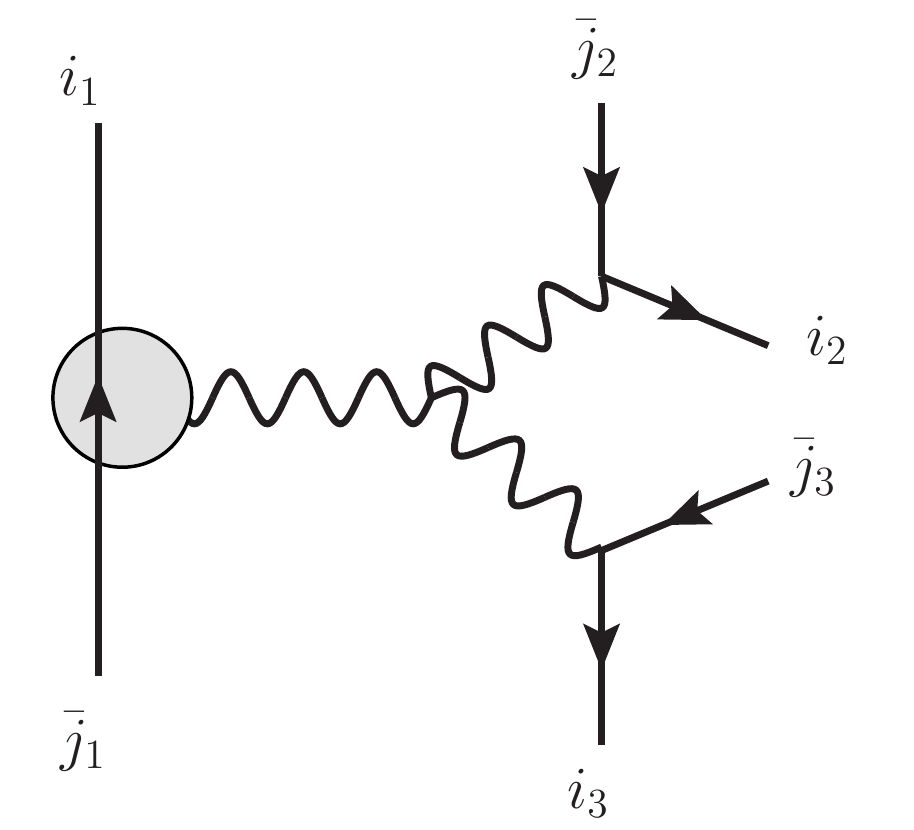}} \hfill
\subfigure[Case 2.2a]{\includegraphics[height=4cm]{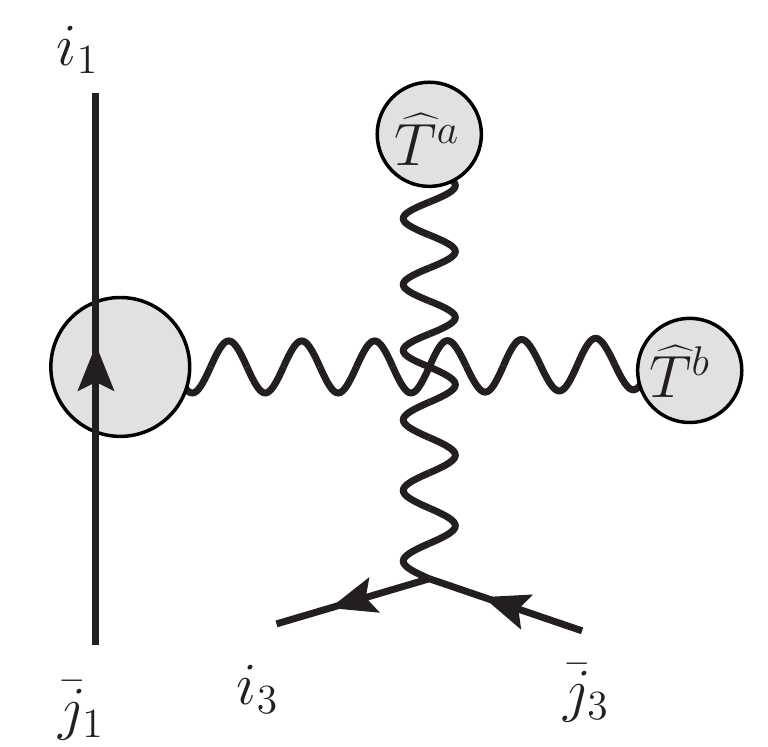}}\\
\hspace*{2cm}
\subfigure[Case 2.2b]{\includegraphics[height=4cm]{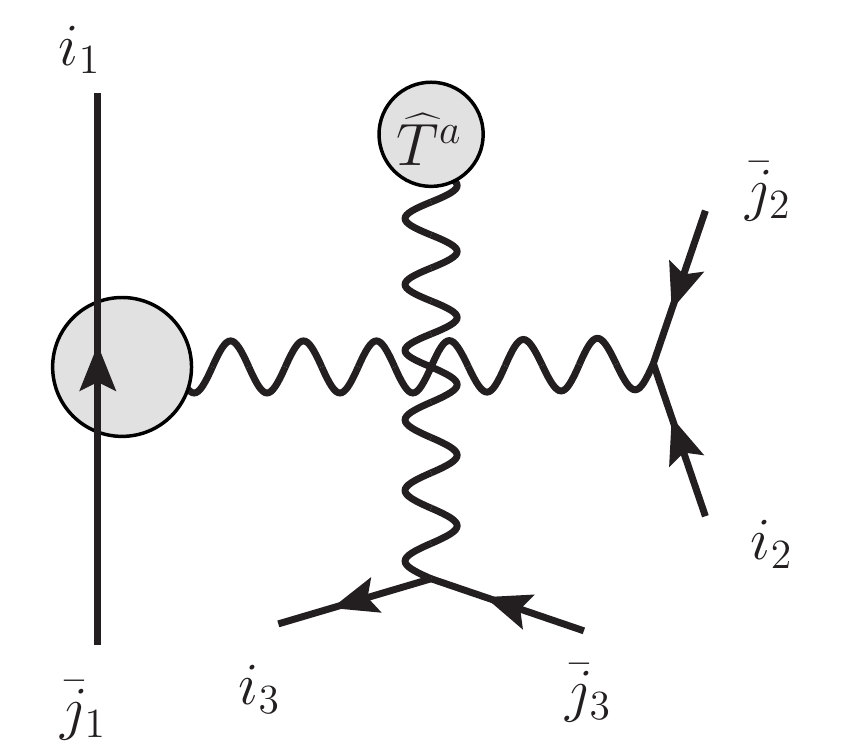}}\hfill
\subfigure[Case 2.2c]{\includegraphics[height=4cm]{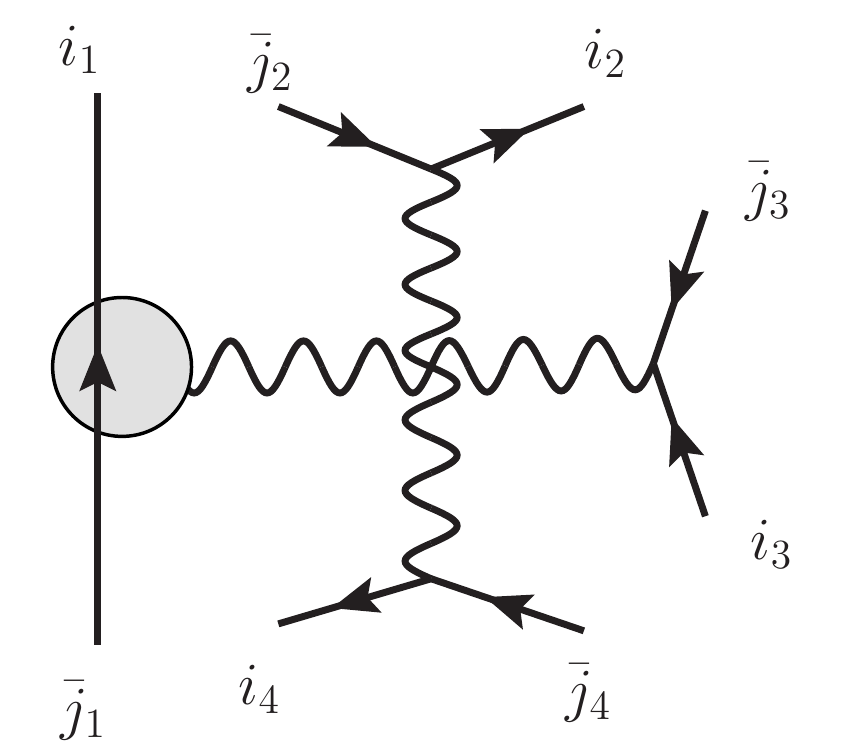}}\hspace*{2cm}
\caption{Determination of the color structure of a quark-gluon Feynman diagram.}\label{fig:colordecompositionQuarks2}
\end{figure}

In order to determine all possible color structures we make use of the simple fact that a given Feynman diagram can be uniquely divided into sub diagrams by cutting the quark lines between quark-gluon vertices. We refer to these sub diagrams as cropped diagrams. An example can be found in \cref{fig:partitionOfDiagrams}. It is obvious that the color structures of a Feynman diagram are obtained by contracting the color structures of its cropped diagrams. Consequently, it is sufficient to know the color structures of an arbitrary cropped Feynman diagram. To determine the color structure of a cropped Feynman diagram we start in one of its quark lines and contract the adjoint indices along the quark-gluon tree attached to it. As shown in \cref{fig:colordecompositionQuarks} the gluon connected to the quark line can be either external (case 1.1), connected to a non-Abelian vertex (cases 1.2a and 1.2b) or directly connected to another quark line (case 1.3).
If the gluon is external, the color structure of the cropped diagram is
\begin{equation}
( T^{a_\alpha})_{i_1\,\bar{j}_1}\,.
\end{equation}
Note that the color indices $i_1$, $\bar{j}_1$ in general do not belong to external quarks but are contracted along the chosen quark line. If the gluon is connected to a three-vertex we get a sum of two color structures
\begin{equation}\label{eq:ConnectGluonTo3Vertex}
 if^{abc}( T^{c})_{i_1\,\bar{j}_1}=[T^a,T^b]_{i_1\,\bar{j}_1}=(T^a,T^b)_{i_1\,\bar{j}_1}-(T^b,T^a)_{i_1\,\bar{j}_1}\,
\end{equation}
and in the case of a four gluon vertex we get a sum of three terms whose color parts read
\begin{align}\label{eq:ConnectGluonTo4Vertex}
 (T^d)_{i_1\,\bar{j}_1}f^{abe}f^{cde}&=[T^c,[T^a,T^b]]_{i_1\,\bar{j}_1}\,,\\
(T^d)_{i_1\,\bar{j}_1}f^{ace}f^{bde}&=[T^b,[T^a,T^c]]_{i_1\,\bar{j}_1}\,,\notag\\
(T^d)_{i_1\,\bar{j}_1}f^{ade}f^{cbe}&=[T^a,[T^c,T^b]]_{i_1\,\bar{j}_1}\,.\notag
\end{align}
Of course these three terms are not independent due to the Jacobi identity
\begin{equation}\label{eq:jacobi}
 [T^a,[T^b,T^c]]+[T^b,[T^c,T^a]]+[T^c,[T^a,T^b]]=0\,.
\end{equation}
Since in the end we will expand all commutators, we do not need to care about that. Finally, if the gluon is connected to another quark line we apply \cref{eq:colorFlow} and get a sum of two color structures for our sub diagram
\begin{align}
\delta_{i_1\bar{j}_2}\delta_{i_2\bar{j}_1}\,,\\
\frac{-1}{N}\delta_{i_1\bar{j}_1}\delta_{i_2\bar{j}_2} \,.\label{eq:ColorDisconnect}
\end{align}
Here \eqref{eq:ColorDisconnect} is a contribution without any color flow between the two quark lines. Hence, it leads to color structures of the whole diagram which are the product of the color structures of the two diagrams we get by removing the gluon connecting the two quarks. In general, if a color structure of a diagram is proportional to $\left(\frac{-1}{N}\right)^p$, than it is a product of $p+1$ terms without color flow between them. As we are going to show now, this QED type gluon exchange between quarks is the only origin of powers of $\frac{-1}{N}$ in the color structures. To further determine the color structure of the cropped diagram we contract all internal gluons of the three- and four-vertex by using \cref{eq:ConnectGluonTo3Vertex} or \cref{eq:ConnectGluonTo4Vertex} if the gluon is connected to another three gluon or another four gluon vertex. Proceeding like this we are eventually left with internal gluons contracted with other fermion lines. In \cref{fig:colordecompositionQuarks2} all 
possible cases are listed how a three gluon and a four gluon vertex can connect to other fermion lines. The blobs represent sub diagrams and the hatted generators $\widehat{T}^a$ denote the nested commutators belonging to them.
In case 2.1a we have a sum of two sub color structures
\begin{equation}
(T^b)_{i_2\,\bar{j}_2}[\widehat{T}^a,T^b]_{i\,\bar{j}}=(\widehat{T}^a)_{i\,\bar{j}_2}\delta_{i_2\,\bar{j}}-\delta_{i\,\bar{j}_2}(\widehat{T}^a)_{i_2\,\bar{j}}\,.
\end{equation}
The $N^{-1}$ term has vanished due to the commutator and the color indices $i$, $\bar{j}$ are contracted inside the nested commutators belonging to the blobs in \cref{fig:colordecompositionQuarks2}. If two fermion lines couple to a three gluon vertex we get a sum of two sub color structures as well
\begin{align}
(T^a)_{i_2\,\bar{j}_2}(T^b)_{i_3\,\bar{j}_3}[T^a,T^b]_{i\,\bar{j}}&=(T^a)_{i_2\,\bar{j}_2}\left((\widehat{T}^a)_{i\,\bar{j}_3}\delta_{i_3\,\bar{j}}-\delta_{i\,\bar{j}_3}(\widehat{T}^a)_{i_3\,\bar{j}}\right)\\
&=\delta_{i\,\bar{j}_2}\delta_{i_2\,\bar{j}_3}\delta_{i_3\,\bar{j}}-\delta_{i\,\bar{j}_3}\delta_{i_3\,\bar{j}_2}\delta_{i_2\,\bar{j}}\,.\notag
\end{align}
A four gluon vertex can be connected to one, two or three other quark lines. In the case of one additional quark line there are two types of contractions
\begin{align}
 (T^c)_{i_2\,\bar{j}_2}[\widehat{T}^a,[\widehat{T}^b,T^c]]_{i\,\bar{j}}&=(\widehat{T}^a\widehat{T}^b)_{i\,\bar{j}_2}\delta_{i_2\,\bar{j}}-(\widehat{T}^a)_{i\,\bar{j}_2}(\widehat{T}^b)_{i_2\,\bar{j}}\\
&\={}+\delta_{i\,\bar{j}_2}(\widehat{T}^b\widehat{T}^a)_{i_2\,\bar{j}}-(\widehat{T}^b)_{i\,\bar{j}_2}(\widehat{T}^a)_{i_2\,\bar{j}}\notag
\end{align}
and
\begin{align}
 (T^c)_{i_2\,\bar{j}_2}[T^c,[\widehat{T}^b,\widehat{T}^a]]_{i\,\bar{j}}&=[\widehat{T}^a,\widehat{T}^b]_{i\,\bar{j}_2}\delta_{i_2\,\bar{j}}-\delta_{i\,\bar{j}_2}[\widehat{T}^a,\widehat{T}^b]_{i_2\,\bar{j}}\,,
\end{align}
leading to the following six sub color structures
\begin{align}
&(\widehat{T}^{\pi(a)}\widehat{T}^{\pi(b)})_{i\,\bar{j}_2}\delta_{i_2\,\bar{j}}\,,\\
&(\widehat{T}^{\pi(a)})_{i\,\bar{j}_2}(\widehat{T}^{\pi(b)})_{i_2\,\bar{j}}\notag\,,\\
&\delta_{i\,\bar{j}_2}(\widehat{T}^{\pi(a)}\widehat{T}^{\pi(b)})_{i_2\,\bar{j}}\notag\,,
\end{align}
where $\pi$ is some permutation.
Connecting two additional quark lines to the four gluon vertex, the six encountered sub color structures read
\begin{align}
&(\widehat{T}^a)_{i\,\bar{j}_{\pi(2)}}\delta_{i_{\pi(2)}\,\bar{j}_{\pi(3)}}\delta_{i_{\pi(3)}\,\bar{j}}\,,\\
&\delta_{i\,\bar{j}_{\pi(2)}}(\widehat{T}^a)_{i_{\pi(2)}\,\bar{j}_{\pi(3)}}\delta_{i_{\pi(3)}\,\bar{j}}\notag\,,\\
&\delta_{i\,\bar{j}_{\pi(2)}}\delta_{i_{\pi(2)}\,\bar{j}_{\pi(3)}}(\widehat{T}^a)_{i_{\pi(3)}\,\bar{j}}\notag\,,
\end{align}
and originate from the contractions
\begin{align}
 (T^b)_{i_3\,\bar{j}_3} (T^c)_{i_2\,\bar{j}_2}[\widehat{T}^a,[T^b,T^c]]_{i\,\bar{j}}&=(\widehat{T}^a)_{i\,\bar{j}_3}\delta_{i_3\,\bar{j}_2}\delta_{i_2\,\bar{j}}-(\widehat{T}^a)_{i\,\bar{j}_2}\delta_{i_2\,\bar{j}_3}\delta_{i_3\,\bar{j}}\\
&\={}+\delta_{i\,\bar{j}_2}(\widehat{T}^a)_{i_2\,\bar{j}_3}\delta_{i_3\,\bar{j}}-\delta_{i\,\bar{j}_3}\delta_{i_3\,\bar{j}_2}(\widehat{T}^a)_{i_2\,\bar{j}}\notag
\end{align}
and
\begin{align}
 (T^b)_{i_3\,\bar{j}_3} (T^c)_{i_2\,\bar{j}_2}[T^c,[T^b,\widehat{T}^a]]_{i\,\bar{j}}&=(\widehat{T}^a)_{i\,\bar{j}_3}\delta_{i_3\,\bar{j}_2}\delta_{i_2\,\bar{j}}-\delta_{i\,\bar{j}_3}(\widehat{T}^a)_{i_3\,\bar{j}_2}\delta_{i_2\,\bar{j}}\\
&\={}-\delta_{i\,\bar{j}_2}(\widehat{T}^a)_{i_2\,\bar{j}_3}\delta_{i_3\,\bar{j}}+\delta_{i\,\bar{j}_2}\delta_{i_2\,\bar{j}_3}(\widehat{T}^a)_{i_3\,\bar{j}}\notag\,.
\end{align}
Finally, if there are three additional quarks we are faced with contractions of the form
\begin{align}
(T^a)_{i_4\,\bar{j}_4} (T^b)_{i_3\,\bar{j}_3} (T^c)_{i_2\,\bar{j}_2}[T^a,[T^b,T^c]]_{i\,\bar{j}}&=\delta_{i\,\bar{j}_4}\delta_{i_4\,\bar{j}_3}\delta_{i_3\,\bar{j}_2}\delta_{i_2\,\bar{j}}-\delta_{i\,\bar{j}_4}\delta_{i_4\,\bar{j}_2}\delta_{i_2\,\bar{j}_3}\delta_{i_3\,\bar{j}}\\
&\={}+\delta_{i\,\bar{j}_2}\delta_{i_2\,\bar{j}_4}\delta_{i_4\,\bar{j}_3}\delta_{i_3\,\bar{j}}-\delta_{i\,\bar{j}_3}\delta_{i_3\,\bar{j}_2}\delta_{i_2\,\bar{j}_4}\delta_{i_4\,\bar{j}}\notag\,.
\end{align}
Hence, the six encountered sub color structures are 
\begin{equation}
 \delta_{i\,\bar{j}_{\pi(2)}}\delta_{i_{\pi(2)}\,\bar{j}_{\pi(3)}}\delta_{i_{\pi(3)}\,\bar{j}_{\pi(4)}}\delta_{i_{\pi(4)}\,\bar{j}}
\end{equation}
where $\pi$ is again some permutation. Expanding all remaining commutators we end up with the color decomposition of the cropped Feynman diagram. Except the color structure $\frac{-1}{N}\delta_{i_1\bar{j}_1}\delta_{i_2\bar{j}_2}$ in the case with $k=2$ quark lines and $n=0$ gluons all color structures of a cropped Feynman diagram are cyclic color structures of the general form
\begin{multline}\label{eq:colorStructureCropped}
 (T^{\sigma(a_1)}\dots T^{\sigma(a_{n_1})})_{i_1\,\bar{j}_{\pi(2)}} (T^{\sigma(a_{n_1+1})}\dots T^{\sigma(a_{n_2})})_{i_{\pi(2)}\,\bar{j}_{\pi(3)}}\times\dots \\\times(T^{\sigma(a_{n_{k-2}+1})}\dots T^{\sigma(a_{n_{k-1}})})_{i_{\pi(k-1)}\,\bar{j}_{\pi(k)}} (T^{\sigma(a_{n_{k-1}+1})}\dots T^{\sigma(a_{n})})_{i_{\pi(k)}\,\bar{j}_{1}}\,.
\end{multline}
Here $0\leq n_1\leq n_2\leq \dots\leq n_{k-1}\leq n$ is some partition of the gluons and $\sigma$, $\pi$ are some permutations. As their name indicates, a cyclic color structure corresponds to a cycle. Since contracting a quark and an anti-quark color index between two cyclic color structures \eqref{eq:colorStructureCropped} yields again a cyclic color structure, we have proven that all color structures proportional to $(\frac{-1}{N})^0$ of a quark gluon amplitude have the form \eqref{eq:colorStructureCropped}. Contracting a cyclic color structure with $\frac{-1}{N}\delta_{i_1\bar{j}_1}\delta_{i_2\bar{j}_2}$ we get a product of two cyclic color structures. Hence, within a quark-gluon amplitude the color structures proportional to $(\frac{-1}{N})^p$ are a product of $p+1$ cyclic color structures. This concludes the proof of \cref{eq:colorstructuresQuarks}, as each of the $p+1$ cycles in the permutation $\tau$ corresponds to one of the $p+1$ cyclic color structures. 

The remaining task is to express the partial amplitudes in terms of color ordered amplitudes. Although the general structure of the color decomposition has in principle been known for a long time, explicit expressions for the partial amplitudes in terms of color ordered amplitudes can be found in the literature only for a small and fixed number of quarks like e.\,g.~in \cite{Mangano:1990by} or more recently in \cite{Ita:2011ar}. Given the above derivation of the color structures appearing in an arbitrary quark-gluon Feynman diagram, it is easy to come up with the color decomposition of a QCD tree amplitude with $k$ quark anti-quark pairs $\{q_i,\bar{q}_i\}$ of distinct flavors and $n$ gluons,
\begin{equation}\label{eq:colorDecomposition}
 \mathcal{A}^{\text{tree}}_{(q \bar{q})^k}=g^{n+2k-2}\sum_{\substack{\sigma \in S_n\\\tau \in S_k}}\sum_{\{n_i\}}\left(\sfrac{-1}{N}\right)^{p}\left(\prod_{\alpha=1}^k\left(n_\alpha\right)_{i_\alpha\,\bar{j}_{\tau(\alpha)}}\right)\sum_{\kappa\in\Gamma(\tau,\sigma,\{n_i\})}A(\kappa)\,.
\end{equation}
The sum over all color structures involves a sum over all permutations $\sigma$ of the $n$ gluons, over all permutations $\tau$ of the $k$ anti-quarks and a sum over all partitions $0=n_0<1\leq n_1\leq n_2\leq\ldots\leq n_k=n$ of the gluons. A given color structure is a product of $p+1$ cyclic color structures and its partial amplitude is given by a sum over all photon exchange permutations $\Gamma(\tau,\sigma,\{n_i\})$, i.\,e.~all possibilities photons can be exchanged between the quarks of different cycles. The number of color ordered amplitudes constituting a partial amplitude only depends on the number of quark lines present, and how the quarks are distributed among the cycles. We have implemented \cref{eq:colorDecomposition} in the \texttt{Mathematica} package \texttt{QCDcolor} described in \cref{appendix:QCDcolor}. 

\begin{figure}[t]
\begin{center}
\includegraphics[width=5cm]{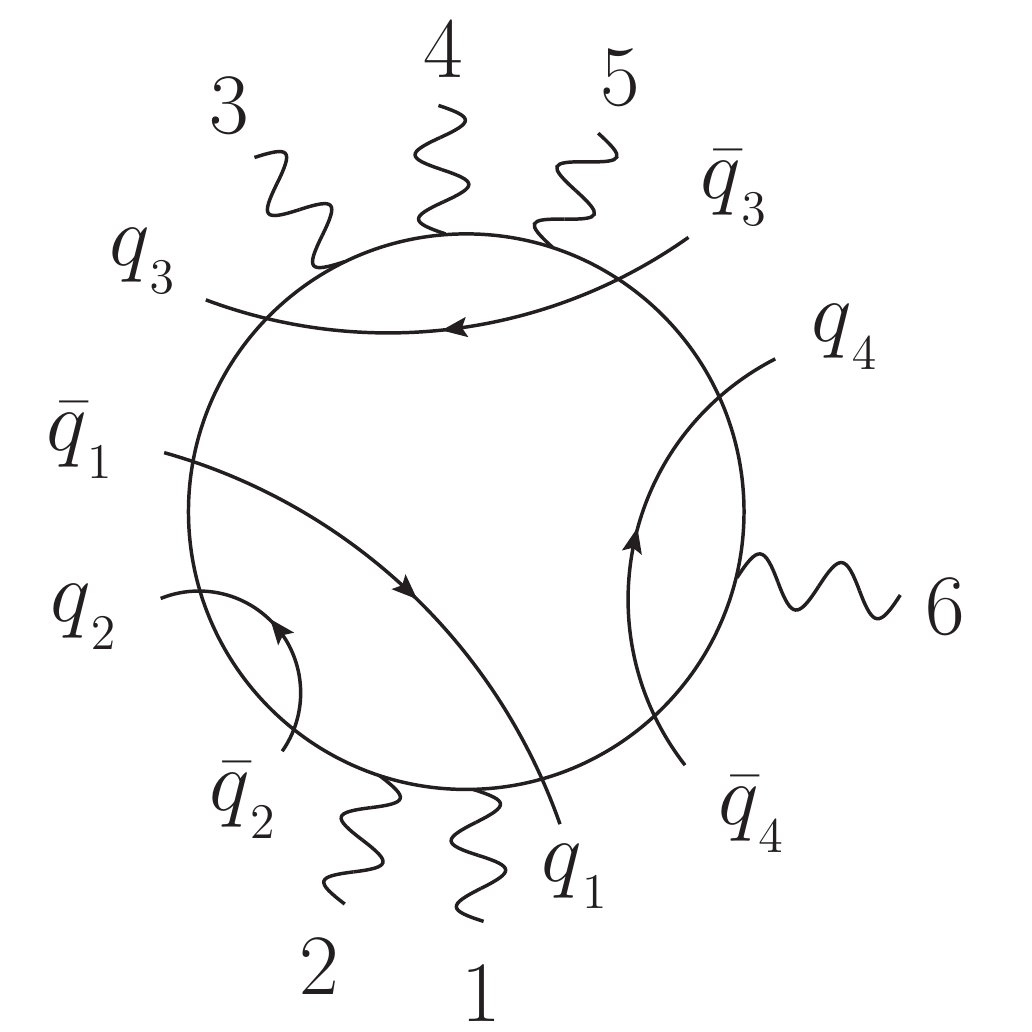}\hfill
\includegraphics[width=5cm]{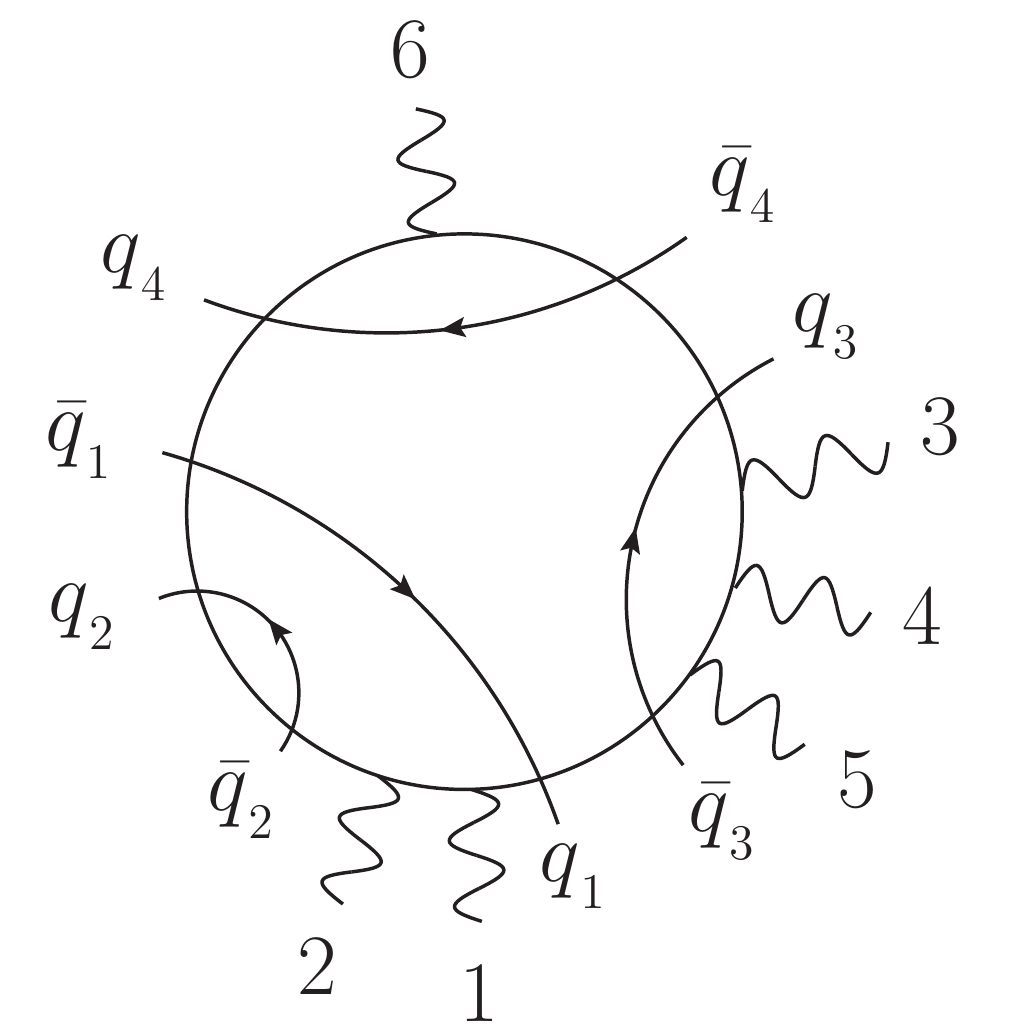}\hfill
\raisebox{0.2cm}{\includegraphics[height=4.7cm]{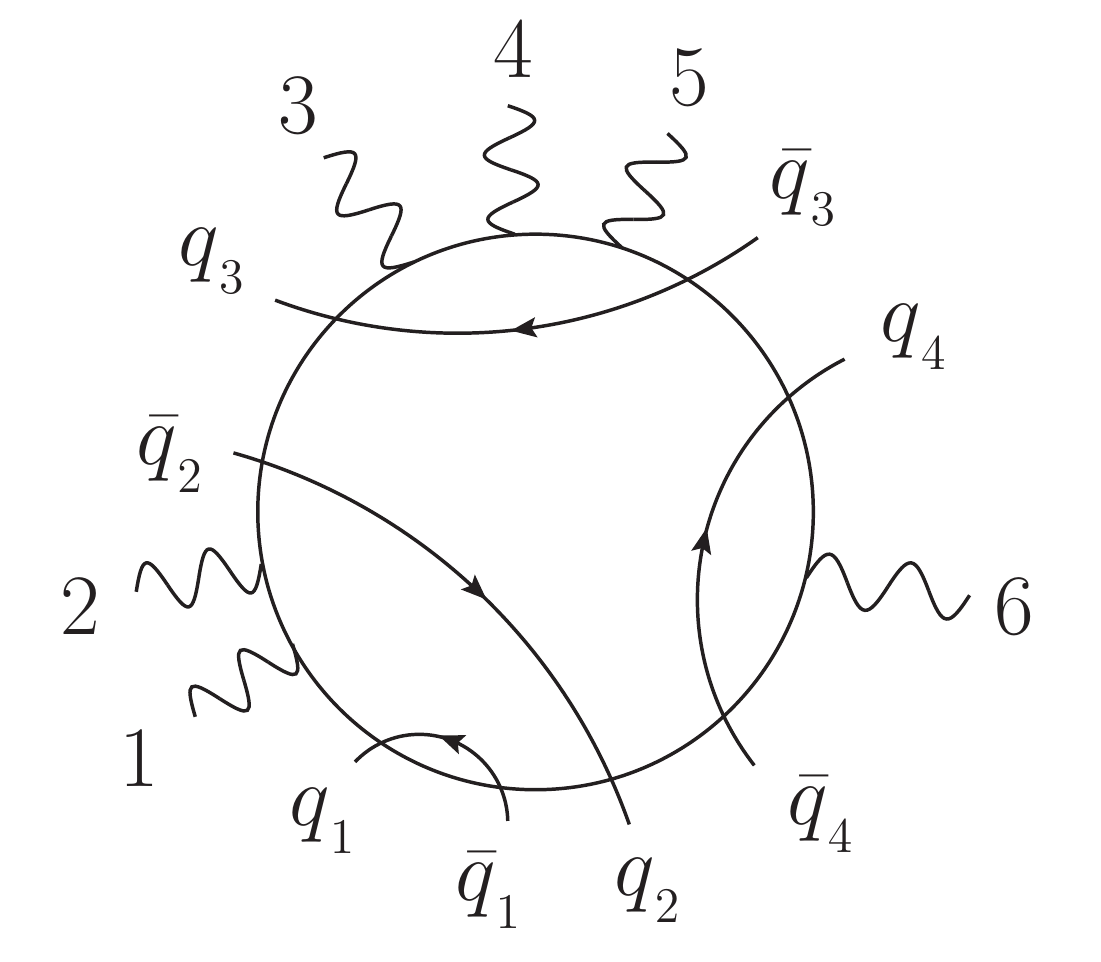}}\\
\raisebox{0.4cm}{\includegraphics[height=4.5cm]{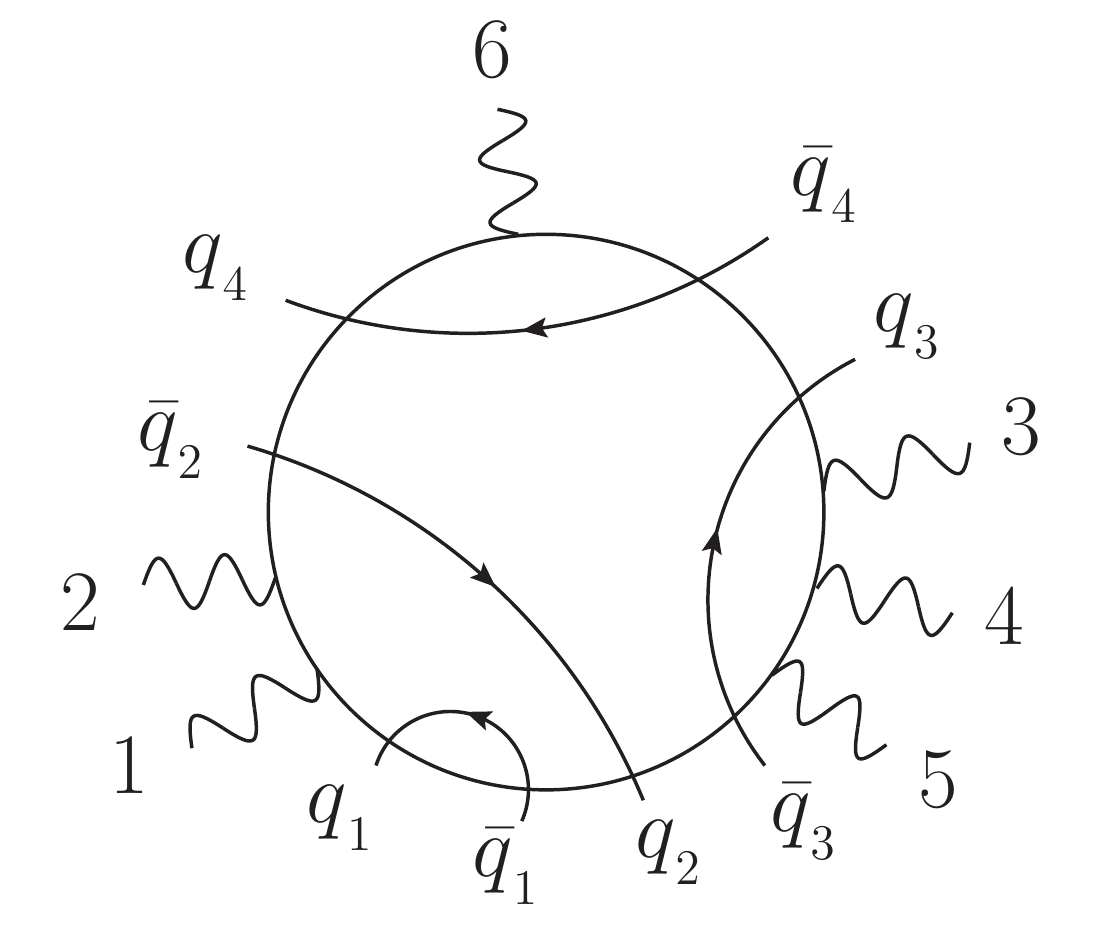}}\hfill
\includegraphics[width=5cm]{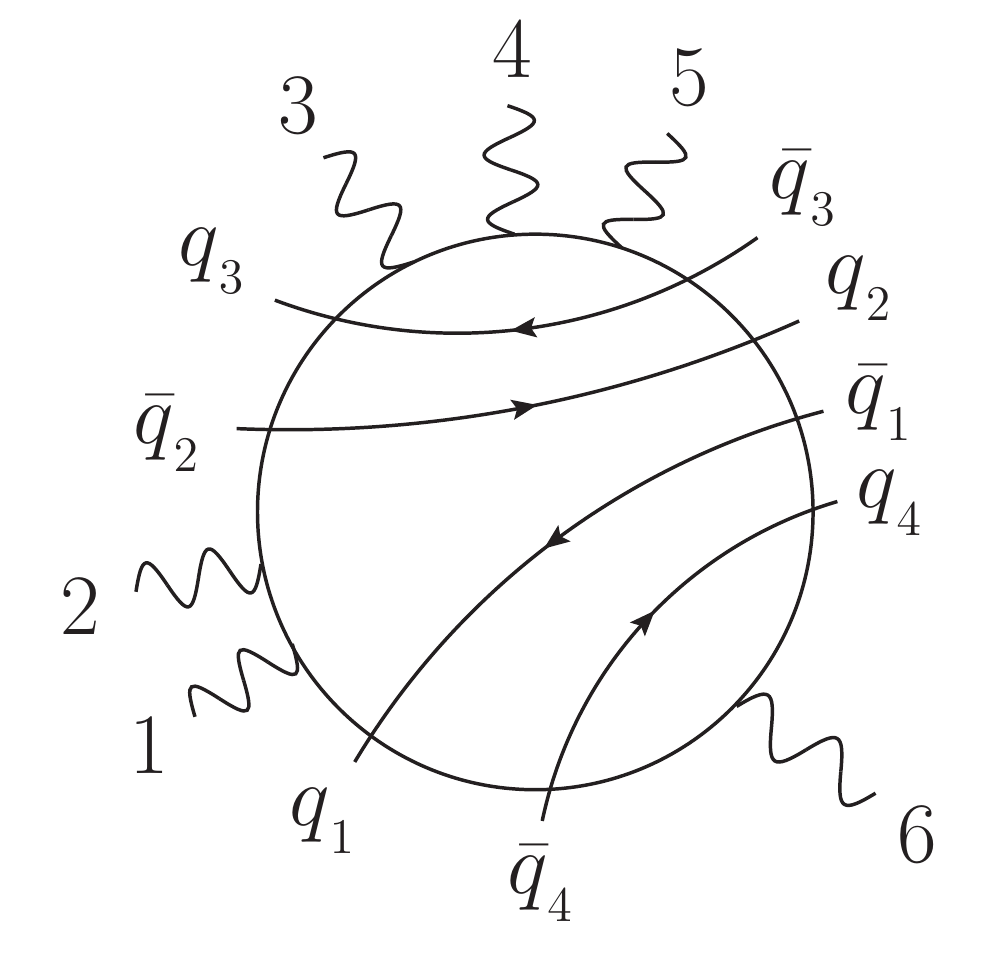}\hfill
\includegraphics[width=5.1cm]{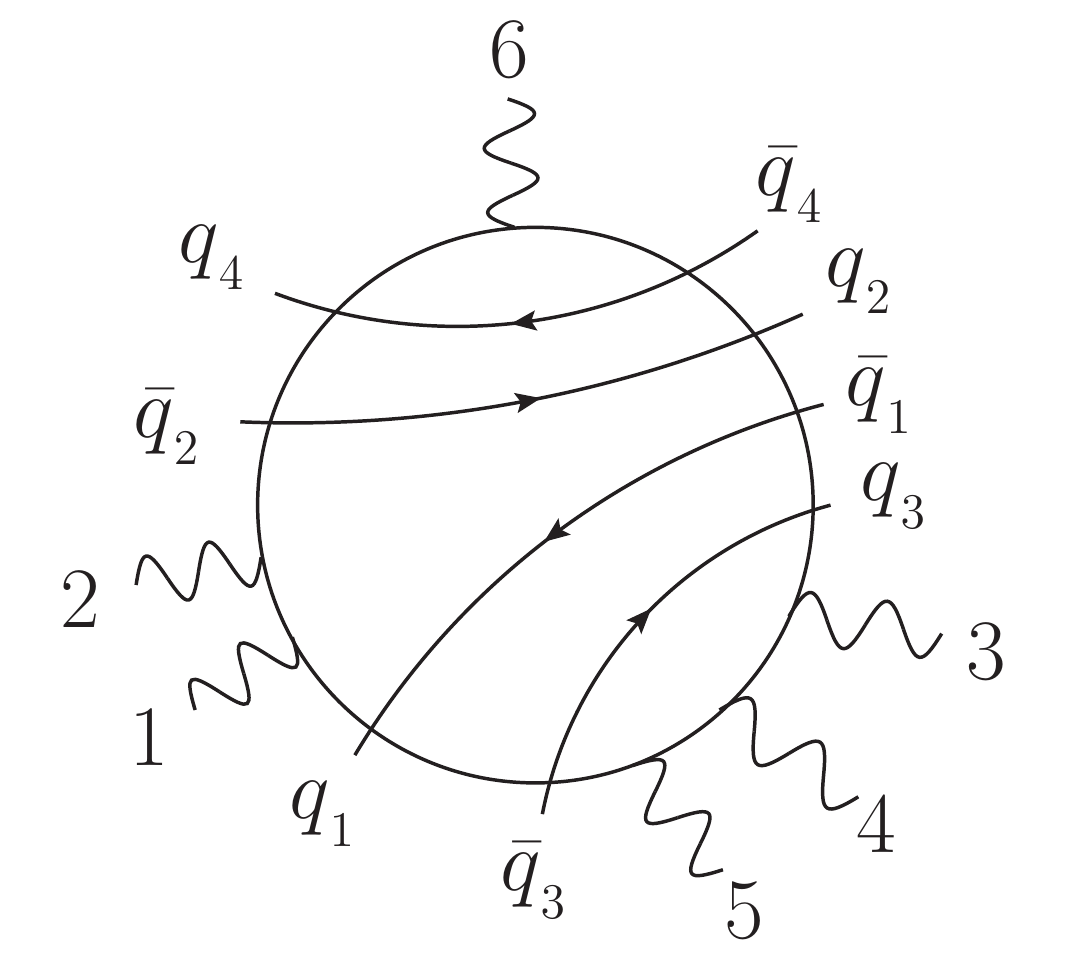}
\end{center}
\caption{Pictorial representation of the six color ordered amplitudes in \cref{eq:example1}.}\label{fig:example1}
\end{figure}

Before properly defining the photon exchange permutations $\Gamma(\tau,\sigma,\{n_i\})$ and thereby proving the general expression for the partial amplitudes, we give the instructive example of the partial amplitude multiplying the color structure
\begin{equation}\label{eq:example1color}
 \left(\frac{-1}{N}\right)^2(T^{a_1}T^{a_2})_{i_1\,\bar{j_2}}\delta_{i_2\,\bar{j_1}}\;(T^{a_3}T^{a_4}T^{a_5})_{i_3\,\bar{j_3}}\;(T^{a_6})_{i_4\,\bar{j_4}}
\end{equation}
in a four quark line, six gluon amplitude. This color structure contains the three cycles  $c_1=\{q_1,1,2,\bar{q}_2,q_2,\bar{q}_1\}$, $c_2=\{q_3,3,4,5,\bar{q}_3\}$ and $c_3=\{q_4,6,\bar{q}_4\}$. According to the derivation of the color structures we have to sum over all possible cyclic subdiagrams whose external legs are ordered according to the three cycles and over all possibilities of exchanging two photons between them. There are four different ways the two photons can be exchanged between the four quark lines in the three cycles. Each of these four contributions is straight forward to express by color ordered amplitudes and the partial amplitude to the color structure \cref{eq:example1color} is given by 
\begin{align}\label{eq:example1}
& \begin{rcase}&\mathrel{\phantom{+}}A(q_1,1,2,\bar{q}_2,q_2,\bar{q}_1,q_3,3,4,5,\bar{q}_3,q_4,6,\bar{q}_4)\\
&+A(q_1,1,2,\bar{q}_2,q_2,\bar{q}_1,q_4,6,\bar{q}_4,q_3,3,4,5,\bar{q}_3)\end{rcase}&&\text{photons between $q_1$, $q_3$, $q_4$}\notag\\[+0.3cm]
&\begin{rcase}
 & +A(q_1,1,2,\bar{q}_2,q_3,3,4,5,\bar{q}_3,q_4,6,\bar{q}_4,q_2,\bar{q}_1)\notag\\
&+A(q_1,1,2,\bar{q}_2,q_4,6,\bar{q}_4,q_3,3,4,5,\bar{q}_3,q_2,\bar{q}_1)\notag
 \end{rcase}&&\text{photons between $q_2$, $q_3$, $q_4$}\notag\\[+0.3cm]
&\begin{rcase}
&+A(q_1,1,2,\bar{q}_2,q_3,3,4,5,\bar{q}_3,q_2,\bar{q}_1,q_4,6,\bar{q}_4)\hspace{-0.3cm}
 \end{rcase}&&\begin{aligned}
               &\text{photon between $q_2$, $q_3$}\\[-0.1cm]
&\text{and between $q_1$, $q_4$}
              \end{aligned}
\notag\\[+0.3cm]
&\begin{rcase}
  &+A(q_1,1,2,\bar{q}_2,q_4,6,\bar{q}_4,q_2,\bar{q}_1,q_3,3,4,5,\bar{q}_3)\hspace{-0.3cm}
 \end{rcase}&&\begin{aligned}
               &\text{photon between $q_2$, $q_4$}\\[-0.1cm]
&\text{and between $q_1$, $q_3$}
              \end{aligned}\,,
\end{align}
where the helicities $h_i$ of the gluons have been suppressed. Note that the first two color ordered amplitudes in \cref{eq:example1} each contain contributions from diagrams with a non-Abelian vertex between $q_1$, $q_3$, $q_4$. These contribution cancel in the sum of the amplitudes due to the anti-symmetry of the three gluon vertex. Similar cancellations appear between the third and fourth amplitude in \cref{eq:example1}. A more intuitive pictorial representation of these color ordered amplitudes can be found in \cref{fig:example1}.  

We start the proof of \cref{eq:colorDecomposition} by investigating the partial amplitudes multiplying the cyclic color structures of the form \cref{eq:colorStructureCropped}, as these are the simplest ones. As already stated before, given a particular Feynman diagram contributing to the cyclic color structure \cref{eq:colorStructureCropped}, it is a simple fact that it can be drawn in planar fashion such that the external legs follow the ordering of the cycle corresponding to the cyclic color structure
\begin{equation}
q_1,\,\sigma(1),\,\dots\,\sigma(n_1),\bar{q}_{\pi(2)},\,q_{\pi(2)},\,\sigma(n_1+1),\dots, \sigma(n_2),\,\bar{q}_{\pi(3)},\,q_{\pi(3)},\,\sigma(n_2+1),\dots, \bar{q}_1
\end{equation}
if we go clockwise around the diagram. Its contribution to the partial amplitude is straightforwardly obtained by applying the color ordered Feynman rules of \cref{fig:ColorOrderedRules} to this planar diagram. Hence, summing up the contributions from all possible Feynman diagrams is equivalent to summing over all possible color ordered Feynman diagrams contributing to the color ordered amplitude
\begin{equation}
 A(q_1,\,\sigma(1),\,\dots\,\sigma(n_1),\bar{q}_{\pi(2)},\,q_{\pi(2)},\,\sigma(n_1+1),\dots, \sigma(n_2),\,\bar{q}_{\pi(3)},\,q_{\pi(3)},\,\sigma(n_2+1),\dots, \bar{q}_1)\,.
\end{equation}
In general the color structure \eqref{eq:colorstructuresQuarks} is a product of $p+1$ cyclic color structures. From the derivation of the color structures we know that each diagram contributing to the color structure \eqref{eq:colorstructuresQuarks} can be composed of $p+1$ planar sub-diagrams whose external legs are ordered according to the cycles in the color structure. These cyclic sub-diagrams are connected only via QED type gluon exchange between various of the quark lines. To get the whole partial amplitude we have to sum over all possible cyclic sub-diagrams as well as over all possibilities of photon exchange between them. Again the contribution of such diagrams to the partial amplitude can be obtained by applying the color ordered Feynman rules of \cref{fig:ColorOrderedRules} to it. However, there is a small sign subtlety. The color ordered quark gluon vertex is anti-symmetric whereas the ordinary quark gluon vertex always comes with a plus sign irrespective of the ordering of its legs. Fortunately, 
the $q\bar{q}g$ vertex appears an even number of times as it is only present in the photon exchange between the cyclic sub-amplitudes.

In order to be able to write down expressions for the partial amplitudes in terms of color ordered amplitudes we need to have control over the photon exchange. In fact, it is straight forward to construct a linear combination of color ordered amplitudes containing the photon exchange between a given number of quark lines. The idea is to let the considered quark lines face against each other and to sum over all non-cyclic permutations of the quarks in order to sum up all possibilities of photon exchange between the quark lines. To be more precise, the photon exchange between $k$ quark lines is given by the linear combination
\begin{equation}\label{eq:photonexchange}
\sum_{\kappa\in S_{k}/Z_k}A(q_{\kappa(1)},R_{\kappa(1)},\bar{q}_{\kappa(1)},q_{\kappa(2)},R_{\kappa(2)},\bar{q}_{\kappa(2)},\dots,q_{\kappa(k)},R_{\kappa(k)},\bar{q}_{\kappa(k)})\,,
\end{equation}
where $R_i$ can be any additional partons connected to the right of the quark line of flavor $i$. The only thing we have to worry about is whether the diagrams containing gluon trees connecting cyclic sub-diagrams cancel within the sum over non-cyclic permutations $\kappa\in S_{k}/Z_k$. However, this cancellation is a direct consequence of the symmetry properties of the color ordered gluon vertices. As can be easily checked using the color ordered Feynman rules of \cref{fig:ColorOrderedRules}, the three gluon vertex is anti-symmetric and the four gluon vertex gives zero when symmetrized over more than two of its legs. For the one-loop amplitudes we need to slightly generalize \cref{eq:photonexchange} to the case where $k$ quark lines couple via photons to one side of the quark line $m$  and additional legs are connected to the same side of the quark line without sharing any subtree with the other legs on this side of the quark line
\begin{align}
& \sum_{\kappa\in S_k}\;\sum_{\sigma\in\operatorname{OP}\{L_m\}\{q_{\kappa(1)},R_{\kappa(1)},\bar{q}_{\kappa(1)},\ldots,q_{\kappa(k)},R_{\kappa(k)},\bar{q}_{\kappa(k)}\}}A(q_{m},R_m,\bar{q}_{m},\sigma)\label{eq:photonexchange1}\;,\\
 &\sum_{\kappa\in S_k}\;\sum_{\sigma\in\operatorname{OP}\{R_m\}\{\operatorname{rev}(q_{\kappa(1)},R_{\kappa(1)},\bar{q}_{\kappa(1)},\ldots,q_{\kappa(k)},R_{\kappa(k)},\bar{q}_{\kappa(k)})\}}A(\bar{q}_m,L_m,q_{m},\sigma)\label{eq:photonexchange2}\,.
\end{align}
Here $L_i$ and $R_i$ denote arbitrary sets of external legs and $\operatorname{OP}\{\alpha_1\}\{\alpha_2\}$ denotes all permutations preserving the order of each of the $\alpha_i$. The reason to reverse the order of the legs $q_{\kappa(1)},R_{\kappa(1)},\bar{q}_{\kappa(1)},\ldots,q_{\kappa(k)},R_{\kappa(k)},\bar{q}_{\kappa(k)}$ in \cref{eq:photonexchange2} is the anti-symmetry of the quark gluon vertex in color ordered diagrams. Since the quark gluon vertex of QCD is symmetric, not reversing these legs would lead to a non uniform relative sign
 between the color ordered diagrams and the corresponding contributions of Feynman diagrams to the partial amplitudes of QCD.

Now we can immediately write down the following contribution to the partial amplitude of the general color structure \cref{eq:colorstructuresQuarks} with $p+1$ cycles of length $l_i$ 
\begin{equation}\label{eq:maxPoly}
 \sum_{\substack{\pi_i\in Z_{l_i}\\\kappa\in S_{\text{\tiny $\scriptscriptstyle p+1$}}/Z_{\text{\tiny $\scriptscriptstyle p+1$}}}}\raisebox{-3.5cm}{\includegraphics[width=8cm]{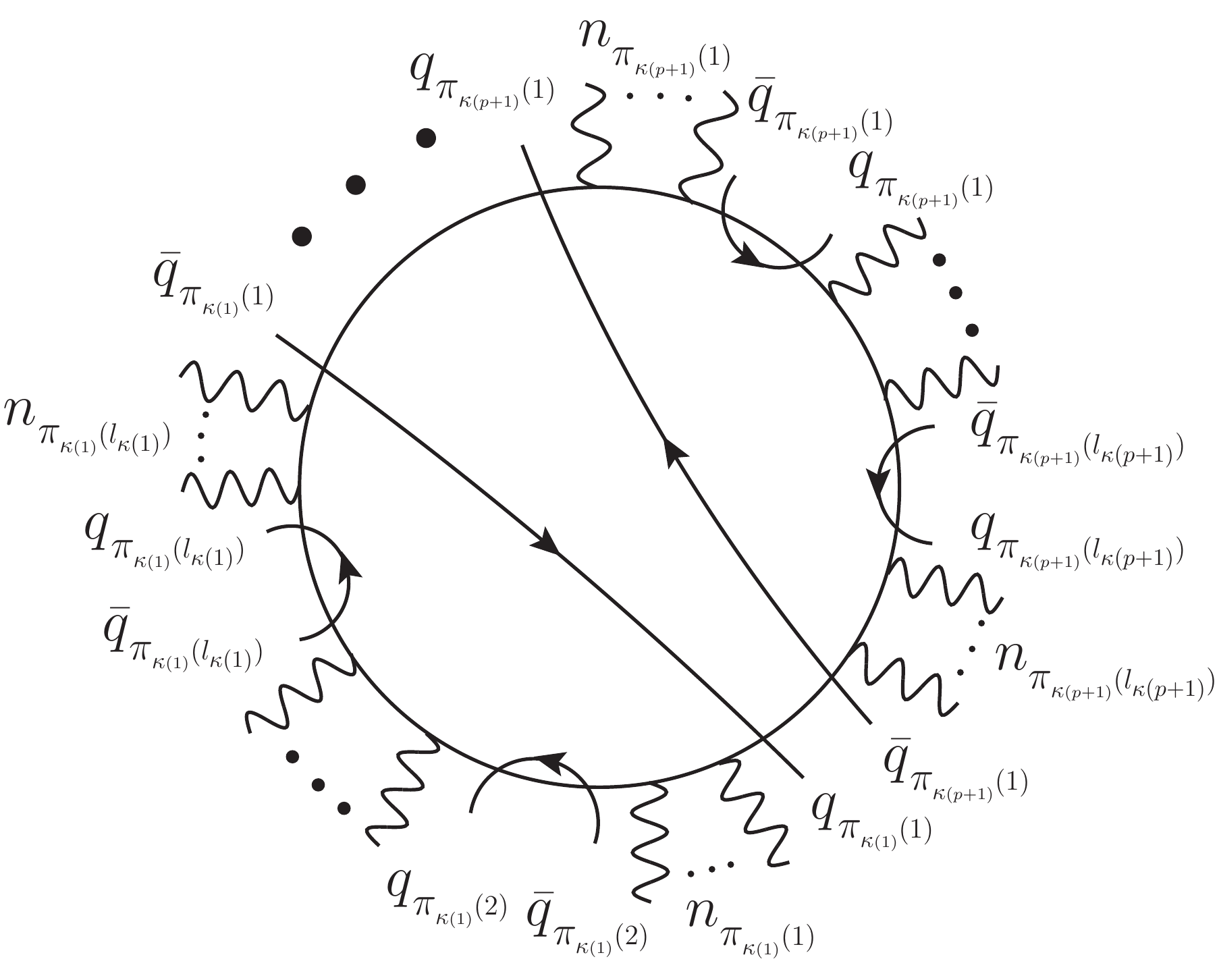}}\,.
\end{equation}
Obviously this sum contains all contributions to the partial amplitude where $p$ photons are exchanged between $p+1$ quark lines. The cyclic permutations $\pi_i\in Z_{l_i}$ fix the quark lines involved in the QED type gluon exchange between the cycles and the non-cyclic permutation $\kappa\in S_{\text{\tiny $\scriptscriptstyle p+1$}}/Z_{\text{\tiny $\scriptscriptstyle p+1$}}$ ensures that we get all possibilities $p$ photons can be exchange between these quark lines. Note that the sum of the first four amplitudes in the partial amplitude \cref{eq:example1} are given by \cref{eq:maxPoly}. 

Whenever the number of quarks in the amplitude exceeds the number of cycles in the color structure, there are additional contributions where $p$ photons are exchanged between up to $2p$ quark lines. A characterization of all these contribution leads to a proper definition of the photon exchange permutations $\Gamma(\tau,\sigma,\{n_i\})$ and will conclude the proof of \cref{eq:colorDecomposition}.

It is convenient to represent the topologically nonequivalent possibilities of photon exchange between cycles by connected planar diagrams. These photon exchange diagrams are assembled from convex polygons, with each pair of polygons sharing at most one vertex. The vertices are indistinguishable and represent cycles. Diagrams which can be related by exchanging subgraphs attached to a polygon are considered equivalent. A convex $k$-gon represents the photon exchange between $k$ quark lines of $k$ cycles. If a vertex is part of $m$ polygons than $m$ of the quarks of a cycle are involved in photon exchange with other cycles. Alternatively, the photon exchange topologies between $k$ cycles can be represented by the set of connected graphs on $k$ unlabeled vertices where every block is a complete graph. The complete graphs, i.\,e.~graphs where every pair of vertices is connected by an edge, take the role of the convex polygons and incorporate the equivalence relation defined above. 
Graphically, both representations differ only by attaching or removing some edges inside the polygons or inside the complete graphs. 

\begin{figure}
 \centering
\subfigure[one cycle]{\hspace{.8cm}\includegraphics[width=.45cm]{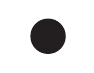}\hspace{.8cm}}\hfill
\subfigure[two cycles]{\includegraphics[width=2.2cm]{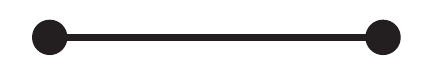}}\hfill
\subfigure[three cycles]{\includegraphics[height=2cm]{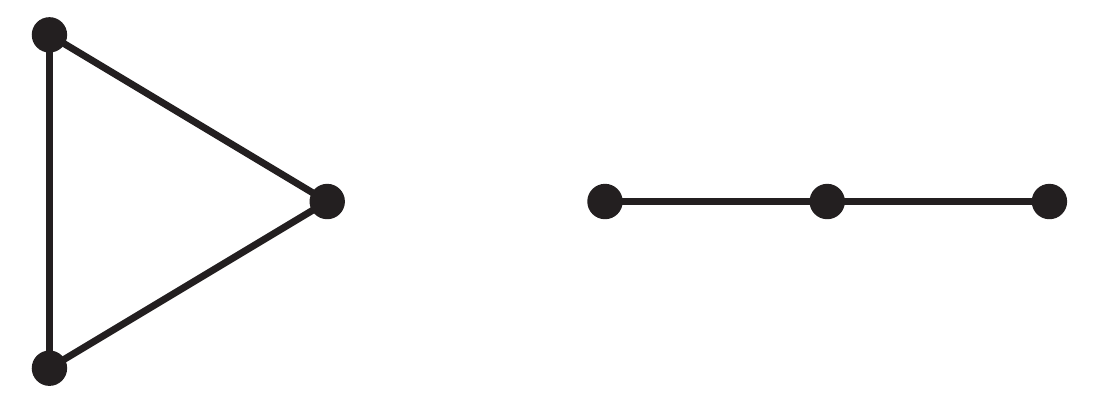}}\hfill
\subfigure[four cycles]{\includegraphics[height=2cm]{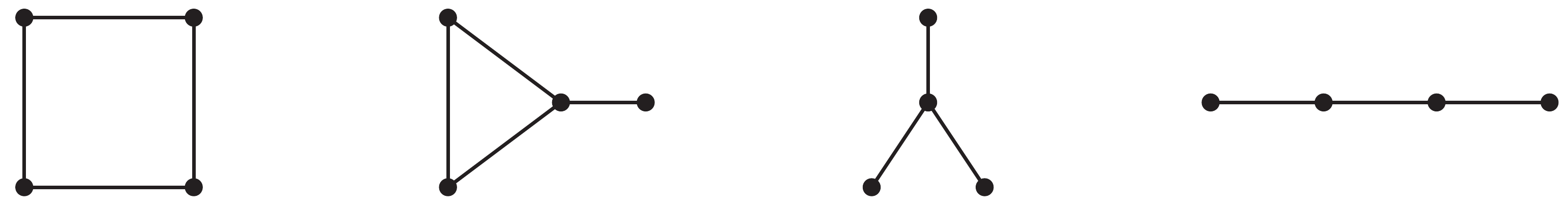}}\hfill
\subfigure[five cycles]{\includegraphics[width=0.45\textwidth]{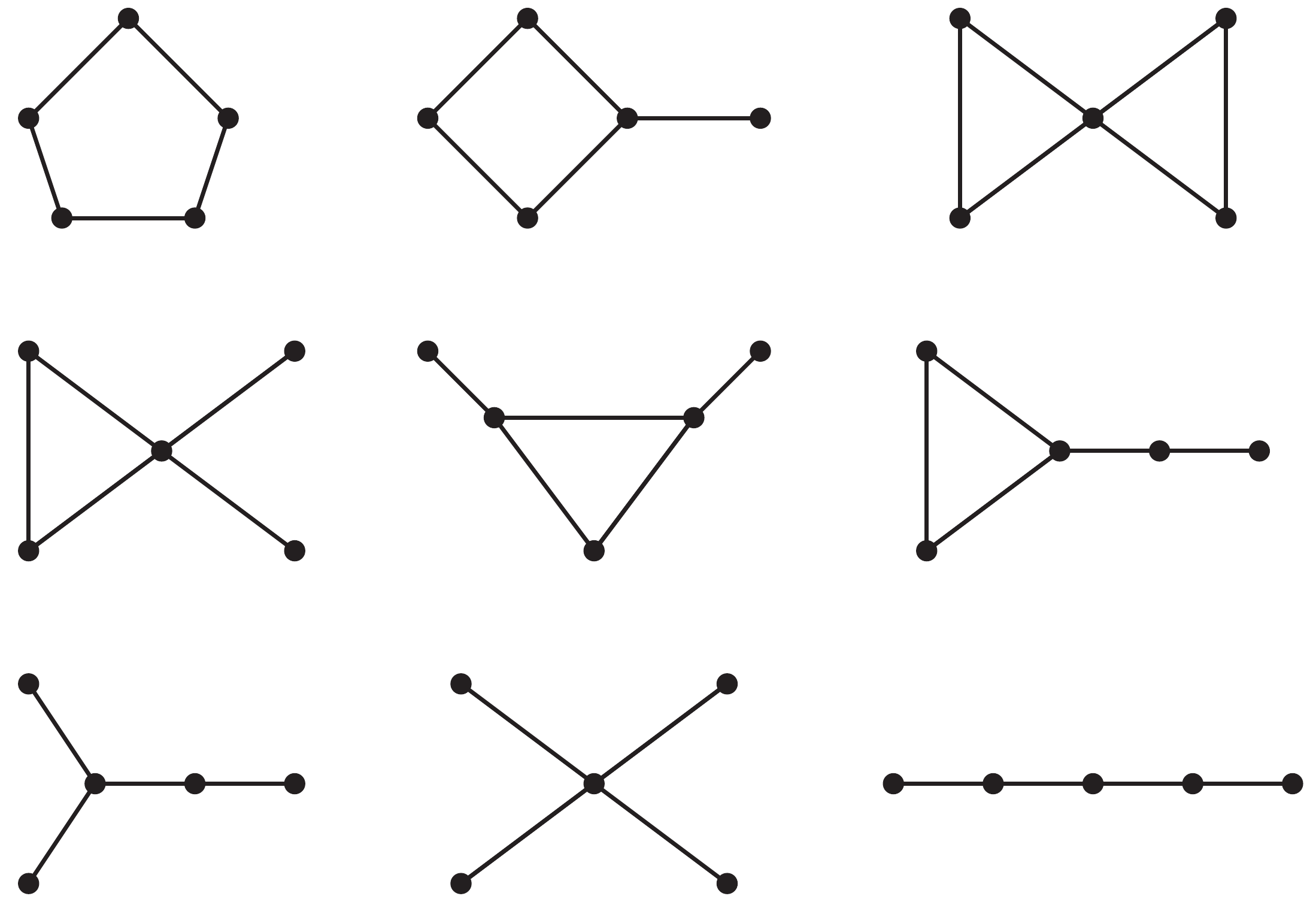}}\hfill
\subfigure[six cycles]{\includegraphics[width=0.45\textwidth]{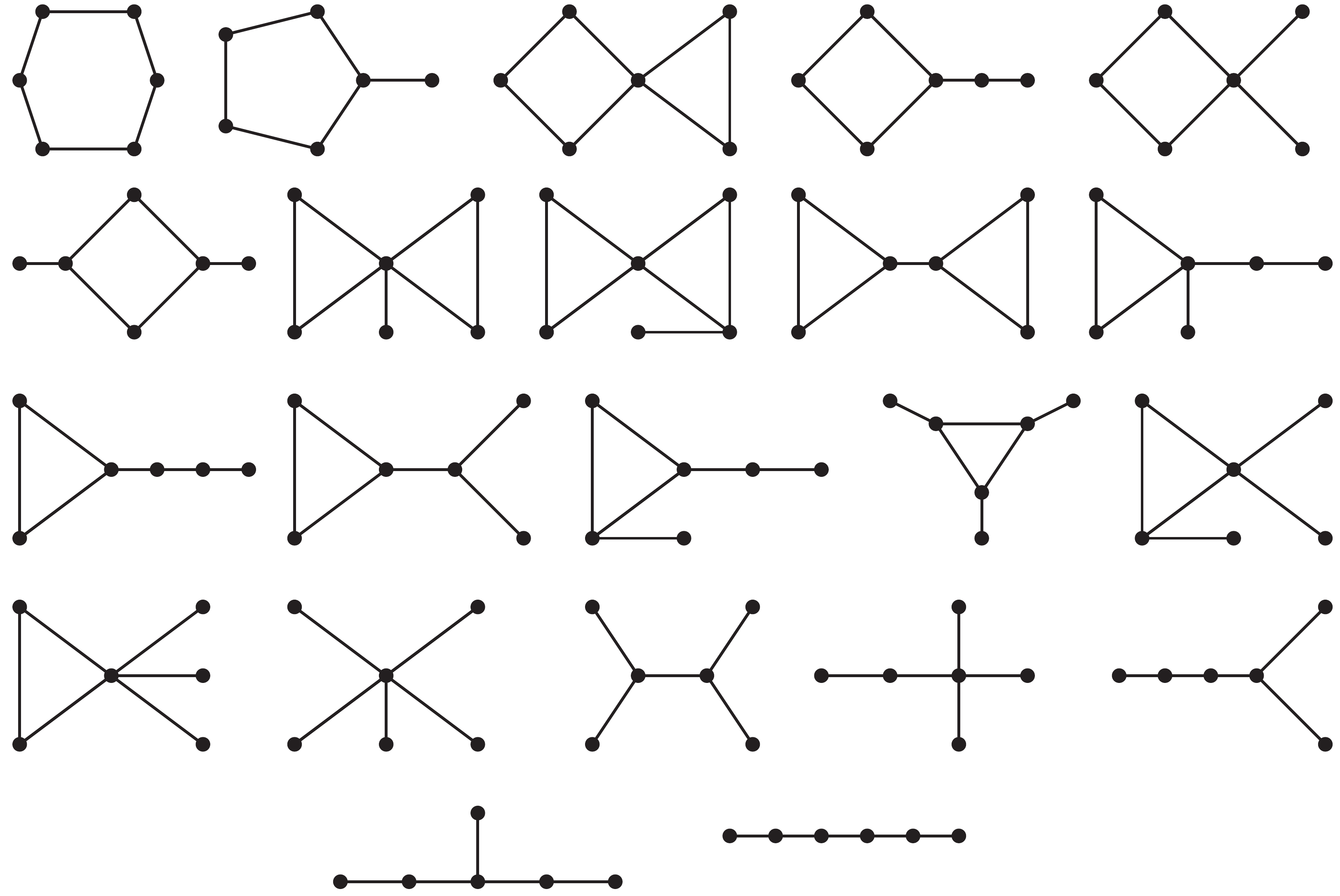}}
\caption{Photon exchange topologies for up to six cycles.}\label{fig:photonTopo}
\end{figure}

In \cref{fig:photonTopo} the photon exchange topologies for up to six cycles are listed. Within a photon exchange diagram consisting of $p+1$ cycles and $m$ polygons there are $p+m$ quark lines involved in the photon exchange.
Given a certain color structure which may be represented by a set of cycles or by permutations $\tau$, $\sigma$ of quarks, gluons and a partition $\{n_i\}$ of the gluons, the photon exchange topologies are straight forward to translate into disjoint subsets of the photon exchange permutations $\Gamma(\tau,\sigma,\{n_i\})=\Gamma(C_1,\dots,C_{p+1})$.  We define $\Gamma(\alpha_1,\dots,\alpha_{p+1})$ to be the set of permutations obtained by applying the following rules to each photon exchange topology:%
\begin{enumerate}
\item sum over all possibilities the $\alpha_i$ can be inserted into the topology,
\item for each of these insertions sum over all possible choices of quark lines involved in the photon exchange,
\item enforce the photon exchange between the chosen quarks according to \cref{eq:photonexchange,eq:photonexchange1,eq:photonexchange2}.
\end{enumerate}
At tree level the $\alpha_i$ are simply the cycles $C_i$, hence only \cref{eq:photonexchange} is necessary to enforce the photons connecting the choice of quarks lines. 
Apparently, the rules given above yield \cref{eq:maxPoly} when applied to a $(p+1)$-gon. It is instructive to recall our initial example given in \cref{eq:example1}. The first four amplitudes in \cref{eq:example1} correspond to the triangle photon exchange diagram in \cref{fig:photonTopo}. Since the amplitudes involves four quark lines, only one of the cycles has two quarks and the second photon exchange topology corresponds to the last two of the color ordered amplitudes in \cref{eq:example1}. The sum over photon exchange permutations is straight forward to implement into a computer algebra system. For details on the \texttt{Mathematica} implementation \texttt{QCDcolor} as well as some more examples of partial amplitudes we refer to \cref{appendix:QCDcolor}.
\subsection{One-Loop level}\label{section:QCDcolorLoop}
At one-loop level the color decomposition is more involved. Despite its increased complexity compared to the tree-level case, it is still possible to directly construct a decomposition of a general QCD one-loop amplitude into color structures and primitive amplitudes. The general idea is again 
\begin{itemize}
\item identify all color structures,
\item characterize all color ordered diagrams contributing to a particular partial amplitude, and
\item construct a linear combination of primitive amplitudes that equals the sum of all those diagrams.
\end{itemize}
Given an arbitrary one-loop Feynman diagram we cut one loop propagator leading to a tree diagram whose color decomposition we already know. Cutting a pure fermion loop we end up with a tree diagram with one additional quark line. From the previous section we know that a color structure in the color decomposition of this tree diagram contains in general up to $k+1$ cycles, where $k$ is the number of quark--anti-quark pairs present in the one-loop amplitude. Depending on the additional partons in the cycle containing the loop quark line we get three different types of color structures when contracting the color indices of the loop quark--anti-quark pair. 

If there are additional quarks in the loop cycle we get the tree level color structures up to a factor of $n_f$
\begin{equation}
 n_f\left(\frac{-1}{N}\right)^{p}\left(\prod_{\alpha=1}^k\left(n_\alpha\right)_{i_\alpha\,\bar{j}_{\tau(\alpha)}}\right)\,,
\end{equation}
and all color ordered diagrams contributing to its partial amplitude are composed of $p+1$ cyclic subdiagrams connected by $p$ photons, with one cyclic subdiagram containing the fermion loop.

If the loop cycle contains only gluons and the loop quark, the one-loop color structures are
\begin{equation}
  n_f \Tr(n_0)\left(\frac{-1}{N}\right)^{p+1}\left(\prod_{\alpha=1}^k\left(n_\alpha\right)_{i_\alpha\,\bar{j}_{\tau(\alpha)}}\right)\,,
\end{equation}
with $\Tr(n_0)=\Tr(T^{\sigma(1)}\dots T^{\sigma(n_0)})$. The corresponding partial amplitude gets contributions from color ordered diagrams composed of $p+1$ cyclic diagrams and a subdiagram containing the fermion loop and the cyclically ordered gluons $\{\sigma(1),\dots,\sigma(n_0)\}$ of the trace. These subdiagrams are connected by $p+1$ photons. 

The last possibility is a loop cycle without additional quarks or gluons, leading to the color structure
\begin{equation}
  -n_f\left(\frac{-1}{N}\right)^{p}\left(\prod_{\alpha=1}^k\left(n_\alpha\right)_{i_\alpha\,\bar{j}_{\tau(\alpha)}}\right)\,,
\end{equation}
which is equal to the tree level color structures up to a factor of $\sfrac{-1}{N} \delta_{i i} n_f=-n_f$. The color ordered diagrams contributing to the corresponding partial amplitude are composed of $p$ cyclic subdiagrams and a fermion loop, all connected by $p+1$ photons.

In order to identify all color structures in the non fermion loop part of the amplitude it is convenient to start with the case of at least one photon in the loop. Since there is no color flowing along the photon line, the color structures are equal to the ones in the tree diagram obtained by removing the photon
\begin{equation}
  \left(\frac{-1}{N}\right)^{p+1}\left(\prod_{\alpha=1}^k\left(n_\alpha\right)_{i_\alpha\,\bar{j}_{\tau(\alpha)}}\right)\,,
\end{equation}
with an additional factor of $\sfrac{-1}{N}$ originating from the loop photon. The derivation of the remaining color structures is now simplified as we can neglect the $\sfrac{-1}{N}$ part in the Fierz identity \cref{eq:colorFlow} when contracting the loop gluons, since it is either not contributing or leads to the case of at least one photon in the loop. 

Contracting the adjoint indices of two gluons in the tree level color structures \cref{eq:colorstructuresQuarks} leads to three different color structures at one-loop level. Contracting adjacent gluons leads to the leading order color structures
\begin{equation}
  N \left(\frac{-1}{N}\right)^{p}\left(\prod_{\alpha=1}^k\left(n_\alpha\right)_{i_\alpha\,\bar{j}_{\tau(\alpha)}}\right)\,.
\end{equation}
A color ordered diagram contributing to the leading order partial amplitude $P_0$ is given by $p+1$ cyclic subdiagrams connected by $p$ photons with one of the cyclic subdiagrams containing the loop.

The leading order color structures are up to the sign equal to the cycle split color structures
\begin{equation}\label{eq:colorStrucCycSplit}
  \left(\frac{-1}{N}\right)^{p-1}\left(\prod_{\alpha=1}^k\left(n_\alpha\right)_{i_\alpha\,\bar{j}_{\tau(\alpha)}}\right)\,
\end{equation}
obtained by contracting two non adjacent gluons $n+1$, $n+2$ separated by quarks of one of the tree level cycles $C_{\text{split}}=\{C_i,n+1,C_j,n+2\}$, thereby splitting it into two cycles $C_i$, $C_j$ of the one-loop color structure. Speaking of diagrams, contracting these gluons in the tree diagram leads to an unconventionally drawn one-loop diagram, since the external legs of one of the split cycles, e.\,g.~$C_j$, and of all the cycles being connected to it by photons, face towards the inside of the loop. Flipping all these subtrees facing inside the loop to the outside of the loop leads to the desired planar way of drawing the Feynman diagram.

Contracting two non adjacent gluons emitted between a quark and the successive anti-quark leads to the color structures
\begin{equation}
 \Tr(n_0) \left(\frac{-1}{N}\right)^{p}\left(\prod_{\alpha=1}^k\left(n_\alpha\right)_{i_\alpha\,\bar{j}_{\tau(\alpha)}}\right)\,
\end{equation}
with $n_0>1$. A color ordered diagram contributing to the trace partial amplitude $P_3$ constitutes of $p+1$ cyclic subdiagrams, one of them containing the loop, and the $n_0$ gluons of the trace being connected to the loop in reversed cyclic ordering with respect to the trace, without sharing a subtree with the remaining external legs. Note that contracting gluons of different cycles of the tree color structure will just fuse them to one larger cycle and leads to a photon loop color structure. 

In summary, the decomposition of an arbitrary one-loop QCD amplitude with $k$ quark--anti-quark pairs and $n$ gluons into color structures and partial amplitudes reads
\begin{align}
 \mathcal{A}^{1\text{-loop}}_{(q \bar{q})^k}&=g^{n+2k}\sum_{\substack{\sigma \in S_n\\\tau \in S_k}}\Biggl[\sum_{\{n_i\}}\left(\sfrac{-1}{N}\right)^{p}\left(\prod_{\alpha=1}^k\left(n_\alpha\right)_{i_\alpha\,\bar{j}_{\tau(\alpha)}}\right)\bigr(N(P_0-P_1)-\sfrac{1}{N}P_2+n_f(P^f_{0}-P^f_{1})\bigl)\notag\\
&\=\phantom{g^{n+2k}\sum_{\substack{\sigma \in S_n\\\tau \in S_k}}\Biggl[}+{\sum_{\{n_i\}}}^\prime\left(\sfrac{-1}{N}\right)^{p}\Tr(n_0)\left(\prod_{\alpha=1}^k\left(n_\alpha\right)_{i_\alpha\,\bar{j}_{\tau(\alpha)}}\right)\bigr(P_3-\sfrac{n_f}{N}P^f_2\bigl)\Biggr]
\,.\label{eq:colorDecomposition1loop}
\end{align}
The partial amplitudes $P_i$ and $P^f_i$ depend on the permutation $\tau$ of the anti-quarks as well as on the partition $\{n_i\}$ and permutation $\sigma$ of the gluons. The integer $p(\tau)$ has been defined in \cref{eq:power}, and the prime on the sum over gluon partitions $\{n_i\}$ in the trace part indicates a sum over $2\leq n_0\leq n_1\leq\dots\leq n_k=n$, compared to the sum over $0=n_0\leq n_1\leq\dots\leq n_k=n$ in the non-trace part.

In the remainder of this section we present a case by case constructions of all the partial amplitudes in \cref{eq:colorDecomposition1loop} as linear combination of primitive amplitudes.  Each of the partial amplitudes is equal to the gauge invariant sum over a well defined set of color ordered diagrams. In all cases these sets can be further decomposed into gauge invariant subsets of color ordered diagrams. The construction of the linear combination of primitive amplitudes equaling a certain gauge invariant set of color ordered diagrams is solely based on the symmetries of the color ordered vertices \cref{fig:ColorOrderedRules} and the fact that the position of the loop can be fixed using the routing of the quarks. 

\subsubsection{The leading order partial amplitudes \texorpdfstring{$\bm{P_0}$ and $\bm{P^f_0}$}{P_0 and P^f_0}}
The partial amplitudes $P_0$ and $P^f_0$ are the only ones contributing to the leading order in a large $N$ expansion with $P^f_0$ being suppressed by a relative factor of $\sfrac{n_f}{N}$. 

The non fermion loop partial amplitude $P_0$ has the most similarities with the tree level partial amplitudes and is given by 
\begin{equation}\label{eq:P_0}
 P_0(\tau,\sigma,\{n_i\})=\sum_{i=1}^{p+1}\;\,\sum_{\kappa\in\Gamma(\tau,\sigma,\{n_i\})}A(\r_i\circ\kappa)\,.
\end{equation}
It involves two ingredients. First of all, we have to sum over all $p+1$ possibilities which cycle is containing the loop by appropriately choosing the routings $r=\{r_1,\dots,r_k\}\in\{L,R\}^{p+1}$ of the quarks, i.\,e.~$\r_i\circ\kappa$ denotes setting $r_j=L$ for the quarks of the loop cycle $C_i$ and remaining routings are fixed accordingly. Second, we have to sum over all possibilities to connect the cycles by $p$ photons by summing over the photon exchange permutations $\Gamma(\tau,\sigma,\{n_i\})$ which have been defined at the end of the previous section.

The leading order fermion loop partial amplitude is more involved. The idea is to split all contributing color ordered diagrams into the $(p+1)2^p$ gauge invariant subsets with one cycle $C_i$ containing the loop, part of the cycles connecting to the fermion loop via photons and all remaining cycles being connected to the quarks of the loop cycle $C_i$ by photons. Let $\mathcal{C}_{i}\defi \{\,\{\alpha,\beta\}\;|\; \alpha\cup\beta=C\;\wedge\;\alpha\cap\beta=\{\}\;\wedge\; C_i\in\alpha\}$ denote all possibilities to split the set of cycles $C=\{C_1,\dots,C_{p+1}\}$ into two disjoint subsets $\alpha$, $\beta$ with $\alpha$ containing the loop cycle $C_i$. We have to sum over all possibilities photons can be exchanged between the cycles $\alpha$ as well as over all possibilities the cycles $\beta$ can be connected to the fermion loop by photons. Similar to the tree level case and \cref{eq:P_0}, the photon exchange between the cycles $\alpha$ is given by a sum over the photon exchange permutations $\Gamma(\alpha)
$. The permutations of the external legs of the cycles $\beta$ are given by the set $\Gamma_{\!\star}(\beta)\defi f_q (\Gamma(\{q,\bar{q}\},\beta))$. The additional cycle $\{q,\bar{q}\}$ represents the fermion loop and the function $f_q$ simply removes it from the permutations $\Gamma(\{q,\bar{q}\},\beta)$ such that the remaining permutations start with the quark that succeeded $\bar{q}$, i.\,e.~$f_q(\{\bar{q},\gamma_1,q\})=\{\gamma_1\}$ and $f_q(\{\gamma_1,q,\bar{q},\gamma_2\})=\{\gamma_2,\gamma_1\}$. If $\kappa_1\in\Gamma(\alpha)$ and $\kappa_2\in\Gamma_{\!\star}(\beta)$ are two such permutations of the external legs in $\alpha$ and $\beta$ we need to ensure that the external legs in $\kappa_2$ only connect directly to the fermion loop without sharing any subtrees with $\kappa_1$. Similar to \cref{eq:PartialAmpAdjoint} this can be accomplished by summing over cyclic ordered permutations of the two sets of external legs. Hence, the partial amplitude is given by
\begin{equation}\label{eq:P^f_0}
 P^f_0(\tau,\sigma,\{n_i\})=\sum_{i=1}^{p+1}\;\sum_{\{\alpha,\beta\}\in \mathcal{C}_{i}}\;\sum_{\substack{\kappa_1\in\Gamma(\alpha)\\\kappa_2\in\Gamma_{\!\star}(\beta)}}\;\;\sum_{\rho\in\text{COP}\{\mathcal{R}_i\circ\,\kappa_1\}\{\mathcal{R}_{\star}\circ\,\operatorname{rev}(\kappa_2)\}}(-1)^{|\kappa_2|}A_f(\rho)\,.
\end{equation}
The order of the external legs $\kappa_2$ has been reversed in order to ensure a uniform relative sign of $(-1)^{|\kappa_2|}$ between the color ordered diagrams of the amplitudes and the contributions to the partial amplitude. The routing $\mathcal{R}_{\star}$ is defined as $\mathcal{R}_{\star}\circ\{\dots,q_l,\dots,\bar{q}_l,\dots\}=\{\dots,q^R_l,\dots,\bar{q}^R_l,\dots\}$, $\mathcal{R}_{\star}\circ\{\dots,\bar{q}_l,\dots,q_l,\dots\}=\{\dots,\bar{q}^L_l,\dots,q^L_l,\dots\}$ and the routing $\mathcal{R}_i$ has been defined below \cref{eq:P_0}.

\subsubsection{The fermion loop partial amplitude \texorpdfstring{$\bm{P_1^f}$}{P^f_1}}
All diagrams where the fermion loop connects via photons to more than one cycle are contributing to the partial amplitude $P^f_1$. It is straightforward to construct a linear combination of primitive amplitudes that equals all these diagrams. Starting point are the photon exchange permutations $\Gamma(\tau,\sigma,\{n_i\})$ ensuring all possible types of photon exchange between the cycles. For each of these permutations we have to sum over all possibilities to put the fermion loop between quark lines that are involved into photon exchange between cycles by appropriately choosing the quark routings. The quarks enclosing the fermion loop all get $r_i=R$ and the routings of the remaining quark are fixed by their orientation with respect to the loop. Denoting the set of all such routings by $\overline{\mathcal{R}}(\kappa)$, the partial amplitude reads
\begin{equation}\label{eq:P^f_1}
 P^f_1(\tau,\sigma,\{n_i\})=\sum_{\kappa\in\Gamma(\tau,\sigma,\{n_i\})}\;\;\sum_{r\in \overline{\mathcal{R}}(\kappa)}A_f(r\circ\kappa)\,.
\end{equation}
\subsubsection{The trace partial amplitudes \texorpdfstring{$\bm{P_3}$ and $\bm{P^f_2}$}{P_3 and P^f_2}}
Building on the construction of the tree level partial amplitudes in \cref{eq:colorDecomposition} and the double trace color structures \cref{eq:PartialAmpAdjoint} the non fermion loop partial amplitude can be written down immediately
\begin{equation}\label{eq:P_3}
 P_3(\tau,\sigma,\{n_i\})=(-1)^{n_0}\sum_{i=1}^{p+1}\;\;\sum_{\kappa\in\Gamma(\tau,\sigma,\{n_i\})}\;\;\sum_{\rho\in\text{COP}\{\sigma(n_0),\dots,\sigma(1)\}\{\kappa\}}A(\r_i\circ\rho)\,.
\end{equation}
It involves a sum over all possibilities which cycle is containing the loop and over all possibilities of photon exchange between the cycles. Furthermore, we have to sum over all cyclic permutations $\text{COP}\{\sigma(n_0),\dots,\sigma(1)\}\{\kappa\}\}$ between the reversed gluons of the trace and each photon exchange permutation $\kappa\in\Gamma(\tau,\sigma,\{n_i\})$. The prefactor of $(-1)^{n_0}$ compensates the relative sign between the color ordered diagrams in the primitive amplitudes and the contributions to the partial amplitude. The routing $\r_i$ has been defined below \cref{eq:P_0} and fixes the loop to lie inside the cycle $C_i$.

The trace partial amplitude with a fermion loop has a similar structure
\begin{equation}\label{eq:P^f_2}
 P^f_2(\tau,\sigma,\{n_i\})=(-1)^{n_0}\sum_{\kappa\in\Gamma(\tau,\sigma,\{n_i\})}\;\;\sum_{\rho\in\text{COP}\{\sigma(n_0),\dots,\sigma(1)\}\{\kappa\}}\;\;\sum_{r\in\overline{\mathcal{R}}(\kappa)\cup \widetilde{\mathcal{R}}(\kappa)}A_f(r\circ\rho)\,,
\end{equation}
differs however by the set of routings we have to sum over. The fermion loop couples via photons to one or several cycles. Hence, for each photon exchange permutation $\kappa\in\Gamma(\tau,\sigma,\{n_i\})$ we have to sum over the routings $\widetilde{\mathcal{R}}(\kappa)$, locating the loop to the left of a quark line not involved into photon exchange between cycles,  as well as over the routings $\overline{\mathcal{R}}(\kappa)$, fixing the fermion loop to be located between cycles.  
\subsubsection{The cycle split partial amplitude \texorpdfstring{$\bm{P_1}$}{P_1}}
A diagram contributing to the cycle split partial amplitude can be categorized according to which cycles $C_i$, $C_j$ are the split cycles and furthermore according to which sets of cycles $\mathcal{C}_{i,j}\defi \{\,\{\alpha,\beta\}\;|\; \alpha\cup\beta=C\;\wedge\;\alpha\cap\beta=\{\}\;\wedge\; C_i\in\alpha\;\wedge\; C_j\in\beta\}$ are connected by photons to either $C_i$ or $C_j$. Let $\{\alpha_i,\alpha_j\}\in\mathcal{C}_{i,j}$ denote one such possibility to split the cycles $C=\{C_1,\dots,C_{p+1}\}$ into two subsets. Summing over cyclic permutations of each pair of photon exchange permutations $\kappa_i\in\Gamma(\alpha_i)$, $\kappa_j\in\Gamma(\alpha_j)$ we obtain the partial amplitude
\begin{equation}\label{eq:P_1}
 P_1(\tau,\sigma,\{n_i\})=\sum_{1\leq i< j\leq p+1}\;\sum_{\{\alpha_i,\alpha_j\}\in\mathcal{C}_{i,j}}\;\sum_{\substack{\kappa_i\in\Gamma(\alpha_i)\\\kappa_j\in\Gamma(\alpha_j)}}\;\sum_{\rho\in\text{COP}\{\mathcal{R}_i\circ\kappa_i\}\{\overline{\mathcal{R}}_j\circ\operatorname{rev}(\kappa_j)\}}\;(-1)^{|\kappa_j|}A(\rho)\,,
\end{equation}
where the photon exchange permutations of $\alpha_j$ have been reversed in order to match the set of color ordered diagrams described below 
\cref{eq:colorStrucCycSplit}. The factor of $(-1)^{|\kappa_j|}$ compensates the relative sign between the color ordered diagrams and the contributions to the partial amplitude. The routings of the quarks are fixed such that the quarks of $C_i$ and $C_j$ enclose the loop, i.\,e.~$\overline{\mathcal{R}}_j\circ\operatorname{rev}(\kappa_j)$ sets $r_m=R$ for all quarks in $C_j$ and fixes the routings of the remaining quarks accordingly. The routing $\mathcal{R}_i$ has been defined below \cref{eq:P_0}.
\subsubsection{The loop photon partial amplitude \texorpdfstring{$\bm{P_2}$}{P_2}}
The partial amplitudes with a photon in the loop are the most intricate. Gaining control over photons in the loop is simply more involved than managing photons in subtrees of loop diagrams. It is reasonable to split the partial amplitude into three gauge invariant pieces
\begin{equation}\label{eq:P_2}
P_2=P_{2,1} +P_{2,2} +P_{2,3}\,,	
\end{equation}
with $P_{2,1}$ corresponding to the set of color ordered diagrams where the loop photon in attached to one of the quark lines, $P_{2,2}$ corresponding to the set of color ordered diagrams where the loop photon connects two quark lines of one of the cycles, and $P_{2,3}$ corresponding to the color ordered diagrams where the loop photon connects quark lines of different cycles. 

The easiest of the three parts is
\begin{equation}
P_{2,1}(\tau,\sigma,\{n_i\})=\sum_{i=1}^{k}\;\sum_{\alpha\in\Gamma(\tau,\sigma,\{n_i\})}\;\sum_{\rho\in\text{Flip}_i(\alpha)}(-1)^{|L_i(\alpha)|}A(\r_{q_i}\circ \rho)\;.
\end{equation}
Whenever there are external legs $L_i(\alpha)$ to the left of the quark line $i$ in the photon exchange permutation $\alpha\in\Gamma(\tau,\sigma,\{n_i\})$, i.\,e. $\alpha=\{q_i,R_i(\alpha),\bar{q}_i,L_i(\alpha)\}$ or $\alpha=\{L_i(\alpha),q_i,R_i(\alpha),\bar{q}_i\}$, than we flip these legs to the right side of the fermion line $i$ by summing over the flip permutations $\text{Flip}_i(\alpha)\defi\{\{q_i,\kappa,\bar{q}_i\}\;|\;\kappa\in\text{OP}\{R_i(\alpha)\}\{\operatorname{rev}(L_i(\alpha))\}$. Here $R_i(\alpha)$ denotes the legs to the right of the quark line $i$ and $\text{OP}\{\alpha_1\}\{\alpha_2\}$ is the set of permutations that preserve the order of $\alpha_1$ and $\alpha_2$ respectively. $\r_{q_i}$ simply sets $r_i=R$ and fixes the other routings accordingly.

Slightly more complicated are the contributions of diagrams with a loop photon connecting two quark lines of one cycle
\begin{equation}
 P_{2,2}(\tau,\sigma,\{n_i\})=\sum_{i=1}^{p+1}\;\sum_{\{s,t\}\in \mathcal{P}_2(Q_i)}\;\sum_{\alpha\in\text{COP}_{s,t}(C_i)}\;\sum_{\rho\in\Gamma(\alpha,C\setminus C_i)}\!\!\!\!\!\!(-1)^{f_{i,s,t}(C,\rho)}\,A(\r_{i,s,t}\circ \rho).
\end{equation}
The construction of the linear combination of primitive amplitudes starts with the loop cycle $C_i$ and one possible choice of quarks $\{s,t\}\in \mathcal{P}_2(Q_i)$ of $C_i$ that are connected by the loop photon, with $Q_i=Q(C_i)$ denoting the set of all quark flavors in the cycle $C_i$ and $\mathcal{P}_k(S)\defi\{\alpha\;|\;\alpha \subset S \;\wedge |\alpha|=k\;\}$ denoting all subsets of cardinality $k$ of the set $S$. The loop cycle gets split into two parts $C_i=\{C_{i,s,t},C_{i,t,s}\}$, where $C_{i,s,t}$ starts with $q_s$ and ends with $\bar{q}_t$ and $C_{i,t,s}$ starts with $q_t$ and ends with $\bar{q}_s$. Taking into account all cyclic permutations of $C_{i,s,t}$ and $\operatorname{rev}(C_{i,t,s})$ ensures that both quark lines are part of the loop and connected by the loop photon. The set of all cyclic permutations is defined as 
\begin{equation}
 \text{COP}_{s,t}(C_i)=\operatorname{cyc}_{s,t}(\text{COP}\{C_{i,s,t}\}\{\operatorname{rev}(C_{i,t,s})\})\;
\end{equation}
with $\operatorname{cyc}_{s,t}$ denoting the cyclic rotation of each of the permutations such that they start in $q_s$ or $\qb_s$ and end in $q_t$ or $\qb_t$. What remains is to sum over all possibilities the loop cycle can be connected by photons to the other cycles. Hence, for each permutation $\alpha\in\text{COP}_{s,t}(C_i)$, we have to sum over the photon exchange permutations $\Gamma(\alpha,C\setminus C_i)$. The factor $(-1)^{f_{i,s,t}(C,\rho)}$ compensates the relative sign of the color ordered diagrams and the contributions to the partial amplitude, with 
\begin{equation}\label{eq:signLoopPhoton}
f_{i,s,t}(C,\rho)=|C_{i,t,s}|+\sum\limits_{\text{reversed cycles in $\rho$}} |C_j|
\end{equation}
simply counting the number of legs whose order has been reversed due to the presence of the loop or due to \cref{eq:photonexchange2}. The routing of the loop quarks is fixed according to
\begin{equation}\label{eq:routingLoopPhoton}
 \begin{split}
 \r_{i,s,t}\circ \{q_s,\dots,\bar{q}_s,\dots,q_t,\dots,\bar{q}_t\}&=\{q^R_s,\dots,\bar{q}^R_s,\dots,q^R_t,\dots,\bar{q}^R_t\}\\
\r_{i,s,t}\circ \{q_s,\dots,\bar{q}_s,\dots,\bar{q}_t,\dots,q_t\}&=\{q^R_s,\dots,\bar{q}^R_s,\dots,\bar{q}^L_t,\dots,q^L_t\}\\
\r_{i,s,t}\circ \{\bar{q}_s,\dots,q_s,\dots,q_t,\dots,\bar{q}_t\}&=\{\bar{q}^L_s,\dots,q^L_s,\dots,q^R_t,\dots,\bar{q}^R_t\}\\
\r_{i,s,t}\circ \{\bar{q}_s,\dots,q_s,\dots,\bar{q}_t,\dots,q_t\}&=\{\bar{q}^L_s,\dots,q^L_s,\dots,\bar{q}^L_t,\dots,q^L_t\}
\end{split}
\end{equation}
The other quarks in $C_{i,s,t}$ and $C_{i,t,s}$ get $r_j=R$ and $r_j=L$ respectively, and the routings of the quarks of the non-loop cycles are fixed accordingly.

Finally we present the contribution of color ordered diagrams with a loop photon connecting quark lines of different cycles. Adding a photon connecting two of the $p+1$ cycles of a tree diagram yields a one-loop diagram with a loop containing $2\leq i\leq p+1$ photons. These $i$ loop photons connect $i$ of the cycles with each cycle having either one or two quark lines that are connected via loop photons to other cycles. For each choice of loop cycles $\alpha=\{C_{\alpha_1},\dots,C_{\alpha_i}\}\in \mathcal{P}_i(C)$, each choice of loop cycles $\beta=\{C_{\beta_1},\dots,C_{\beta_j}\}\in \mathcal{P}_j(\alpha)$ with two loop quarks and each choice of loop quarks $\pi=\{\pi_1,\dots,\pi_{i-j}\}$ for the cycles $\bar{\beta}=\alpha\setminus\beta=\{C_{\bar{\beta}_1},\dots,C_{\bar{\beta}_{i-j}}\}$ and $\kappa=\{\{\kappa_{1,1},\kappa_{1,2}\},\dots,\{\kappa_{j,1},\kappa_{j,2}\}\}$ for the cycles $\beta$, we get a gauge invariant subset of color ordered diagrams whose sum we denote by $p_{2,3}(C,\bar{\beta},\beta,\pi,\
kappa)$. Hence, the contribution $P_{2,3}$ to the partial amplitude $P_2$ is given by
\begin{equation}\label{eq:P_{2,3}}
 P_{2,3}(\tau,\sigma,\{n_i\})=\sum_{i=2}^{p+1}\;\sum_{j=0}^{i}\;\sum_{\alpha\in \mathcal{P}_i(C)}\;\sum_{\substack{\beta\in \mathcal{P}_j(\alpha)\\\bar{\beta}=\alpha\setminus\beta}}\;\sum_{\substack{\pi_s\in Q(\bar{\beta}_s)\\1\leq s\leq i-j}}\;\sum_{\substack{\{\kappa_{t,1},\kappa_{t,2}\}\in \mathcal{P}_2(Q(\beta_t))\\1\leq t\leq j}}p_{2,3}(C,\bar{\beta},\beta,\pi,\kappa)\,.
\end{equation}
In order to write down an expression for $p_{2,3}(C,\bar{\beta},\beta,\pi,\kappa)$ we need to be able to construct linear combinations of primitive amplitudes that single out diagrams with a loop containing a predefined number of photons connecting a predefined set of quarks. As a first step we consider the special case of photon loop, i.\,e.~a loop containing only quark gluon vertices. The sum over all possible photon loops between $m+1$ quark lines $l_i=\{q_i,R_i,\bar{q}_i\}$ is given by
\begin{equation}\label{eq:photonloop}
 \sum_{f\in\{0,1\}^{m}} \sum_{\tau\in S_{m}}(-1)^{\sum\limits_{k=1}^m f_k |l_{\tau(k)}|}A(\operatorname{rev}(f_1,\r_{q_{\tau(1)}}\circ l_{\tau(1)}),\dots,\operatorname{rev}(f_{m},\r_{q_{\tau(m)}}\circ l_{\tau(m)}),\r_{q_{m+1}}\circ l_{m+1}),
\end{equation}
where 
\begin{equation}
\operatorname{rev}(\theta,\alpha)=\begin{cases}
                                   \alpha&\quad\text{if}\quad \theta=0\\
				  \operatorname{rev}(\alpha)&\quad\text{if}\quad \theta=1\;,
                                  \end{cases}
\end{equation}
and the reversal of $\alpha$ is meant to include an inversion of the routings of the quarks, if present.  
Obviously this linear combination contains all possible photon loop diagrams. All diagrams without a photon loop, that are contributing to individual primitive amplitudes, cancel out. For a fixed $f$ all diagrams with several quark lines connecting to the loop via a gluon tree cancel in the sum over $\tau$ due to the symmetries of the color ordered vertices. On the other hand, for a fixed $\tau$ all diagrams with some of the quarks connecting to the loop by a gluon cancel in the sum over $f$ due to the anti symmetry of the color ordered quark gluon vertex if we compensate the signs introduced by reversing individual quark lines.
\Cref{eq:photonloop} covers the special case $\kappa=\{\}$ of all loop cycles $\alpha$ only contributing one loop quark, and has to be generalized in order to cover the case of cycles with two loop quarks as well. First of all we define the loop permutations of some set $\omega=\{\omega_1,\dots,\omega_{m+1}\}$ of sequences $\omega_i$ of external legs
\begin{equation}
 \text{LP}(\omega)\defi \left\{\sigma\in S(\omega)\;\left|\begin{aligned}&\;\exists\,\tau\in S_m,\;f\in\{0,1\}^m\,:\\&\;\qquad\sigma=\{\operatorname{rev}(f_1,\omega_{\tau(1)}),\dots,\operatorname{rev}(f_{m},\omega_{\tau(m)}),\omega_{m+1}\}\end{aligned}\right.\right\}\;.
\end{equation}
For the loop cycles $\bar{\beta}$ with one loop quark we know from \cref{eq:photonloop} that the sequences of external legs entering the loop permutations are given by $\pi\circ\bar{\beta}$, where the action of $\pi$ on $\bar{\beta}$ indicates a rotation of the cycles such that they start with their loop quark. In case of a cycle $C_{\beta_i}$ with two loop quarks $q_{\kappa_{i,1}}$, $q_{\kappa_{i,2}}$ the permutations $\delta_i\in\text{COP}_{\kappa_{i,1},\kappa_{i,2}}(C_{\beta_i})$ ensure that both quarks are part of the loop. Hence, we have to take loop permutations of each possible set of sequences $\delta=\{\delta_1,\dots,\delta_j\}$  and the sequences $\pi\circ\bar{\beta}$. What remains is to connect the loop part via photons to the non-loop cycles $C\setminus(\beta\cup\bar{\beta}))$ by summing over all photon exchange permutations, as well as to fix the routings and signs according to \cref{eq:signLoopPhoton,eq:routingLoopPhoton,eq:photonloop}. Consequently, the missing piece in \cref{eq:P_{2,3}} is 
given by
\begin{equation}
 p_{2,3}(C,\bar{\beta},\beta,\pi,\kappa)=\sum_{\substack{\delta_i\in\text{COP}_{\kappa_{i,1},\kappa_{i,2}}(C_{\beta_i})\\1 \leq i\leq j}}\;\sum_{\gamma\in\text{LP}(\delta,\pi\circ\bar{\beta})}\;\sum_{\rho\in\Gamma(\gamma,C\setminus(\beta\cup\bar{\beta}))}\!\!\!\!\!\!\!(-1)^{f(C,\beta,\kappa,\rho)}A(\r(\beta,\pi,\kappa)\circ \rho),
\end{equation}
with
\begin{equation}
 f(C,\beta,\kappa,\rho)=\sum_{i=1}^{j}(\theta(\rho,C_{\beta_j,\kappa_{j,1},\kappa_{j,2}})+\theta(\rho,C_{\beta_j,\kappa_{j,2},\kappa_{j,1}}))+\sum_{\alpha\in C\setminus\beta} \theta(\rho,\alpha)
\end{equation}
and
\begin{equation}
\theta(\alpha,\beta)=\begin{cases}
 |\beta|                     &\text{if $\text{rev}(\beta)$ is a sub sequence of $\alpha$}\\
0&\text{else}\,.
                   \end{cases}
 \end{equation}
The routing function $\r(\beta,\pi,\kappa)$ acts as follows on the permutation $\rho$. The routings of the loop quarks $\pi$ are set to $r_{\pi_i}=L$ if their cycle got reversed and $r_{\pi_i}=R$ if not. For each cycle $C_{\beta_i}$ in $\beta$ the quark routings are fixed as described in and below \cref{eq:routingLoopPhoton} whenever $\text{rev}(C_{\beta_i,\kappa_{i,2},\kappa_{i,1}})$ is a sub sequence of $\rho$. If $C_{\beta_i,\kappa_{i,2},\kappa_{i,1}}$ is a sub sequence of $\rho$ instead, a reversed version of \cref{eq:routingLoopPhoton} applies. The routings of the quarks in the non-loop cycles $C\setminus(\beta\cup\bar{\beta})$ are fixed according to their orientations with respect to the loop. 
\section{Identities among primitive amplitudes}\label{section:SymPrim}
Primitive amplitudes fulfill a large number of identities. A detailed understanding of these identities can be used to significantly speed up the numerical evaluation of a scattering amplitude, as the number of primitive amplitudes constituting the QCD amplitude can be reduced. Besides the obvious symmetry under cyclic permutations $\tau\in Z_n$ of the external legs
\begin{equation}\label{eq:cycSym}
A(1,\dots,n)=A(\tau(1),\dots,\tau(n))
\end{equation}
and the reflection symmetry
\begin{equation}\label{eq:refSym}
A(1,\dots,n)=(-1)^n A(n,\dots,1)\,,
\end{equation}
there are additional identities that can be easily understood on the level of color ordered Feynman diagrams and color ordered vertices. 

\begin{figure}[t]
\begin{center}
\includegraphics[width=0.5\textwidth]{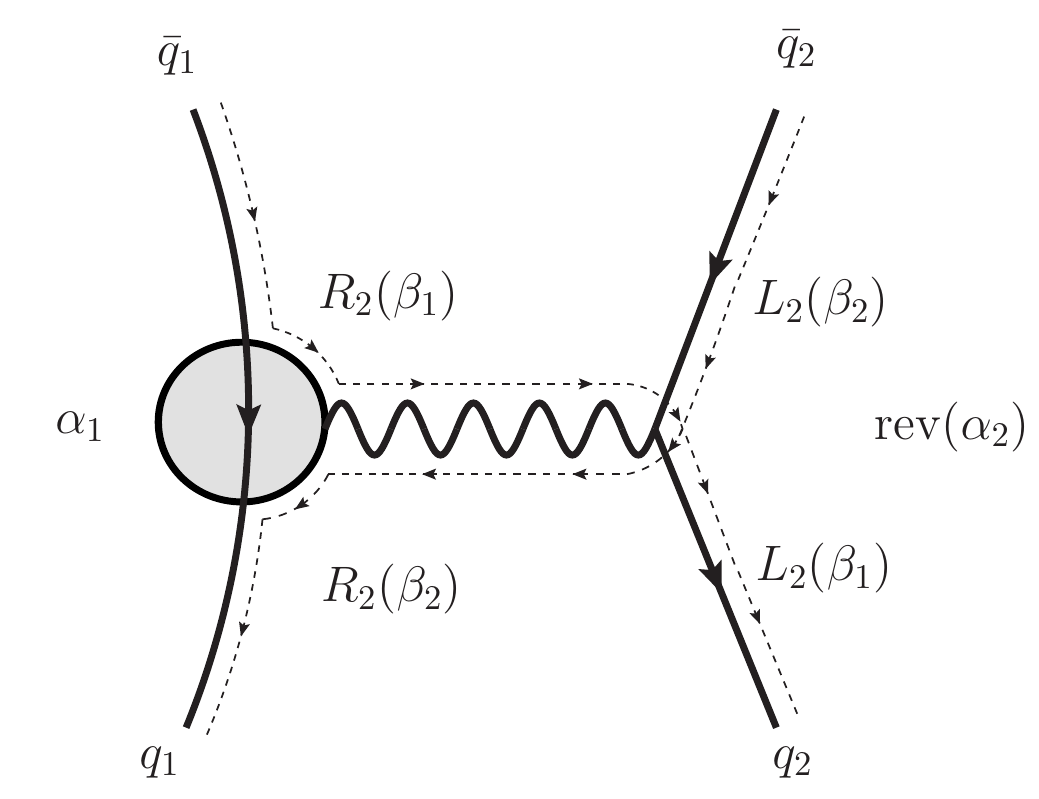}
\end{center}
\caption{Schematic form of the diagrams contributing to the right side of the fermion flip identity \cref{eq:FFId}. All external legs except $q_1$, $q_2$, $\qb_1$, $\qb_2$ are omitted. The dotted lines indicate to which part of the diagram the legs $\beta_1$ and $\beta_2$ can be attached. $R_2(\beta_i)$ and $L_2(\beta_i)$ denote the legs of $\beta_i$ to the right or to the left of the quark line $\{\qb^L_2,q^L_2\}$.}\label{fig:FFId}
\end{figure}

All of these additional identities of the primitive amplitudes that rely on symmetries of the color ordered Feynman diagrams can be written as linear combinations of fermion flip identities, which  are basically tree level identities relying on the symmetries of the color ordered vertices \cref{fig:ColorOrderedRules} and the resulting reversion properties of the color ordered diagrams. Fermion flip identities have been first observed in \cite{Ita:2011ar}, however without stating their general form. In general, fermion flip identities allow to flip one quark line with respect to another one by taking a well defined linear combination of amplitudes. Depending on the position of the loop with respect to the two quark lines, there are two different types of flip identities. At first we are going to present the general fermion flip identity for the case of a quark line $l_2$, on the non-loop side of the loop quark line $l_1$, gets flipped. We are going to present the identity for the non-fermion loop part of the 
primitive amplitudes, but emphasize that due to its tree level nature the flip identity equally holds for the fermion loop part and for the tree level color ordered amplitudes. Let $l_1$ be a quark line with legs $\alpha_1$ on its loop side and $l_2$ be a quark line with legs $\alpha_2$ on its non-loop side, i.\,e.~$l_1=\{q^L_1,\alpha_1,\qb^L_1\}$ or $l_1=\{\qb^R_1,\alpha_1,q^R_1\}$,  and $l_2=\{q^R_2,\alpha_2,\qb^R_2\}$ or $l_2=\{\qb^L_2,\alpha_2,q^R_2\}$. Let further $\beta_1$, $\beta_2$ denote two sequences of external legs. The general fermion flip identity is given by
\begin{equation}\label{eq:FFId}
A(l_1,\beta_1,l_2,\beta_2)= (-1)^{|l_2|+1}\sum_{\sigma\in\text{FOP}\{\beta_1\}\{\operatorname{rev}(\overline{l_2})\}\{\beta_2\}}A(l_1,\sigma)\,,
\end{equation}
where $\overline{l_2}$ indicates the inversion of the quark routings in $l_2$ and the flip ordered permutations $\text{FOP}\{\gamma_1\}\{\gamma_2\}\{\gamma_3\}\subset\text{OP}\{\gamma_1\}\{\gamma_2\}\{\gamma_3\}$ of three sequences $\gamma_i$ 
are the subset of ordered permutations where the last entry of $\gamma_1$ is always before the last entry of $\gamma_2$ and the first entry of $\gamma_3$ is always after the first entry of $\gamma_2$.  Obviously \cref{eq:FFId} generalizes the examples of flip identities given in \cite{Ita:2011ar}. In fact, there are no diagram based identities among the non-fermion loop parts of primitive amplitudes that cannot be written as linear combinations of the fermion flip identities \cref{eq:FFId}.
The reasoning behind the flip ordered permutations is quite simple. The diagrams contributing to the right side of \cref{eq:FFId} can be categorized according to which parts $L_2(\beta_i)$, $R_2(\beta_i)$ of $\beta_1$ and $\beta_2$ are to the left and the right of the fermion line $l_2$. Within each of the diagrams there is a quark gluon vertex connecting the quark line $l_2$ to the loop part of the diagram. To simplify the discussion we specify the orientation of the quark lines to be $l_1=\{q^L_1,\alpha_1,\qb^L_1\}$ and $l_2=\{q^R_2,\alpha_2,\qb^R_2\}$. The schematic form of a diagram contributing to the sum over flip permutations is depicted in \cref{fig:FFId}, with all external legs except $q_1$, $q_2$, $\qb_1$, $\qb_2$ being omitted.
As indicated by the dotted lines, the legs $\beta_i$ can not be attached to every part of the diagram, e.\,g.~diagrams where part of the legs $R_2(\beta_1)$ connect to the left of the connection vertex cancel within the permutations against diagrams where part of the legs $L_2(\beta_1)$ are attached to the right of the connection vertex. Furthermore, the flip permutations assure that there are no shared trees between the legs $L_2(\beta_i)$ and $\alpha_2$. Consequently each diagram contributing to the sum over flip permutations is up to a factor of $(-1)^{|l_2|+1}$ equal to a diagram contributing to $A(l_1,\beta_1,l_2,\beta_2)$.

The simplest of the fermion flip identities \cref{eq:FFId} are the ones with $\beta_i=\{\}$
\begin{equation}\label{eq:flip2}
 A(l_1,l_2)+(-1)^{|l_2|}A(l_1,\operatorname{rev}(l_2))=0\,,
\end{equation}
as they involve only two amplitudes. The three term flip identities have the form
\begin{equation}
 A(l_1,\beta_1,q^R_2,\qb^R_2)+A(l_1,\beta_1,\qb^L_2,q^L_2)+A(l_1,\qb^L_2,\beta_1,q^L_2)=0\,,
\end{equation}
with $\beta_1$ either being a single gluon or a quark line of the form $\beta_1=\{\qb^L_3,\dots,q^L_3\}$ or $\beta_1=\{q^R_3,\dots,\qb^R_3\}$.
An example of \cref{eq:FFId} with $\beta_1\neq\{\}$ and $\beta_2\neq\{\}$ is
\begin{equation}
\begin{split}
A(q_1^L,\qb_1^L,1,q_2^R, \qb_2^R,q_3^R,\qb_3^R)&={}-A(q_1^L,\qb_1^L,1,\qb_2^L, q_2^L,q_3^R,\qb_3^R)-A(q_1^L,\qb_1^L,\qb_2^L,1, q_2^L,q_3^R,\qb_3^R)\\
&\={}-A(q_1^L,\qb_1^L,1,\qb_2^L,q_3^R,\qb_3^R, q_2^L)-A(q_1^L,\qb_1^L,\qb_2^L,1,q_3^R,\qb_3^R, q_2^L)\\
&\={}-A(q_1^L,\qb_1^L,\qb_2^L,q_3^R,1,\qb_3^R, q_2^L)-A(q_1^L,\qb_1^L,\qb_2^L,q_3^R,\qb_3^R,1, q_2^L)\,.\end{split}
\end{equation}
The fermion flip identities \cref{eq:FFId} are implemented in the \texttt{Mathematica} package \texttt{QCDcolor} described in \cref{appendix:QCDcolor}.

We remark, that \cref{eq:FFId} holds even in the case of $\beta_1$ and $\beta_2$ being connected by one or several quark lines. In this case the routings of the quarks in each of the flip permutations $\sigma\in\text{FOP}\{\beta_1\}\{\operatorname{rev}(l_2)\}\{\beta_2\}$ need to be fixed relative to the loop quark line $l_1$. This special set of flip identities is reducible. It can be written as a sum of a flip identity \cref{eq:FFId} with $\beta_1$ and $\beta_2$ not being connected by a quark line and identities of the form
\begin{equation}\label{eq:subtree}
 \sum_{\sigma\in\text{OP}\{\beta\}\{\gamma\}}A(l_1,\sigma)=0
\end{equation}
where $\beta$ and $\gamma$ are connected by at least one quark line. In general, the sum over $\text{OP}\{\beta\}\{\gamma\}$ implies that all diagrams with subtrees containing legs of $\beta$ and of $\gamma$ cancel. Since $\beta$ and $\gamma$ are connected by a quark line, the sum over $\text{OP}\{\beta\}\{\gamma\}$ gives zero.

\begin{figure}[t]
\begin{center}
 \subfigure[]{\includegraphics[width=0.3\textwidth]{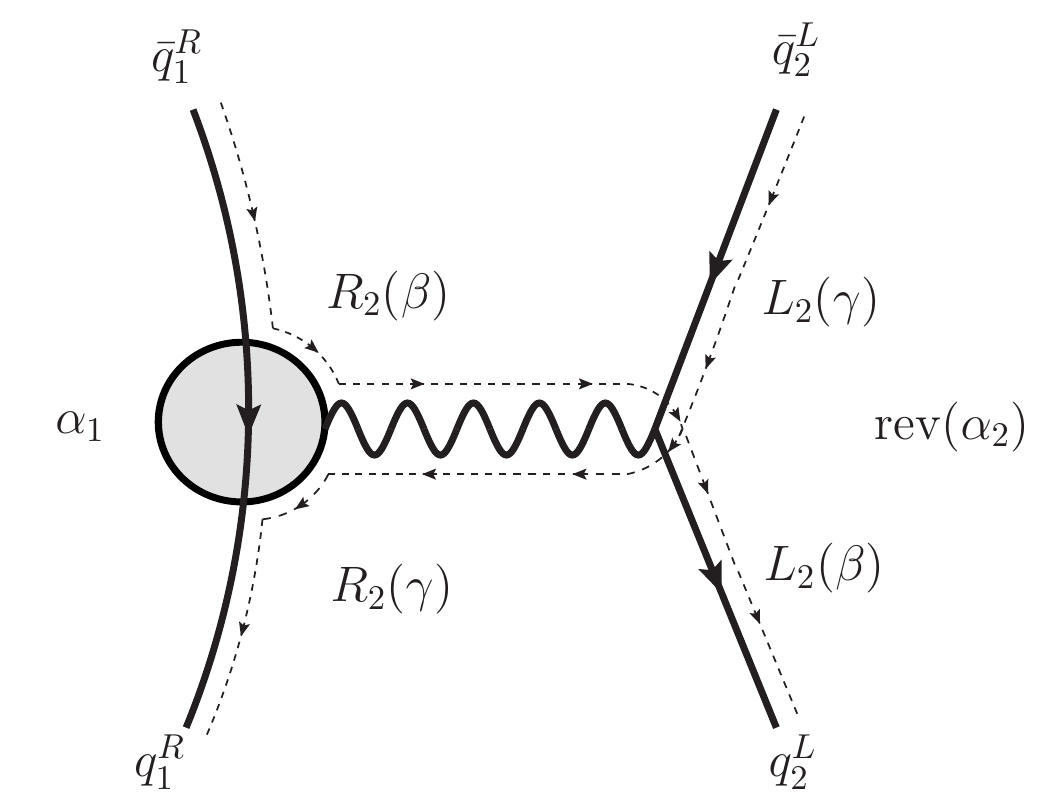}}\hfill
\subfigure[]{\includegraphics[width=0.3\textwidth]{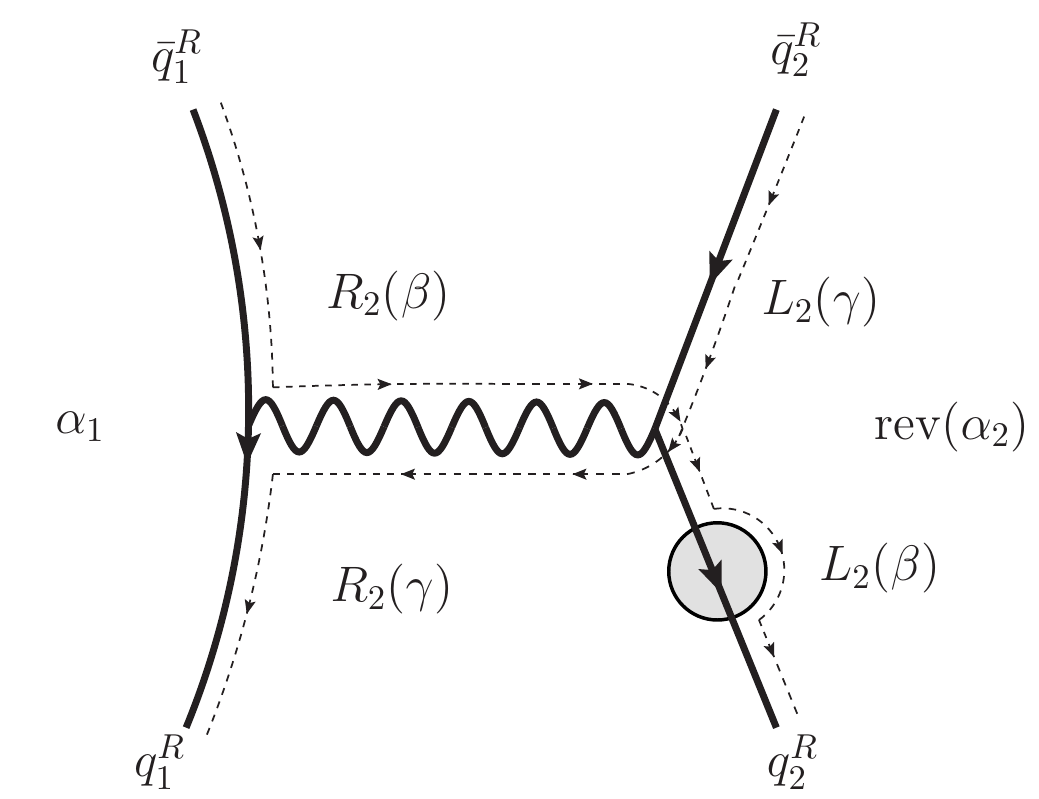}}\hfill
\subfigure[]{\includegraphics[width=0.3\textwidth]{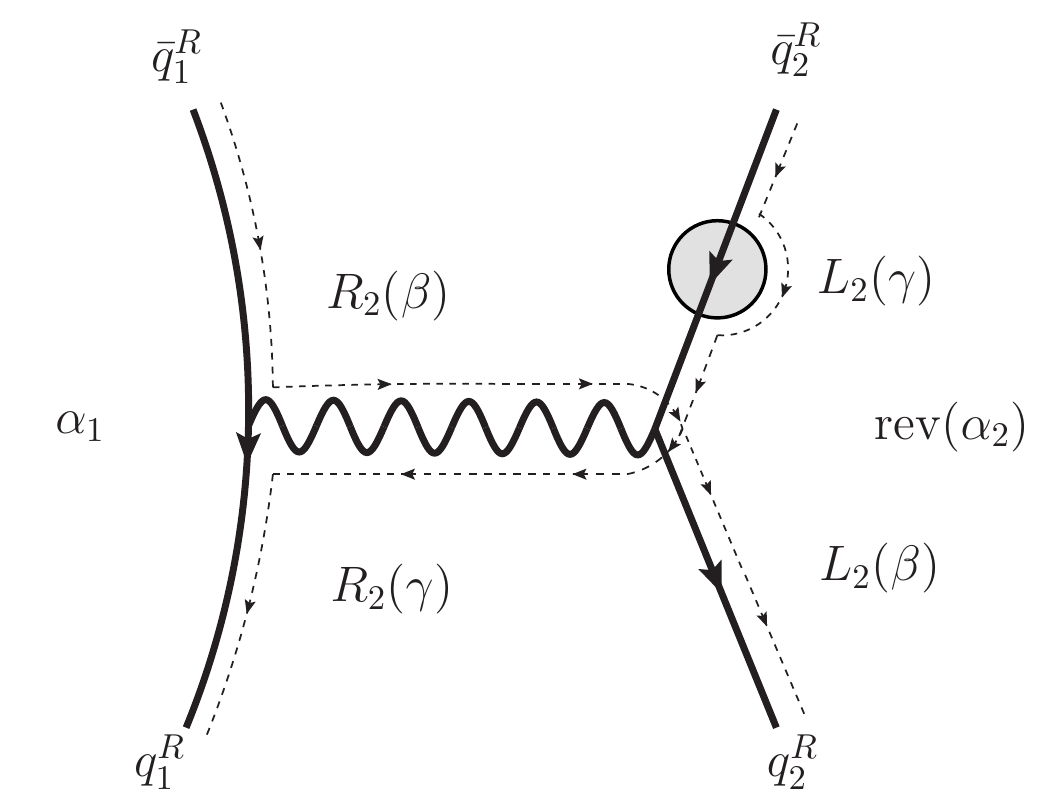}}
\end{center}
\caption{Schematic representations of flipped versions of the diagrams contributing to $A_f(q_1^R,\alpha_1,\qb^R_1,\beta,q_2^R,\alpha_2,\qb_2^R,\gamma)$. All external legs except $q_1$, $q_2$, $\qb_1$, $\qb_2$ are omitted. The circle indicates the position of the loop.  The dotted lines indicate to which part of the diagram the legs $\beta$ and $\gamma$ can be attached. $R_2(\beta)$, $R_2(\gamma)$ and $L_2(\beta)$, $L_2(\gamma)$ denote the legs of $\beta$, $\gamma$ to the right or to the left of the flipped quark line $\{\qb_2,q_2\}$.}\label{fig:FFId3}
\end{figure}

In the case of the fermion loop part of the primitive amplitudes, there are additional identities that are not captured by \cref{eq:FFId}. The reason being the simpler structure of the diagrams with a fermion loop, the furry identity and the fact that massless tadpoles as well as loop corrections to massless external legs vanish
\begin{align}\label{eq:tadpoles}
 A_f(q_1^L,\qb_1^L,\dots)&=0\\
 A_f(q_1^L,1,\qb_1^L,\dots)&=0
\end{align}
In fact it is straight forward to write down additional non-trivial identities for the fermion loop part of the primitive amplitudes. As a first example we consider the class of color ordered diagrams with a quark line $\{q_1^R,\alpha_1,\qb_1^R\}$ coupling by a photon to the fermion loop and the remaining external legs having a fixed cyclic ordering $\beta$. Reversing this quark line in one such diagram leads to a relative factor of $(-1)^{|\alpha_1|+1}$. Hence, the following identity holds
\begin{equation}\label{eq:FFId2}
 \sum_{\sigma\in\text{COP}\{q_1^R,\alpha_1,\qb_1^R\}\{\beta\}}A_f(\sigma)=(-1)^{|\alpha_1|+1} \sum_{\sigma\in\text{COP}\{\qb_1^L,\text{rev}(\overline{\alpha_1}),q_1^L\}\{\beta\}}A_f(\sigma)\,,
\end{equation}
where the cyclic permutations ensure that the quark $q_1$ couples via a photon to the fermion loop. Again $\overline{\alpha_1}$ indicates the inversion of the quark routings in $\alpha_1$. Apparently the flip identity \cref{eq:FFId2} cannot span the whole null space due to the special nature of the involved diagrams. However, it is possible to perform a flip of a quark line similar to the one in \cref{eq:FFId} in the cases where the loop is between the quark lines $l_1$ and $l_2$. In order to simplify the notation we exploit the reflection symmetry to fix the flipped quark line to the form $l_2=\{q_2^R,\alpha_2,\qb_2^R\}$ whereas the fixed quark line $l_1$ can be either of the form $l_1=\{q_1^R,\alpha_1,\qb^R_1\}$ or $l_1=\{\qb_1^L,\alpha_1,q^L_1\}$. If $\beta=\{\beta_1,\dots,\beta_s\}$ and $\gamma=\{\gamma_1,\dots,\gamma_t\}$ denote sequences of external legs than
\begin{equation}\label{eq:FFId3}
\begin{split}
 A_f(l_1,\beta,l_2,\gamma)&= (-1)^{|\alpha_2|+1}\sum_{\sigma\in\text{FOP}\{\beta\}\{\mathrm{rev}(\overline{l_2})\}\{\gamma\}}A_f(l_1,\sigma)\\
&\=+\sum_{i=0}^{|\beta|}\sum_{j=0}^{|\gamma|}(-1)^{j+1}\sum_{\sigma\in\text{OP}\{\alpha_1\}\{\text{rev}(\overline{\gamma_{1,j}})\}\{\gamma_{j+1,t},l_1,\beta_{1,i}\}}A_f(\sigma,\qb_2^R,\beta_{i+1,s},q_2^R)\\
&\=+\sum_{i=0}^{|\beta|}\sum_{j=0}^{|\gamma|}(-1)^{|\beta|-i+1}\sum_{\sigma\in\text{OP}\{\alpha_1\}\{\text{rev}(\overline{\beta_{i+1,s}})\}\{\gamma_{j+1,t},l_1,\beta_{1,i}\}}A_f(\sigma,\qb_2^R,\gamma_{1,j},q_2^R)\,,
\end{split}
\end{equation}
where $\beta_{i,j}=\{\beta_i,\dots,\beta_j\}$ denotes the sub sequence of $\beta$ starting with $\beta_i$ and ending in $\beta_j$. The reasoning behind \cref{eq:FFId3} is similar to the one that led to the fermion flip identity \cref{eq:FFId}. Depending on the position of the fermion loop, the color ordered diagrams contributing to $A_f(l_1,\beta,l_2,\gamma)$ can be divided into three categories, as depicted in \cref{fig:FFId3} for the case $l_1=\{q_1^R,\alpha_1,\qb^R_1\}$.
By summing over the flip permutations $\text{FOP}\{\beta\}\{\mathrm{rev}(l_2)\}\{\gamma\}$, we sum over all flipped diagrams where the fermion loop is to the left of the flipped quark line $\{\qb_2,q_2\}$, which corresponds to \cref{fig:FFId3} (a). The relative sign between flipped and non-flipped diagrams is $(-1)^{|\alpha_2|+1}$.
All flipped diagrams where the fermion loop is to the right of the flipped quark line $\{\qb_2,q_2\}$ can be further divided into diagrams where either legs of $\beta$ or legs of $\gamma$ are connected to the fermion loop, depicted in \cref{fig:FFId3} (b) and \cref{fig:FFId3} (c). In order to express these diagrams by a linear combination of primitive amplitudes, we need to flip the legs that are not allowed to be connected to the fermion loop, i.\,e.~$\text{rev}(\alpha_2)$ as well as either $L_2(\beta)$ or $L_2(\gamma)$, to the non loop-side of the flipped quark line. This is straight forward to accomplish by summing over ordered permutations $\sigma\in\text{OP}\{\alpha_1\}\{\text{rev}(L_2(\beta))\}\{R_2(\gamma),l_1,R_2(\beta)\}$ or $\sigma\in\text{OP}\{\alpha_1\}\{\text{rev}(L_2(\gamma))\}\{R_2(\gamma),l_1,R_2(\beta)\}$ of the legs on the non-loop side of the flipped quark line $\{\qb_2,q_2\}$. The relative sign between flipped and non-flipped diagrams is $(-1)^{|L_2(\beta)|+1}$ or $(-1)^{|L_2(\gamma)|+1}$.

We emphasize that in the special case of either $\beta$ or $\gamma$ containing a quark line $l_3=\{q_3^L,\alpha_3,\qb_3^L\}$ or $l_3=\{\qb_3^R,\alpha_3,q_3^R\}$ that separates $l_1$ and $l_2$ from the loop, \cref{eq:FFId3} holds for the mixed loop part of the primitives as well. 

Beside the identities \cref{eq:FFId,eq:FFId3} that are solely based on the symmetries of the color ordered Feynman rules \cref{fig:ColorOrderedRules}, there are additional identities relying on the Furry identity as well. On the level of Feynman diagrams the Furry identity is the simple observation that a fermion loop with $n$ off-shell gluons connecting to it and a fermion loop with the same $n$ off-shell gluons connecting to it in reversed cyclic ordering are equal up to a relative sign of $(-1)^n$. These properties of the fermion loop allow to reverse the ordering of the legs in every color ordered sub-diagram by properly adjusting the overall sign of the diagram. Before showing how this translates into identities of the primitive amplitudes we present a more straight forward application of the furry identity. Let $\alpha_i$ be either a single gluon or a fermion line of the form $\{q_j^R,\dots,\qb_j^R\}$ or $\{\qb_j^L,\dots,q_j^L\}$, than the following  identity holds
\begin{equation}\label{eq:furry1}
\sum_{\sigma\in\text{COP}\{\alpha_1\}\dots\{\alpha_{2k+1}\}}A_f(\sigma)=0\,.
\end{equation}
The cyclically ordered permutations of the $\alpha_i$ ensure that only diagrams with each of the $\alpha_i$ coupling directly to the fermion loop survive in the sum. Since the number of $\alpha$'s is odd all these diagrams cancel pairwise. 

\begin{figure}[ht]\begin{center}
                    \subfigure[]{\includegraphics[width=0.4\textwidth]{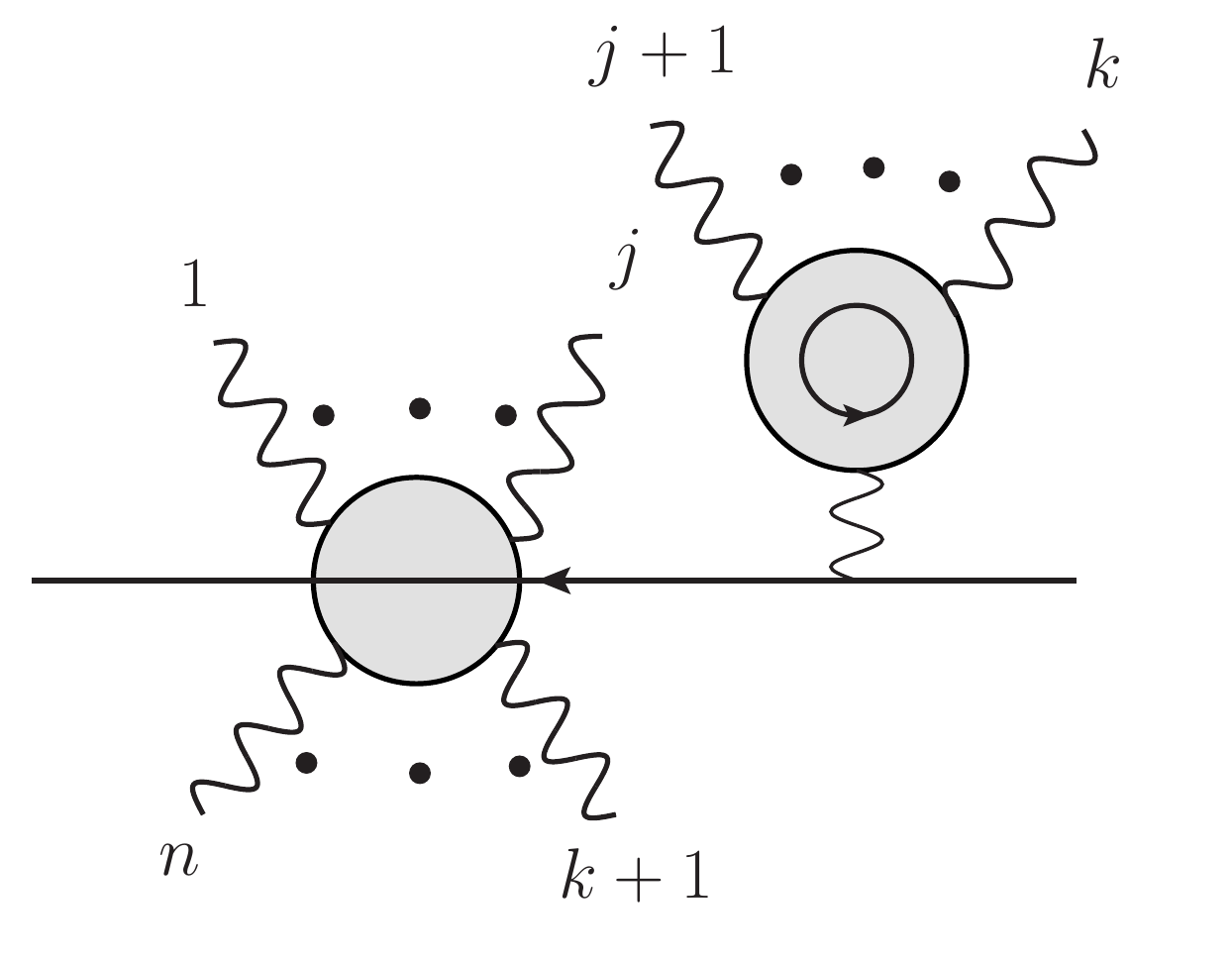}}
\subfigure[]{\includegraphics[width=0.4\textwidth]{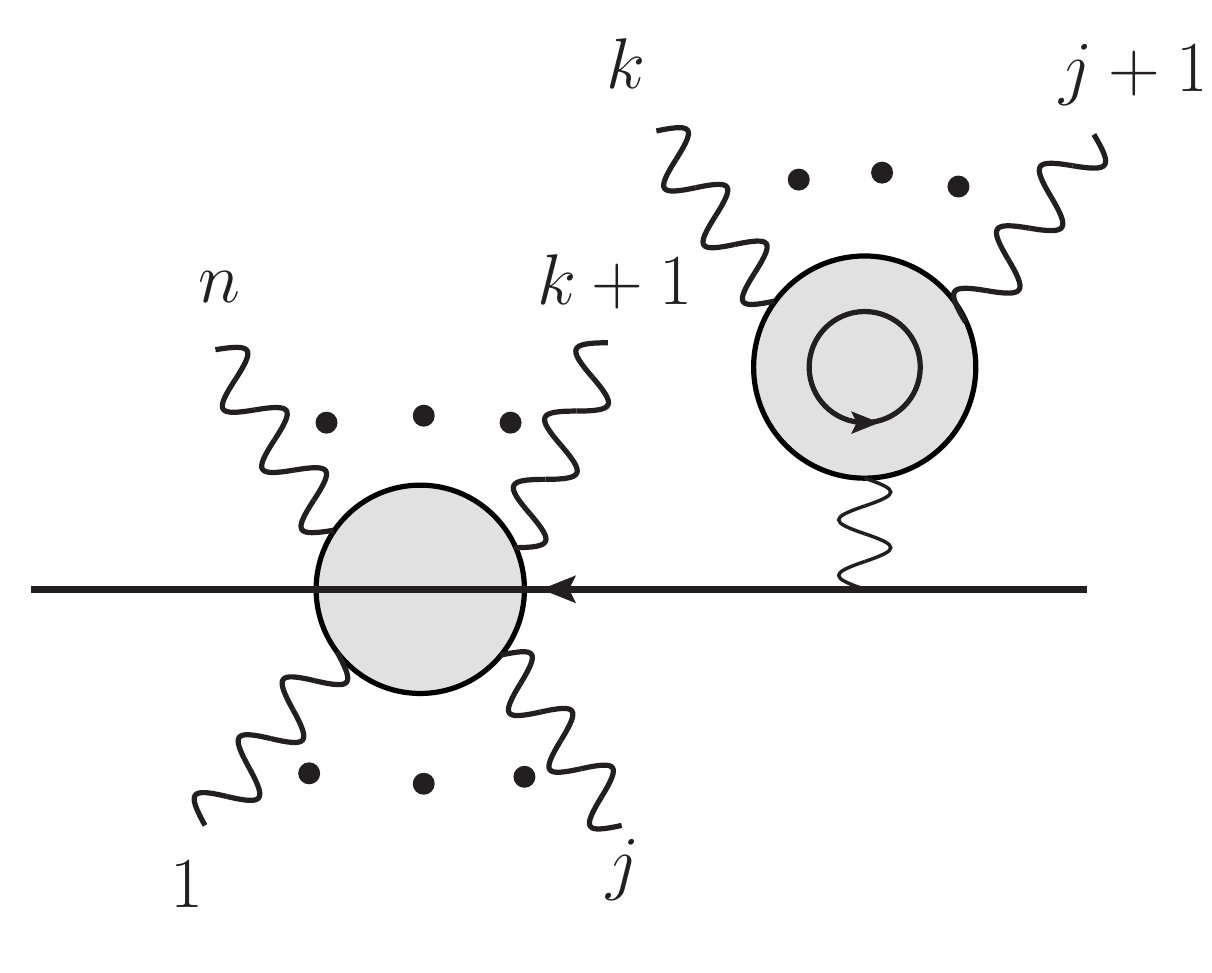}}
                  \end{center}
\caption{Pictorial representation of the diagrams contributing to the first (a) and second term (b) in \cref{eq:furry2} for $\beta=\{1,\dots,n\}$ containing only gluons. The external legs $\alpha_1$ are omitted as they are not relevant for the pairwise cancellation between (a) and (b), which are equal up to a sign of $(-1)^{n+1}$.}\label{fig:furry2}
\end{figure}

Out of all the identities derived so far, only \cref{eq:FFId2} and \cref{eq:furry1} apply to the fermion loop part of primitive amplitudes with only one quark line. However, both identities do not span the whole null space of the two quark primitive amplitudes implying that there are additional identities between them.  The missing piece is the general reversion identity. Let $\beta$, and $\alpha_1$ be some arbitrary sequences of external legs than
\begin{equation}\label{eq:furry2}
 \sum_{\sigma\in\text{OP}\{\beta\}\{\qb_1^L,\alpha_1\}}A_f(q_1^L,\sigma)+(-1)^{|\beta|}\sum_{\sigma\in\text{OP}\{\text{rev}(\overline{\beta})\}\{\qb_1^L,\alpha_1\}}A_f(q_1^L,\sigma)=0\,.
\end{equation}
In \cref{fig:furry2} we present a pictorial representation of the diagrams contributing to the first and second term in \cref{eq:furry2} in the case of $\beta=\{1,\dots,n\}$ containing only gluons. The shaded circles represent some color ordered subdiagrams. Obviously, there is a one to one correspondence between the diagrams (a) and (b). Reversing the order of the legs $j+1,\dots, k$ in the diagrams of \cref{fig:furry2} (a), by exploiting the symmetries of the color ordered vertices and the Furry identity, leads to a relative sign of $(-1)^{k-j+1}$. Additionally flipping the legs $1,\dots,j$ and $k+1,\dots,n$ to the other side of the quark line, as well as reversing the legs on each flipped subtree, we end up in the diagrams (b) with a relative sign of $(-1)^{k-j+1}(-1)^{j+n-k}=(-1)^{n+1}$. The inclusion of quarks in $\beta$ does not alter the logic presented above. 
In case of $\beta$ containing a quark line $l_2=\{q_2^L,\alpha_2,\qb_2^L\}$ or $l_2=\{\qb_2^R,\alpha_2,q_2^R\}$ separating $l_1=\{\qb_1^L,\alpha_1,q_1^L\}$ from the loop, \cref{eq:furry2} is true even for the mixed loop part of the primitive amplitudes. Indeed, \cref{eq:furry2} spans the null space of the mixed loop as well as the fermion loop part of the primitive amplitudes and all the identities \cref{eq:FFId,eq:subtree,eq:FFId2,eq:FFId3} can be written as linear combinations of reversion identities. In fact, we checked that the null space of the mixed loop part of the primitive amplitudes is spanned by either the fermion flip identity \cref{eq:FFId}, the identity \cref{eq:subtree} or the reversion identity \cref{eq:furry2}, whereas the null space of the mixed loop part of the primitive amplitudes is spanned by the  reversion identity \cref{eq:furry2}. Hence, the general reversion identity \cref{eq:furry2} provides full analytical control over expressions containing primitive amplitudes. 

\section{Checks and remarks}\label{section:checks}
 
Since the primitive amplitudes are not all linear independent, the color decomposition of a QCD amplitude is not unique. However, given the reversion identities \cref{eq:furry2} it is very easy to analytically check the equivalence of two different decompositions of a particular amplitude. Indeed, we analytically checked that our formulas for the partial amplitudes \cref{eq:colorDecomposition1loop} agree with all the QCD partial amplitudes presented in the ancillary files of reference \cite{Ita:2011ar}. If the diagram based algorithm for the determination of the partial amplitudes incorporates the Furry identities among the color ordered diagrams, as described in \cite{Badger:2012pg}, the obtained representations of the amplitudes contain only linear 
independent primitive 
amplitudes. This is not the case for the representation of the QCD amplitudes obtained from \cref{eq:colorDecomposition1loop}. However, given a particular QCD one-loop amplitude, it is straightforward to reduce the number of primitive amplitudes in \cref{eq:colorDecomposition1loop} to the number of linear independent primitive amplitudes by applying the reversion identities \cref{eq:furry2}. Consequently, both approaches can provide the same answer.  Exploiting the identities we were able to determine the number $N(k,n)$ of independent mixed loop parts of primitive amplitudes with up to $k=4$ quark lines, as well as the ratio $\kappa(k)=\frac{N(k,n)}{\mathcal{N}(k,n)}$ of the number of independent and the overall number $\mathcal{N}(k,n)$ of   
mixed loop parts of primitive amplitudes.
\begin{equation}
 \begin{aligned}
  N(1,n)&=(n+1)!\qquad\qquad&\kappa(1)&=1\\
N(2,n)&=\frac{2}{3}(n+3)!\qquad\qquad&\kappa(2)&=\frac{2}{3}\\
N(3,n)&=\frac{4}{15}(n+5)!\qquad\qquad&\kappa(3)&=\frac{2}{5}\\
N(4,n)&=\frac{8}{105}(n+7)!\qquad\qquad&\kappa(4)&=\frac{8}{35}
 \end{aligned}
\end{equation}
In general the number of independent primitive amplitudes with a mixed loop seems to be 
\begin{equation}
N(k,n)=\frac{2^{k-1}}{(2k-1)!!}(n+2k-1)!\,,
\end{equation}
Given the reversion identity \cref{eq:furry2} it is straightforward to determine the number $N_f(k,n)$ of independent fermion loop parts of primitive amplitudes with a particular number of quarks and gluons, like e.\,g.~ $N_f(1,n)=\sfrac{1}{2}(n-1)n!$. In contrast to the mixed loop part, it is however possible to find minimal representations of the $(q\qb)^k (g)^n$ QCD amplitude containing less than $N_f(k,n)$ fermion loop primitive amplitudes. Despite the fact that finding such minimal representations is not straightforward, this diminishes the relevance of the numbers $N_f(k,n)$. For up to seven external legs the number $\bar                     {N}_f(k,n)$ of fermion loop primitive amplitudes in such a minimal representation are
\begin{center}
\begin{tabular}[t]{c|c|c}
$n$ &$N_f(1,n)$& $\bar{N}_f(1,n)$\\
\hline
2 &1&1\\
3 &6&6\\
4 &36&33\\
5 &240&230
 \end{tabular}\hfill
\begin{tabular}[t]{c|c|c}
$n$ &$N_f(2,n)$& $\bar{N}_f(2,n)$\\
\hline
0 &1&1\\
1 &3&3\\
2 &15&13\\
3 &96&75
 \end{tabular}
\hfill
\begin{tabular}[t]{c|c|c}
$n$ &$N_f(3,n)$& $\bar{N}_f(3,n)$\\
\hline
0 &7&4\\
1 &36&20
 \end{tabular}
\end{center}
and agree with the numbers presented in \cite{Badger:2012pg}. Considering the rapid growth of the number of primitive amplitudes constituting the amplitudes it is tempting to apply a leading color approximation keeping only the terms in \cref{eq:colorDecomposition1loop}, which are proportional to either $N$ or $n_f$, which reduces the number of primitives to $2\cdot(n+k-1)!$. While a leading color approximation for $N=3$ at a strong coupling of $\alpha_s\approx 0.1$ seems very questionable in theory, keeping in mind that $N=3$ is not a large number and $N^2\alpha_s=\mathcal{O}(1)$ is of order one, it seems to work quite well in practice \cite{Ita:2011ar}.

\section[From \texorpdfstring{$\mathcal{N}=4$}{N=4} SYM to QCD]{From \texorpdfstring{$\bm{\mathcal{N}=4$}}{N=4} SYM to QCD}\label{section:FromN=4toQCD}

Based on the color decomposition of QCD tree amplitudes \cref{eq:colorDecomposition} and the fermion flip identities \cref{eq:FFId} we prove the conjecture made in reference \cite{Dixon:2010ik} that every color ordered tree amplitude of massless QCD can be written as a linear combination of gluon-gluino amplitudes of ${\cal N}=4$ super Yang-Mills theory. The proof will includes a general construction of these linear combinations.

\begin{figure}[t]
\begin{center}
\begin{equation*}
\raisebox{-1.8cm}{\includegraphics[width=0.38\textwidth]{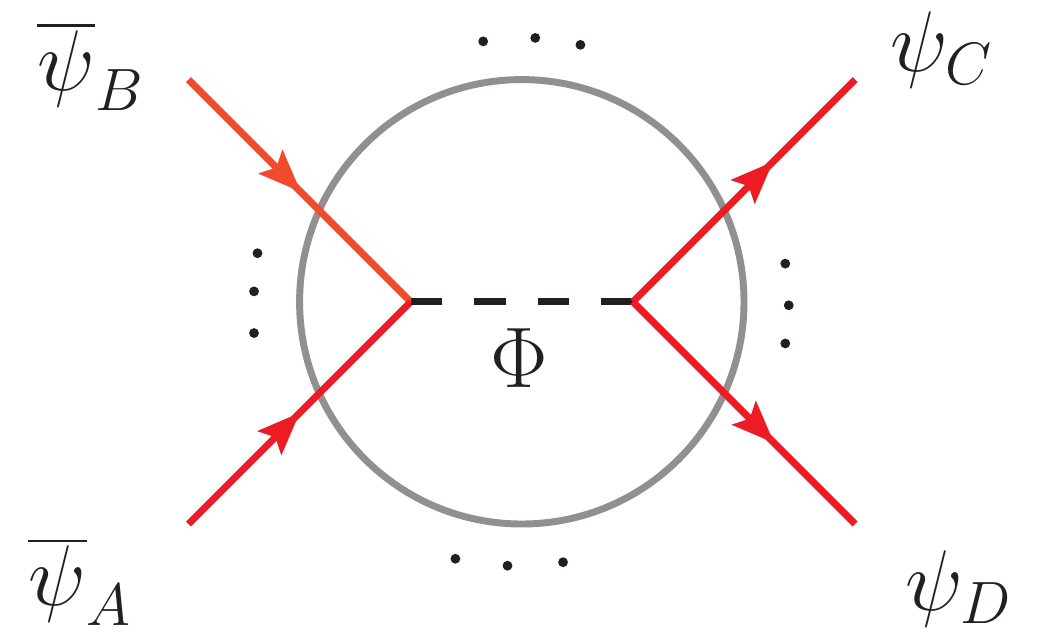}}\quad \sim\quad \delta_{A C}\delta_{B D}-\delta_{A D}\delta_{B C}   
\end{equation*}
\end{center}
\caption{Schematic representation of the simplest contributions from internal scalars in color ordered SYM amplitudes.}\label{fig:scalarExchange}
\end{figure}

From the point of view of tree amplitudes, there are two
principal differences between $\mathcal{N}=4$ SYM and massless QCD.
First of all, the fermions in $\mathcal{N}=4$ SYM, the gluinos,
are in the adjoint representation of $SU(N)$, rather than the 
fundamental representation, and come in four flavors.
Secondly, the $\mathcal{N}=4$ SYM theory contains six massless scalars in the
adjoint representation.
As discussed in
\ref{section:ColorDecomposition}, decomposing gauge theory amplitudes into color structures and color ordered amplitudes, the first difference is fairly unimportant. At the level of color ordered amplitudes there is no difference between gluinos and quarks apart from the differing number of flavors. However, a generic color ordered gluon-gluino SYM amplitude contains contributions from internal scalars, as depicted in \cref{fig:scalarExchange}. Whilst making the transition to QCD, these internal scalars have to be either avoided or subtracted.
\begin{figure}[t]
\begin{align*}
\raisebox{-1.9cm}{\includegraphics[width=0.3\textwidth]{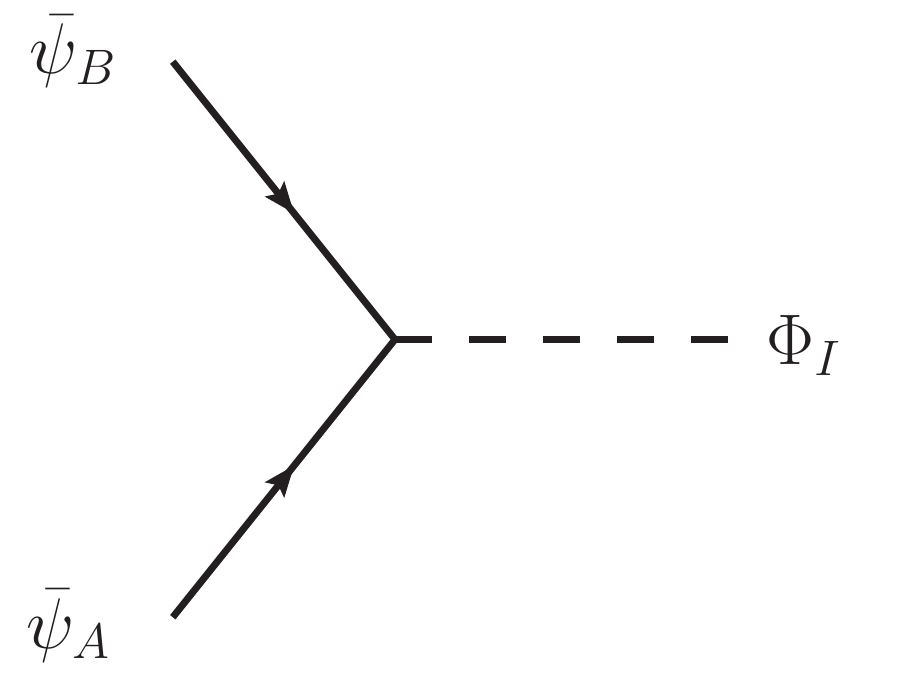}}&\sim \;\Sigma^I_{AB} \qquad\quad\;& \raisebox{-1.9cm}{\includegraphics[width=0.3\textwidth]{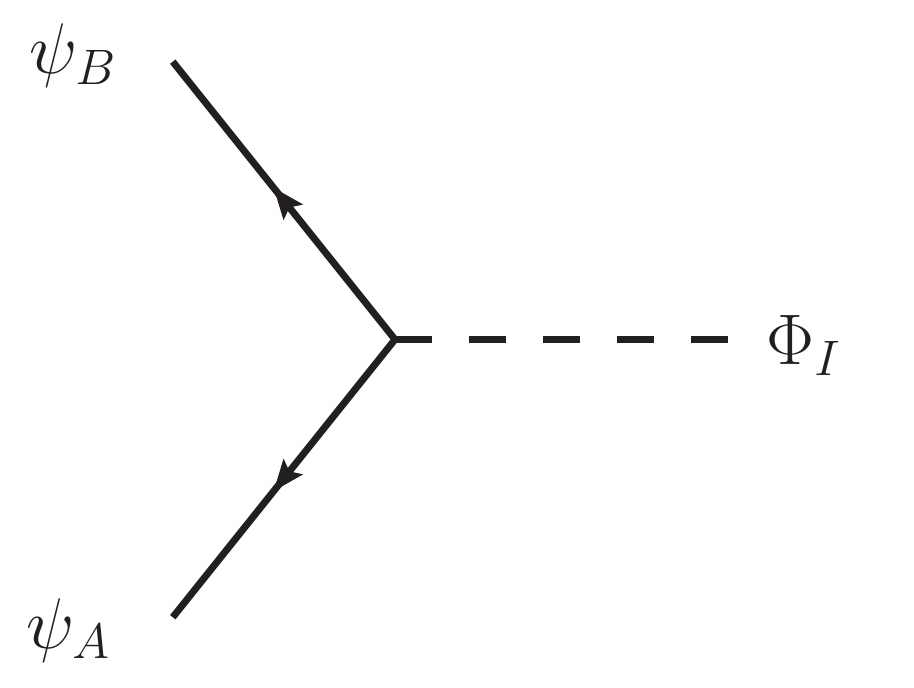}}&\sim\;\widetilde{\Sigma}^I_{AB}
\end{align*}
\caption{The gluino-scalar vertices in $\mathcal{N}=4$ SYM are proportional to the antisymmetric six-dimensional Pauli-matrices.}\label{fig:YukawaCoupling}
\end{figure}
In any gauge theory, tree amplitudes that contain only external gluons
are independent of the matter states in the
theory~\cite{Parke:1985pn,Kunszt:1985mg}; hence they
are identical between $\mathcal{N}=4$ SYM and QCD.
The reason is simply that the vertices that couple gluons to
the other states in the theory always produce the fermions and scalars
in pairs. There are no vertices that can destroy all the fermions and
scalars, once they have been produced.  If a fermion or scalar
were produced at any point in a tree diagram, it would have to
emerge from the diagram, which would no longer have only external
gluons.  In other words, the pure-gluon theory forms a closed
subsector of $\mathcal{N}=4$ SYM.

Another closed subsector of $\mathcal{N}=4$ SYM is $\mathcal{N}=1$
SYM, which contains a gluon and a single gluino.  Let $g$ denote the
gluon, $\psi_A$, $A=1,2,3,4$, denote the four gluinos, and
$\Phi_{I}$ denote the six real scalars of $\mathcal{N}=4$
SYM.  Then the $\mathcal{N}=1$ SYM subsector is formed by
$(g,\psi_1)$.  The reason it is closed is similar to the
pure-gluon case just discussed: There are vertices that produce states
other than $(g,\psi_1)$, but they always do so in pairs.  For
example, the Yukawa coupling  $\phi_I\Sigma^{I\,AB} \psi_A \psi_B$,
$A\neq B$, can convert $\psi_1$ into a scalar and a gluino each
carrying an index $B\neq1$.  However, this index cannot be destroyed
by further interactions.
The fact that $\mathcal{N}=1$ SYM forms a closed subsector of
$\mathcal{N}=4$ SYM, in addition to color ordering,
immediately implies that any
color-ordered QCD tree amplitude for gluons, plus arbitrarily many
quarks of a single flavor, is given directly by the corresponding
amplitude (with $\psi_1$ replacing the single quark flavor)
evaluated in $\mathcal{N}=4$ SYM.  

The less trivial QCD amplitudes to
extract are those for multiple fermion flavors, primarily because of
the potential for intermediate scalar exchange induced by the Yukawa
coupling.
Looking at the master formula for the color decomposition of QCD at tree level \cref{eq:colorDecomposition}, it contains only color ordered amplitudes were the quark helicities are alternating. The subset of color ordered SYM amplitudes with alternating gluino helicities is special as these amplitudes do not contain internal scalars. The reason being that any two gluinos of equal helicity and unequal flavor that could potentially be connected to a gluino-scalar vertex, divide the remaining gluinos into two odd sets. However there is no way to connect an odd number of gluinos by a diagram. This immediately proves that all QCD amplitudes with up to four quark lines can be directly obtained from SYM amplitudes by substituting quarks by gluinos in the color ordered amplitudes in \cref{eq:colorDecomposition}. Increasing the number of quark lines beyond four we run into the problem that in contrast to the number of quark flavors, the number of gluino flavors is not a free parameter. In general, a color ordered 
amplitude with several quarks of equal flavor is given by a sum over all fermion line configurations compatible with the color ordering and the choice of flavors. Each of these fermion line configurations is equal to a color ordered amplitudes with distinct flavors. These relations can be inverted, making it is possible to express an arbitrary fermion line configuration by a linear combination of color ordered amplitudes containing only four different quark flavors. Indeed, as we checked for up to 10 quark lines, even two different flavors would be sufficient. For a given amplitude with more than four quark lines it is straightforward to transform \cref{eq:colorDecomposition} into a representation using only four flavors. Consequently, all QCD tree amplitudes can be obtained from color ordered SYM amplitudes.

When calculating the cut constructable part of QCD loop amplitudes using a maximal number of cuts, color ordered tree amplitudes with non-alternating quark helicities are necessary as well. However, the corresponding gluon gluino amplitudes are not free of internal scalars. Since every color ordered QCD amplitude can be written as a linear combination of color ordered QCD amplitudes with alternating quark helicities, it is possible to obtain all color ordered QCD amplitudes from color ordered SYM amplitudes. In order to prove this statement we will explain how such a linear combination can be obtained by repeated application of the fermion flip identity \cref{eq:FFId}. Exploiting the cyclic symmetry, an arbitrary color ordered amplitude has the form $A(\qb_1,\alpha_1,q_1,\alpha_2)$, where $\alpha_1$ and $\alpha_2$ denote the external legs to the left and to the right of the quark line $\{\qb_1,q_1\}$. Repeatedly flipping quark lines with respect to $\{\qb_1,q_1\}$, the amplitude $A(\qb_1,\alpha_1,q_1,\alpha_
2)$ is given by a linear combination of amplitudes of the form 
\begin{equation}
A(\qb_1,\beta_{i_1},q_{i_1},\gamma_{i_1},\qb_{i_1},\beta_{i_2},\dots,q_1,\beta_{j_1},\qb_{j_1},\gamma_{j_1},q_{j_1},\beta_{j_2},\dots) 
\end{equation}
 where the $\beta$'s contain only gluons. Successively performing analogous flips with respect to each of the other quark lines $\{\qb_i,q_i\}$, yields a representation in terms of color ordered amplitudes with alternating quark helicities, i.\;e.~ a representation in terms of SYM amplitudes. Despite being straightforward to implement the described construction of color ordered QCD tree amplitudes is of limited practical relevance. The obtained representations are far from being minimal as, due to the permutations in the flip identity \cref{eq:FFId}, the number of SYM amplitudes constituting the color ordered QCD amplitude depends on the number of gluons in the amplitude. For practical applications it is more convenient to construct minimal representations of color ordered QCD amplitudes by summing over different gluino flavor assignments to a fixed helicity configuration as described in \cite{Dixon:2010ik}.

\section{Summary and Conclusions}
In \cref{section:QCDcolorTree,section:QCDcolorLoop,section:SymPrim} we derived the color decomposition of an arbitrary QCD amplitude at tree- and one-loop level as well as general  fermion flip and reversion identities of the primitive one-loop amplitudes. The obtained results are implemented in the freely available {\tt Mathematica} package {\tt QCDcolor} described in the  appendix and shall provide an alternative to the diagram based algorithm for the determination of the color decomposition of a particular QCD amplitude \cite{Ellis:2008qc,Ellis:2011cr,Ita:2011ar,Badger:2012pg}. Exploiting the fermion flip identities we were able to prove that all color ordered tree amplitudes of massless QCD can be written as linear combinations of color ordered tree amplitudes of $\mathcal{N}=4$ super Yang-Mills theory. As a remarkable consequence all tree amplitudes as well as the cut constructable part of all QCD loop amplitudes can be obtained from $\mathcal{N}=4$ super Yang-Mills theory.

Despite the indisputable powers of the color ordered approach to calculating QCD scattering amplitudes it has obvious limitations. The number of primitive amplitudes constituting the QCD amplitude of multiplicity $n+2k$ grows as $(n+2k-1)!$,  which results in some hard cut off for the multiplicities that allow for a numerical evaluation of a one-loop QCD amplitude on a particular computer. Since the factorial growth in complexity seems to be intrinsic to the one loop corrections, the goal should be to either dampen it by refining the computational methods or by using approximations like e.\;g.~monte carlo methods or a leading color approximation. 

In an independent work published recently in reference \cite{Reuschle:2013qna}, C. Reuschle and S. Weinzierl found a tree level and a one-loop color decomposition of QCD similar to our results by using shuffle relations. A detailed comparison to our results would be an interesting future project.
\subsection*{Acknowledgments}

We thank S. Badger, B. Biedermann, and V. Yundin for helpful discussions and are greateful to J. Plefka for support, encouragement and valuable comments on the manuscript.   Several figures in this paper were
made with {\sc Jaxodraw}~\cite{Binosi:2003yf,Binosi:2008ig}, based on
{\sc Axodraw}~\cite{Vermaseren:1994je}. 
This work was supported by the Deutsche Forschungsgemeinschaft through the Research Training Group (GK1504) ``Mass, Spectrum, Symmetry, Particle Physics in the Era
of the Large Hadron Collider'' and the SFB 647 ``Raum - Zeit - Materie''.

\appendix
\section{The {\tt Mathematica} package {\tt QCDcolor}}
\label{appendix:QCDcolor}
The {\tt Mathematica} package {\tt QCDcolor} provides the color decomposition of QCD at tree (\cref{section:QCDcolorTree}) and one-loop level (\cref{section:QCDcolorLoop}) as well as implementations of all the identities of primitive amplitudes derived in \cref{section:SymPrim}. Accompanying the package is a notebook file {\tt QCDcolor.nb} demonstrating its usage. 

Within the package {\tt QCDcolor} gluons are specified by distinct integers. 
Quarks are represented by {\tt Q[i,R]}, {\tt Q[i,L]}, where {\tt i} is some integer labeling different flavors and the routing label {\tt R,L} is absent at
tree level. Anti-quarks are represented by {\tt Qb[i,R]}, {\tt Qb[i,L]}, where i is some integer labeling different flavors and the routing label {\tt R}, {\tt L} is absent
at tree level. Integer labels of gluons and quarks can overlap. Color ordered tree amplitudes are represented by the function {\tt Atree[...]}.
The part of the primitive amplitude containing no fermion loop is represented by the function {\tt A[...]}, whereas the fermion loop part is represented by {\tt Af[ ...]}. In contrast to the conventions in the rest of this paper (compare \cref{eq:defColorStructure,eq:colorDecomposition,eq:colorDecomposition1loop}) the dependence of the amplitudes on the number of different quark flavors {\tt nf} and the number of colors {\tt Nc} is not absorbed into the definition of the color structures, leading to partial amplitudes with explicit {\tt nf}, {\tt Nc} dependence. In this way a color structure can be uniquely represented by lists of cycles. A cycle is represented by a list of the form {\tt \{Q[i],...,Qb[k],Q[k],..., Qb[i]\}}. A trace in the color structure is represented by a list of integers. Traces have to be in the last entries of the list representing the color structure.

Just type
\begin{flushleft}
\begin{tt}
<< QCDcolor.m
\end{tt}
\end{flushleft}
To load the package. A list of all stored in the variable  {\tt \$QCDcolorFunction}. The documentations of the functions can be accessed by typing e.\,g.
\begin{flushleft}
\begin{tt}
?QCDnice
\end{tt}
\end{flushleft}
will return
\begin{flushleft}
\begin{tt}
 Typing: QCDnice (shift+enter), leads to nicely formatted output using the Notation package.
\end{tt}
\end{flushleft}
Evaluating {\tt QCDbasic} it is possible to return to basic input formatting of the output.  The functions {\tt  ColorS[k,n]} and {\tt  ColorSLoop[k,n]} generate all color structures in a tree or a one-loop amplitude with {\tt k} quark lines and {\tt n} gluons. For example evaluation of {\tt  ColorS[3,1]} gives the list
\begin{align}
&\{\{\{Q_1,1,\bar{Q}_1\},\{Q_2,\bar{Q}_2\},\{Q_3,\bar{Q}_3\}\},\{\{Q_1,1,\bar{Q}_1\},\{Q_2,\bar{Q}_3,Q_3,\bar{Q}_2\}\},\notag\\
&\{\{Q_1,1,\bar{Q}_2,Q_2,\bar{Q}_1\},\{Q_3,\bar{Q}_3\}\},\{\{Q_1,1,\bar{Q}_2,Q_2,\bar{Q}_3,Q_3,\bar{Q}_1\}\},\notag\\
&\{\{Q_1,1,\bar{Q}_3,Q_3,\bar{Q}_2,Q_2,\bar{Q}_1\}\},\{\{Q_1,1,\bar{Q}_3,Q_3,\bar{Q}_1\},\{Q_2,\bar{Q}_2\}\},\{\{Q_1,\bar{Q}_1\},\{Q_2,1,\bar{Q}_2\},\{Q_3,\bar{Q}_3\}\},\notag\\
&\{\{Q_1,\bar{Q}_1\},\{Q_2,1,\bar{Q}_3,Q_3,\bar{Q}_2\}\},\{\{Q_1,\bar{Q}_2,Q_2,1,\bar{Q}_1\},\{Q_3,\bar{Q}_3\}\},\{\{Q_1,\bar{Q}_2,Q_2,1,\bar{Q}_3,Q_3,\bar{Q}_1\}\},\notag\\
&\{\{Q_1,\bar{Q}_3,Q_3,\bar{Q}_2,Q_2,1,\bar{Q}_1\}\},\{\{Q_1,\bar{Q}_3,Q_3,\bar{Q}_1\},\{Q_2,1,\bar{Q}_2\}\},\{\{Q_1,\bar{Q}_1\},\{Q_2,\bar{Q}_2\},\{Q_3,1,\bar{Q}_3\}\},\notag\\
&\{\{Q_1,\bar{Q}_1\},\{Q_2,\bar{Q}_3,Q_3,1,\bar{Q}_2\}\},\{\{Q_1,\bar{Q}_2,Q_2,\bar{Q}_1\},\{Q_3,1,\bar{Q}_3\}\},\{\{Q_1,\bar{Q}_2,Q_2,\bar{Q}_3,Q_3,1,\bar{Q}_1\}\},\notag\\
&\{\{Q_1,\bar{Q}_3,Q_3,1,\bar{Q}_2,Q_2,\bar{Q}_1\}\},\{\{Q_1,\bar{Q}_3,Q_3,1,\bar{Q}_1\},\{Q_2,\bar{Q}_2\}\}\}  \,,
\end{align}
and evaluation of {\tt  ColorS[2,2]} generates
\begin{align}
 &\{\{\{Q_1,1,2,\bar{Q}_1\}, \{Q_2, \bar{Q}_2\}\}, \{\{Q_1,1,2,\bar{Q}_2,Q_2,\bar{Q}_1\}\}, \{\{Q_1,1,\bar{Q}_1\}, \{Q_2,2,\bar{Q}_2\}\}, \notag\\
&\{\{Q_1,1,\bar{Q}_2, Q_2,2,\bar{Q}_1\}\},\{\{Q_1,    \bar{Q}_1\}, \{Q_2, 1, 2,  \bar{Q}_2\}\},\{\{Q_1,\bar{Q}_2,Q_2,1,2,\bar{Q}_1\}\},\notag\\
 &\{\{Q_1,2,1,\bar{Q}_1\},\{Q_2,\bar{Q}_2\}\},\{\{Q_1,2,1,\bar{Q}_2,Q_2,\bar{Q}_1\}\},\{\{Q_1, 2,    \bar{Q}_1\}, \{Q_2, 1,    \bar{Q}_2\}\}, \notag\\
&\{\{Q_1,2,\bar{Q}_2, Q_2,1,\bar{Q}_1\}\}, \{\{Q_1, \bar{Q}_1\}, \{Q_2,2,1, \bar{Q}_2\}\}, \{\{Q_1,\bar{Q}_2, Q_2,2,1,\bar{Q}_1\}\},\notag\\
&\{\{Q_1, \bar{Q}_1\}, \{Q_2, \bar{Q}_2\}, \{1, 2\}\}, \{\{Q_1, \bar{Q}_2, Q_2, \bar{Q}_1\}, \{1, 2\}\}\}\,.
\end{align}
The tree-level partial amplitudes multiplying the color structure {\tt C} are implemented in the function {\tt PartialAmplitudeTree[C]}. Make sure {\tt C} is in cycle notation and consistent or simply use the output of {\tt  ColorS[k,n]}. Evaluating 
\begin{flushleft}
{\tt PartialAmplitudeTree[\{\{Q[1],1,2,Qb[1]\},\{Q[2],Qb[2]\},\{Q[3],Qb[3]\},\{Q[4],Qb[4]\}\}]} 
\end{flushleft}
 generates the partial amplitude 
\begin{equation}
\begin{split}
  -\frac{1}{N_c^3}\Bigl(&A_{\text{tree}}\left[Q_1,1,2,\bar{Q}_1,Q_2,\bar{Q}_2,Q_3,\bar{Q}_3,Q_4,\bar{Q}_4\right]+A_{\text{tree}}\left[Q_1,1,2,\bar{Q}_1,Q_2,\bar{Q}_2,Q_4,\bar{Q}_4,Q_3,\bar{Q}_3\right]\\
&+A_{\text{tree}}\left[Q_1,1,2,\bar{Q}_1,Q_3,\bar{Q}_3,Q_2,\bar{Q}_2,Q_4,\bar{Q}_4\right]+A_{\text{tree}}\left[Q_1,1,2,\bar{Q}_1,Q_3,\bar{Q}_3,Q_4,\bar{Q}_4,Q_2,\bar{Q}_2\right]\\
&+A_{\text{tree}}\left[Q_1,1,2,\bar{Q}_1,Q_4,\bar{Q}_4,Q_2,\bar{Q}_2,Q_3,\bar{Q}_3\right]+A_{\text{tree}}\left[Q_1,1,2,\bar{Q}_1,Q_4,\bar{Q}_4,Q_3,\bar{Q}_3,Q_2,\bar{Q}_2\right]\Bigr)\,,
\end{split}
\end{equation}
multiplying the color structure $(T^{a_1}T^{a_2})i_1\bar{j}_1\delta_{i_2\bar{j}_2}\delta_{i_2\bar{j}_2}\delta_{i_3\bar{j}_3}\delta_{i_4\bar{j}_4}$ within the eight quark two gluon amplitude.
Evaluation of 
\begin{flushleft}
 \begin{tt}
PartialAmplitudeTree[\{\{Q[1],1,2,Qb[2],Q[2],Qb[1]\},\{Q[3],3,4,5,Qb[3]\}, \{Q[4],6,Qb[4]\}\}]  
 \end{tt}
\end{flushleft}
yields the partial amplitude
\begin{equation}
 \begin{split}
\frac{1}{N_c^2}\Bigl(&A_{\text{tree}}\left[Q_1,1,2,\bar{Q}_2,Q_2,\bar{Q}_1,Q_3,3,4,5,\bar{Q}_3,Q_4,6,\bar{Q}_4\right]\\
&+A_{\text{tree}}\left[Q_1,1,2,\bar{Q}_2,Q_2,\bar{Q}_1,Q_4,6,\bar{Q}_4,Q_3,3,4,5,\bar{Q}_3\right]\\
&+A_{\text{tree}}\left[Q_1,1,2,\bar{Q}_2,Q_3,3,4,5,\bar{Q}_3,Q_2,\bar{Q}_1,Q_4,6,\bar{Q}_4\right]\\
&+A_{\text{tree}}\left[Q_1,1,2,\bar{Q}_2,Q_3,3,4,5,\bar{Q}_3,Q_4,6,\bar{Q}_4,Q_2,\bar{Q}_1\right]\\
&+A_{\text{tree}}\left[Q_1,1,2,\bar{Q}_2,Q_4,6,\bar{Q}_4,Q_2,\bar{Q}_1,Q_3,3,4,5,\bar{Q}_3\right]\\
&+A_{\text{tree}}\left[Q_1,1,2,\bar{Q}_2,Q_4,6,\bar{Q}_4,Q_3,3,4,5,\bar{Q}_3,Q_2,\bar{Q}_1\right]\Bigr)\,,
\end{split}
\end{equation}
which is multiplying $(T^{a_1}T^{a_2})_{i_1\bar{j}_2}\delta_{i_2\bar{j}_1}(T^{a_3}T^{a_4}T^{a_5})_{i_3\bar{j}_3}(T^{a_6})_{i_4\bar{j}_4}$. The one-loop partial amplitudes are implemented in the  function {\tt PartialAmplitudeLoop}, which generates the coefficients of the color structure {\tt C}. As explained at the beginning of this section, {\tt PartialAmplitudeLoop} contains the $N$ and $n_f$ dependence of the amplitude and is related to  the $N$, $n_f$ independent partial amplitude defined in \cref{eq:colorDecomposition1loop} by
\begin{equation}
\text{\tt PartialAmplitudeLoop[C]}= \begin{cases}
\left(\frac{-1}{N}\right)^p\bigr(P_3-\sfrac{n_f}{N}P^f_2\bigl)& \begin{gathered}[t]
                                                                 \text{if {\tt C} contains}\\\text{a trace}
                                                                \end{gathered}
\\
 \left(\frac{-1}{N}\right)^p\bigr(N(P_0-P_1)-\sfrac{1}{N}P_2+n_f(P^f_{0}-P^f_{1})\bigl)&\text{else}
 \end{cases}
\end{equation}

 The cyclic symmetry, reversion identity and the two term fermion flip and furry identities \cref{eq:cycSym,eq:refSym,eq:flip2,eq:FFId3,eq:furry1} are applied in order to canonicalize the arguments of the primitives generated by \cref{eq:P_0,eq:P_1,eq:P_2,eq:P_3,eq:P^f_0,eq:P^f_1,eq:P^f_2}, e.\,g.~the first argument of all primitives is always {\tt Q[1,L]}. Furthermore, tadpoles and loop corrections to massless legs are removed, \cref{eq:tadpoles}. This canonicalization of expressions of primitives is implemented in the function {\tt canonical[...]} and removes the most basic redundancies. \Cref{eq:P_0,eq:P_1,eq:P_2,eq:P_3,eq:P^f_0,eq:P^f_1,eq:P^f_2} are all implemented separately, the map of the function names is
\begin{equation}
\begin{aligned}
P_0\qquad&\rightarrow \qquad\text{\tt LOpartial}\\
P_1\qquad&\rightarrow \qquad\text{\tt CycleSplit}\\
P_2\qquad&\rightarrow\qquad \text{\tt PhotonLoop}\\
P_3\qquad&\rightarrow\qquad \text{\tt TracePart}\\
P_0^f\qquad&\rightarrow\qquad \text{\tt FermionLoop}\\
P_1^f\qquad&\rightarrow\qquad \text{\tt FermionLoopPhoton}\\
P_2^f\qquad&\rightarrow\qquad \text{\tt FermionLoopTrace}\,.
\end{aligned}
\end{equation}
Evaluation of
\begin{flushleft}
 \begin{tt}
  PartialAmplitudeLoop[\{\{Q[1],1,Qb[1]\},\{Q[2],2,Qb[2]\}\}]
 \end{tt}
\end{flushleft}
gives the partial amplitude
\begin{equation}
 \begin{split}
&-A\left[Q_1^L,1,2,Q_2^R,\bar{Q}_2^R,\bar{Q}_1^L\right]+A\left[Q_1^L,1,2,\bar{Q}_1^L,\bar{Q}_2^L,Q_2^L\right]-A\left[Q_1^L,1,Q_2^R,\bar{Q}_2^R,2,\bar{Q}_1^L\right]\\
&-A\left[Q_1^L,1,Q_2^R,\bar{Q}_2^R,\bar{Q}_1^L,2\right]-A\left[Q_1^L,1,\bar{Q}_1^L,2,Q_2^R,\bar{Q}_2^R\right]-A\left[Q_1^L,1,\bar{Q}_1^L,Q_2^R,\bar{Q}_2^R,2\right]\\
&-A\left[Q_1^L,1,\bar{Q}_1^L,\bar{Q}_2^L,2,Q_2^L\right]-A\left[Q_1^L,1,\bar{Q}_2^R,2,Q_2^R,\bar{Q}_1^L\right]-A\left[Q_1^L,2,1,Q_2^R,\bar{Q}_2^R,\bar{Q}_1^L\right]\\
&+A\left[Q_1^L,2,1,\bar{Q}_1^L,\bar{Q}_2^L,Q_2^L\right]-A\left[Q_1^L,2,Q_2^R,1,\bar{Q}_2^R,\bar{Q}_1^L\right]-A\left[Q_1^L,2,Q_2^R,\bar{Q}_2^R,1,\bar{Q}_1^L\right]\\
&+A\left[Q_1^L,Q_2^L,1,2,\bar{Q}_2^L,\bar{Q}_1^L\right]+A\left[Q_1^L,Q_2^L,2,1,\bar{Q}_2^L,\bar{Q}_1^L\right]-A\left[Q_1^L,Q_2^L,2,\bar{Q}_2^L,\bar{Q}_1^L,1\right]\\
&-A\left[Q_1^L,Q_2^R,1,\bar{Q}_2^R,2,\bar{Q}_1^L\right]-A\left[Q_1^L,Q_2^R,1,\bar{Q}_2^R,\bar{Q}_1^L,2\right]-A\left[Q_1^L,Q_2^R,\bar{Q}_2^R,1,2,\bar{Q}_1^L\right]\\
&-A\left[Q_1^L,Q_2^R,\bar{Q}_2^R,1,\bar{Q}_1^L,2\right]-A\left[Q_1^L,Q_2^R,\bar{Q}_2^R,2,1,\bar{Q}_1^L\right]-A\left[Q_1^L,\bar{Q}_2^R,2,Q_2^R,1,\bar{Q}_1^L\right]\\
&+\frac{1}{N_c^2}\Bigl(\begin{aligned}[t]&-A\left[Q_1^L,2,\bar{Q}_2^R,Q_2^R,\bar{Q}_1^L,1\right]+A\left[Q_1^L,Q_2^L,\bar{Q}_2^L,\bar{Q}_1^L,1,2\right]+A\left[Q_1^L,Q_2^L,\bar{Q}_2^L,\bar{Q}_1^L,2,1\right]\\
&-A\left[Q_1^L,Q_2^R,2,\bar{Q}_2^R,\bar{Q}_1^L,1\right]-A\left[Q_1^L,\bar{Q}_1^L,1,Q_2^R,2,\bar{Q}_2^R\right]-A\left[Q_1^L,\bar{Q}_1^L,Q_2^R,2,\bar{Q}_2^R,1\right]\\
&+A\left[Q_1^L,\bar{Q}_1^L,\bar{Q}_2^L,1,2,Q_2^L\right]+A\left[Q_1^L,\bar{Q}_1^L,\bar{Q}_2^L,2,1,Q_2^L\right]+A\left[Q_1^L,\bar{Q}_2^L,2,Q_2^L,\bar{Q}_1^L,1\right]\\
&-A\left[Q_1^L,\bar{Q}_2^R,Q_2^R,2,\bar{Q}_1^L,1\right]\Bigr)\end{aligned}\\
&-\frac{1}{N_c}n_f\Bigl(\begin{aligned}[t]&-A_f\left[Q_1^L,1,\bar{Q}_2^L,2,Q_2^L,\bar{Q}_1^L\right]-A_f\left[Q_1^L,2,\bar{Q}_2^L,Q_2^L,\bar{Q}_1^L,1\right]-A_f\left[Q_1^L,\bar{Q}_2^L,1,2,Q_2^L,\bar{Q}_1^L\right]\\
&-A_f\left[Q_1^L,\bar{Q}_2^L,2,1,Q_2^L,\bar{Q}_1^L\right]-A_f\left[Q_1^L,\bar{Q}_2^L,2,Q_2^L,1,\bar{Q}_1^L\right]-A_f\left[Q_1^L,\bar{Q}_2^L,2,Q_2^L,\bar{Q}_1^L,1\right]\\
&-A_f\left[Q_1^L,\bar{Q}_2^L,Q_2^L,2,\bar{Q}_1^L,1\right]-A_f\left[Q_1^L,\bar{Q}_2^L,Q_2^L,\bar{Q}_1^L,1,2\right]-A_f\left[Q_1^L,\bar{Q}_2^L,Q_2^L,\bar{Q}_1^L,2,1\right]\Bigr)\,,  \end{aligned}
 \end{split}
\end{equation}
which multiplies the color structure $(T^{a_1})_{i_1\bar{j}_1}(T^{a_2})_{i_2\bar{j}_2}$ within the four quark two gluon one-loop QCD amplitude.

The identities among the primitive amplitudes are implemented in the package {\tt QCDcolor} as well. The general reversion identity \cref{eq:furry2} for the mixed loop primitives {\tt A[...]} and the fermion loop primitives {\tt Af[...]} is implemented as 
{\tt RevID[l1,a1,A or Af]}. Modulo cyclic and reflection symmetry all other identities are linear combinations of reversion identities. {\tt l1} is a list of the form {\tt\{Q[1,L],Qb[i,L],...\}} or {\tt\{Qb[1,R],Q[i,R],...\}} and {\tt a1} is a list of external legs. In the case of the mixed loop primitives {\tt A[...]}, {\tt a1} has to be of the form {\tt\{...,Q[i,L],...,Qb[i,L],...\}} or {\tt\{...,Qb[i,R],...,Q[i,R],...\}}. The output is canonicalized. Two examples are
\begin{flushleft}
 \begin{tt}
  RevID[{Q[1,L],Qb[1,L],1},{2,3,4},Af] 
 \end{tt}
\end{flushleft}
 returning the identity
\begin{equation}
 \begin{split}
 & A_f\left[Q_1^L,2,3,4,\bar{Q}_1^L,1\right]+A_f\left[Q_1^L,2,3,\bar{Q}_1^L,1,4\right]+A_f\left[Q_1^L,2,3,\bar{Q}_1^L,4,1\right]\\
&+A_f\left[Q_1^L,3,4,\bar{Q}_1^L,1,2\right]+A_f\left[Q_1^L,3,4,\bar{Q}_1^L,2,1\right]-A_f\left[Q_1^L,4,3,2,\bar{Q}_1^L,1\right]
 \end{split}
\end{equation}
and
\begin{flushleft}
 \begin{tt}
  RevID[{Q[1,L],Qb[1,L],1},{2,Q[2,L],3,4,Qb[2,L]},A]
 \end{tt}
\end{flushleft}
 returning the identity
\begin{equation}
 \begin{split}
 & A\left[Q_1^L,2,Q_2^L,3,4,\bar{Q}_2^L,\bar{Q}_1^L,1\right]+A\left[Q_1^L,Q_2^L,3,4,\bar{Q}_2^L,\bar{Q}_1^L,1,2\right]+A\left[Q_1^L,Q_2^L,3,4,\bar{Q}_2^L,\bar{Q}_1^L,2,1\right]\\
&-A\left[Q_1^L,\bar{Q}_2^R,4,3,Q_2^R,2,\bar{Q}_1^L,1\right]\,.
 \end{split}
\end{equation}
{\tt FFId[Amplitude,flip quark,loop quark]} gives the fermion flip identity \cref{eq:FFId}
for the one loop amplitude  {\tt Amplitude} where the flip quark gets flipped with respect to the loop quark. For example
\begin{flushleft}
 \begin{tt}
FFId[A[Q[1,L],Qb[1,L],2,Q[2,R],Qb[2,R],1,Q[3,R],Qb[3, R]],Q[2,R],Q[1, L]]  
 \end{tt}
\end{flushleft}
returns the linear combination
\begin{equation}
 \begin{split}
&A\left[Q_1^L,\bar{Q}_1^L,2,Q_2^R,1,Q_3^R,\bar{Q}_3^R,\bar{Q}_2^R\right]+A\left[Q_1^L,\bar{Q}_1^L,2,Q_2^R,1,\bar{Q}_2^R,Q_3^R,\bar{Q}_3^R\right]\\
+&A\left[Q_1^L,\bar{Q}_1^L,2,Q_2^R,\bar{Q}_2^R,1,Q_3^R,\bar{Q}_3^R\right]+A\left[Q_1^L,\bar{Q}_1^L,2,\bar{Q}_2^L,Q_2^L,1,Q_3^R,\bar{Q}_3^R\right]\\
+&A\left[Q_1^L,\bar{Q}_1^L,Q_2^R,1,2,Q_3^R,\bar{Q}_3^R,\bar{Q}_2^R\right]+A\left[Q_1^L,\bar{Q}_1^L,Q_2^R,1,2,\bar{Q}_2^R,Q_3^R,\bar{Q}_3^R\right]\\
+&A\left[Q_1^L,\bar{Q}_1^L,Q_2^R,1,Q_3^R,2,\bar{Q}_3^R,\bar{Q}_2^R\right]+A\left[Q_1^L,\bar{Q}_1^L,Q_2^R,1,Q_3^R,\bar{Q}_3^R,2,\bar{Q}_2^R\right]\\
+&A\left[Q_1^L,\bar{Q}_1^L,Q_2^R,2,1,Q_3^R,\bar{Q}_3^R,\bar{Q}_2^R\right]+A\left[Q_1^L,\bar{Q}_1^L,Q_2^R,2,1,\bar{Q}_2^R,Q_3^R,\bar{Q}_3^R\right]\\
+&A\left[Q_1^L,\bar{Q}_1^L,Q_2^R,2,\bar{Q}_2^R,1,Q_3^R,\bar{Q}_3^R\right] \,,
 \end{split}
\end{equation}
which equals zero. An alternative implementation of \cref{eq:FFId} is given by {\tt FFId1[l1,a1,l2,a2]}, where {\tt l1} is the loop quark line {\tt l2} the flipped quark line and  {\tt a1},  {\tt a2} lists of external legs, e.\,g.~{\tt FFId1[{Q[1,L],Qb[1,L]},{2},{Qb[2,L],Q[2, L]},{1,Q[3,R],Qb[3,R]}]} reproduces the flip identity given above. {\tt FFId2[l1,b1]}  is an implementation of \cref{eq:FFId2}. {\tt l1} is the flipped fermion line and {\tt b1} is a list of external legs. The output is canonicalized. 
\begin{flushleft}
 \begin{tt}
FFId2[\{Q[1,L],Qb[1,L]\},\{2,Qb[2,L],Q[2,L]\}]
 \end{tt}
\end{flushleft}
will return the identity
\begin{equation}
\begin{split}
  &-A_f\left[Q_1^L,2,Q_2^R,\bar{Q}_2^R,\bar{Q}_1^L\right]+A_f\left[Q_1^L,2,\bar{Q}_2^L,Q_2^L,\bar{Q}_1^L\right]-A_f\left[Q_1^L,Q_2^R,\bar{Q}_2^R,2,\bar{Q}_1^L\right]\\
&+A_f\left[Q_1^L,\bar{Q}_2^L,Q_2^L,2,\bar{Q}_1^L\right]+2 A_f\left[Q_1^L,\bar{Q}_2^L,Q_2^L,\bar{Q}_1^L,2\right]\,
\end{split}
\end{equation}
{\tt FFId3[l1,a1,l2,a2]} is an implementation of a fermion flip identity \cref{eq:FFId3} that applies to the fermion loop primitives. The fermion line {\tt  l2} is flipped with respect to the fermion line {\tt l1}. {\tt a1} and {\tt a2} are lists of external legs. The quark routing have to be such that the loop in {\tt Af[l1,a1,l2,a2]} is in between {\tt l1} and {\tt l2}.  The output is canonicalized. 
 \begin{flushleft}
 \begin{tt}
FFId3[\{Q[1,R],Qb[1,R]\},\{2\},\{Qb[2,L],Q[2,L]\},\{1,Q[3,R],Qb[3,R]\}]
 \end{tt}
\end{flushleft}
will return the identity
\begin{equation}
\begin{split}
&-A_f\left[Q_1^L,Q_2^L,1,Q_3^R,\bar{Q}_3^R,\bar{Q}_2^L,\bar{Q}_1^L\right]-A_f\left[Q_1^L,\bar{Q}_2^L,\bar{Q}_3^L,Q_3^L,1,Q_2^L,\bar{Q}_1^L\right]\\
&-A_f\left[Q_1^L,\bar{Q}_3^L,Q_3^L,1,Q_2^R,\bar{Q}_2^R,\bar{Q}_1^L\right]-A_f\left[Q_1^L,\bar{Q}_3^L,Q_3^L,1,\bar{Q}_2^L,Q_2^L,\bar{Q}_1^L\right]\\
&-A_f\left[Q_1^L,\bar{Q}_3^L,Q_3^L,\bar{Q}_2^L,1,Q_2^L,\bar{Q}_1^L\right]\,.
\end{split}
\end{equation}
{\tt Furry[list]} gives the furry identity for a list of gluons and quark-lines. Gluons are specified by {\tt \{integer\}}, and a quark line by {\tt \{Q[i,R],...,Qb[i,R]\}}, or {\tt \{Qb[i,L],...,Q[i,L]\}}. The output is canonicalized. 
\begin{flushleft}
 \begin{tt}
Furry[\{\{Qb[1,L],Q[1,L]\},\{Q[2,R],2,Qb[2,R]\},\{1\}\}]
 \end{tt}
\end{flushleft}
returns the identity
\begin{equation}
\begin{split}
&A_f\left[Q_1^L,1,Q_2^R,2,\bar{Q}_2^R,\bar{Q}_1^L\right]+A_f\left[Q_1^L,Q_2^R,2,\bar{Q}_2^R,1,\bar{Q}_1^L\right]-A_f\left[Q_1^L,\bar{Q}_2^L,1,2,Q_2^L,\bar{Q}_1^L\right]\\&-A_f\left[Q_1^L,\bar{Q}_2^L,2,1,Q_2^L,\bar{Q}_1^L\right]+A_f\left[Q_1^L,\bar{Q}_2^L,2,Q_2^L,\bar{Q}_1^L,1\right]\,.
\end{split}
\end{equation}

\providecommand{\href}[2]{#2}\begingroup\raggedright\endgroup

\end{document}